\documentclass[a4paper,12pt]{article}
\usepackage[T1]{fontenc}
\usepackage[utf8]{inputenc}
\let\articleNormal\normalsize \let\articleSmall\small \let\articleFoot\footnotesize
\let\articleLarge\large \let\articleLARGE\Large
\usepackage{PRIMEarxiv}
\let\normalsize\articleNormal \let\small\articleSmall \let\footnotesize\articleFoot
\let\large\articleLarge \let\Large\articleLARGE \normalsize
\AtBeginDocument{\newgeometry{left=23mm,right=23mm,top=23mm,bottom=24mm,
  headheight=15pt,headsep=10pt,footskip=17pt}\raggedbottom}
\usepackage{amsmath,amssymb,amsthm}
\usepackage{array,booktabs,longtable,tabularx,graphicx}
\usepackage{microtype}
\usepackage{tikz,pgfplots}
\pgfplotsset{compat=1.18}
\usepgfplotslibrary{groupplots}
\usetikzlibrary{arrows.meta,shapes.geometric}
\usepackage[section]{placeins}
\usepackage{hyperref}
\newcommand{\PaperTitle}{Parameterized Complexity and Geometric Approximation of Riesz Energy Subset Selection}
\hypersetup{colorlinks=true,linkcolor=blue!42!black,citecolor=blue!42!black,
  urlcolor=blue!42!black,bookmarksnumbered=true,
  pdftitle={\PaperTitle},
  pdfauthor={Michael Emmerich and Soumen Atta},
  pdfsubject={Parameterized approximation, geometric subset selection and Riesz energy},
  pdfkeywords={Riesz energy, subset selection, computational geometry, parameterized approximation}}
\newcommand{\R}{\mathbb R}

\newcommand{\OPT}{\operatorname{OPT}}

\newcolumntype{Y}{>{\raggedright\arraybackslash}X}
\newtheorem{theorem}{Theorem}
\newtheorem{lemma}[theorem]{Lemma}
\newtheorem{proposition}[theorem]{Proposition}
\newtheorem{corollary}[theorem]{Corollary}
\title{\PaperTitle}
\author{\href{https://orcid.org/0000-0002-7342-2090}{Michael Emmerich}\qquad
        \href{https://orcid.org/0000-0001-7492-5602}{Soumen Atta}\\[4pt]
        \normalfont Faculty of Information Technology\\
        \normalfont University of Jyv\"askyl\"a, Finland}
\date{2 December 2026}
\begin{document}
\maketitle
\begin{abstract}
Given a finite set of $n$ sites or a compact domain such as a filled polygon, Riesz-energy subset selection seeks $k$ distinct points that minimize the sum of the inverse $s$th powers of their pairwise distances. For finite sets of sites, exact selection is polynomial on a line but NP-hard in the plane, already at $s=1$. For fixed dimension $d$ and exponent $s>0$, generic diversity proxies yield a deterministic efficient parameterized approximation scheme (EPAS) in $k$ and $1/\varepsilon$. We sharpen its geometric compression bound by averaging the cost of replacing one optimal site over feasible alternatives among $2k$ separated centers. For planar inverse-square energy, this improves the representative bound from $O(k^2\log(2k)/\varepsilon^2)$ under a total-energy packing analysis to $O(k\log(2k)/\varepsilon^2)$. In general, we retain $O_{d,s}(k(H_{2k}^{(s/d)})^{d/s}/\varepsilon^d)$ sites, where $H_m^{(p)}=\sum_{j=1}^m j^{-p}$, independently of $n$ and input spacing. A rational planar construction shows that the underlying separation bound is sharp even for one-swap local optima. For rational planar sites at $s=2$, a $(1+\varepsilon)$ approximation takes $O(nk+n\log n)+(C\log(2k)/\varepsilon^2)^k$ operations of polynomial bit cost. The scheme extends to metrics of bounded doubling dimension and to compact domains with effective geometric routines, including filled simple polygons. On metrics with distances $1$ and $2$, exact decision at $s=2$ is W[1]-complete in $k$, and a gap reduction excludes a deterministic EPAS unless $\mathrm{FPT}=\mathsf{W[1]}$. For rational planar sites at $s=2$, minimizing the energy after omitting interactions beyond distance $R$ gives a lower bound on the full optimum with additive error at most $\binom{k}{2}/R^2$. Synthetic benchmarks show how clustering and $\varepsilon$ affect the number of retained sites and the cost of search. We also evaluate refinement on a Swiss facility-dispersion instance and on triangulated domains shaped like numerals. On the numeral shaped domains with $2\le k\le10$ and $\varepsilon=1$, separate instance-specific bounds certify that each refined solution has energy at most $11.78\%$ above optimal.
\end{abstract}
\keywords{Riesz energy \and geometric subset selection \and parameterized approximation \and representative sets \and parameterized hardness \and certified refinement}
\section{Introduction}\label{sec:introduction}
Riesz energy models configurations of mutually repelling particles or points. It has applications, often in two or three dimensions, in facility location~\cite{atta2026,attaEmmerich2026}, in the representation of continuous Pareto fronts in multiobjective optimization~\cite{falcon2026,pereverdieva2025}, and in physical models of interacting particles~\cite{hardinSaff2005,borodachov2019}, including the Coulomb crystals studied by Navarro and Emmerich~\cite{navarroEmmerich2026}.

For a compact metric space $(P,d)$, exponent $s>0$ and cardinality $k$, we minimize the Riesz energy over all $k$-point subsets $S\subseteq P$. The energy and its minimum are given by
\begin{equation}\label{eq:problem}
 E_s(S)=\sum_{\{x,y\}\subseteq S}d(x,y)^{-s},
 \qquad
 \OPT_s(P,k)=\min_{\substack{S\subseteq P\\ |S|=k}} E_s(S).
\end{equation}

Riesz energy differs from other measures of diversity and dispersion in several respects. For a finite selection $S$, it sums contributions over \textit{all} unordered pairs of distinct points. Consequently, two selections that tie under max--min $p$-dispersion can have different Riesz energies~\cite[Proposition 1]{metric2026}. Dropping distant pairs changes the objective, and merging close sites risks a singularity. Moreover, $-E_s$ is submodular and nonincreasing, so the standard greedy approximation guarantee for nondecreasing submodular maximization does not apply~\cite{falconDiversity2025}. For a broad class of compact $d$-dimensional manifolds and $s>d$, Hardin and Saff showed that Riesz energy minimizers become equidistributed with respect to normalized $d$-dimensional Hausdorff measure as the number of selected points grows~\cite{hardinSaff2005,borodachov2019}.

Exact selection from a finite set is polynomial on lines~\cite{line2026} and ultrametrics~\cite{metric2026}. Two recent preprints establish hardness in the Euclidean plane. Emmerich~\cite[Theorem 1.1]{plane2026} proved that the decision problem on rational planar sites is NP-complete at $s=2$, and Navarro and Emmerich~\cite[Theorem 2]{navarroEmmerich2026} proved NP-hardness already at $s=1$. On continuous domains in $d \geq 2$, the selection problem poses a $k d$-dimensional non-convex optimization problem.

Generic diversity proxies already imply an EPAS; the issue is how many sites must remain. Our contribution is a stronger bound on the separation of an \emph{optimal selection}. Choose $2k$ farthest-first centers and let $r$ be their last insertion distance. Each of an optimal site's $k-1$ companions excludes at most one center closer than $r/2$. Averaging the feasible replacement costs bounds that site's incident energy by a harmonic sum (Lemma~\ref{lem:exchange}). In the plane at $s=2$, every optimum is $\Omega(r/\sqrt{\log(2k)})$-separated. A rational lattice square with a remote close pair attains this order for every one-swap local optimum (Theorem~\ref{thm:sharp-planar}).

Rounding retains $O(k\log(2k)/\varepsilon^2)$ original sites. A total-energy packing estimate instead gives $O(k^2\log(2k)/\varepsilon^2)$. At fixed accuracy the respective search bounds are $2^{O(k\log\log(2k))}$ and $2^{O(k\log k)}$. For $s>d$ the representative bound is linear in $k$ (Table~\ref{tab:regimes}). These are worst-case upper bounds, independent of $n$ and aspect ratio. The actual number $M$ of occupied cells is computed before search; $\binom Mk$ determines whether enumeration fits the budget. Sharpness of the separation lemma does not establish a lower bound on representative-set size.

\begin{table}[t]
\caption{Theorem~\ref{thm:exchange-epas}, for fixed $d,s$ and $0<\varepsilon\le1$. Add $O(nk+n\log n)$ preprocessing to the search bound. Constants depend on $d,s$. Bounds use the exact-operation model; rational inputs and fixed positive even $s$ give polynomial bit cost per operation.}\label{tab:regimes}
\centering\small
\begin{tabular}{lll}
\hline
Exponent & Retained candidates & Search over cardinality-$k$ subsets\\\hline
$0<s<d$ & $O_{d,s}(k^{d/s}/\varepsilon^d)$ & $(C_{d,s}k^{d/s-1}/\varepsilon^d)^k$\\
$s=d$ & $O_d(k\log(2k)/\varepsilon^d)$ & $(C_d\log(2k)/\varepsilon^d)^k$\\
$s>d$ & $O_{d,s}(k/\varepsilon^d)$ & $(C_{d,s}/\varepsilon^d)^k$\\\hline
\end{tabular}
\end{table}

Two antecedents explain both the scheme and the improvement. Farthest-first covering goes back to Gonzalez~\cite[Section 2]{gonzalez1985}. For a single input block, the proxy lemma of Ceccarello, Pietracaprina, Pucci and Upfal~\cite[Section 5, Lemma 5]{ceccarello2017} gives displacement at most $\varepsilon'\rho/2$, where $\rho$ is the best $k$-point minimum distance. A best-packing selection costs at most $N\rho^{-s}$, $N=\binom{k}{2}$, so every pair of a Riesz optimum is at distance at least $\rho/N^{1/s}$. Taking $\varepsilon'=\varepsilon/(2qN^{1/s})$, $q=\lceil\max(1,s)\rceil$, gives distinct proxies and the pairwise guarantee of Lemma~\ref{lem:rounding}. This already implies an EPAS. Our improvement comes from bounding the cost incident to \emph{one} site rather than the energy of the entire selection. For $s\ge d$, Hardin and Saff~\cite[Theorem 2.3, Section 6, equations (59)--(64)]{hardinSaff2005} average replacement costs over a region of positive measure. Here the averaging set consists of feasible input sites, even for an arbitrarily irregular finite input. The critical harmonic sum is classical~\cite[Lemma 3.1]{hardinSaff2005}. Continuous subcritical separation results impose regularity hypotheses~\cite[Theorems 2.3, 2.5]{hardinEtAl2019}; positive-power diversity schemes~\cite[Theorem 4]{cevallos2018} address a different kernel.

Atta's fixed-cardinality grid scheme~\cite[Sections 6.2--6.4]{atta2026} optimizes a capped objective in $O(n+g^k k^2)$ time for $g$ occupied cells. Its original-energy guarantee assumes minimum input spacing $\tau$ and uses cells of diameter below $\tau$; then $g=n$. Our scheme permits arbitrarily close candidates and bounds $g$ independently of $n$. It is an EPAS jointly in $k$ and $1/\varepsilon$. It does not give a PTAS polynomial in $n$ for fixed accuracy and unrestricted $k$.

A net gives the corresponding harmonic bound for finite metrics of fixed doubling dimension (Theorem~\ref{thm:doubling}). Without that restriction, exact decision on two distances is W[1]-complete, and the reduction gives a gap excluding an EPAS at $s=2$ unless $\mathrm{FPT}=\mathsf{W[1]}$ (Theorem~\ref{thm:metric-barrier}). For continuously movable sites, Corollary~\ref{cor:domains} requires routines to construct the cover and find feasible cell representatives. Filled simple polygons admit both.

Proofs of all main results are given below; Appendix~\ref{sec:supp-extra} collects complete details; Appendices~\ref{sec:swiss-density}--\ref{app:numerals} give the experimental protocols. Code, inputs and recorded results are available in the public \href{https://github.com/emmerichmtm/PCGA-Riesz}{PCGA-Riesz repository} and its \href{https://github.com/emmerichmtm/PCGA-Riesz/releases/tag/arxiv-v2}{versioned reproducibility release} (Appendix~\ref{app:reproduce}). Cutoff and binned proxy lower bounds are discussed briefly in Section~\ref{sec:experiments}, with the treewidth algorithm in Appendix~\ref{sec:structure-short}.

\section{Model and computational assumptions}\label{sec:model}
The $n$ candidates are distinct. Handle $k\le1$ and $k=n$ directly; replace $\varepsilon>1$ by $1$. Dimension $d$ and exponent $s>0$ are fixed. General-$s$ operation bounds use an exact-real model with arithmetic, comparisons, positive powers and roots, and integer floor as primitives. They do not automatically imply bit bounds for arbitrary real exponents.

For rational Euclidean coordinates and fixed positive even $s$, all weights are rational. The algorithm below uses rational arithmetic, comparisons, binary lengths, and integer division, with polynomial bit cost in total input length $L$ (including rational accuracy). It avoids constructing all $\binom n2$ input weights. A deterministic EPAS returns a feasible $(1+\varepsilon)$ approximation in $f(k,1/\varepsilon)L^c$ bit time for an absolute constant $c$. A bound $n^{g(k)}$ establishes only XP. Our representative set contains at least one near-optimal selection for the requested $k$; it need not preserve every subset or bound coordinate encoding length. It is not a bounded-bit kernel.

For exact decision, the input also includes a rational threshold. We use parameterized reductions and the classes $\mathrm{FPT}\subseteq\mathsf{W[1]}\subseteq\mathsf{W[P]}$~\cite{cygan2015}. W[1]-hardness provides conditional evidence against exact FPT; completeness additionally requires membership. Approximation hardness will require a separate gap argument.
\section{Geometric compression by oversampling and exchange}\label{sec:epas}
Put $H_m^{(p)}=\sum_{j=1}^m j^{-p}$. Throughout this section $d\ge1$ and $s>0$ are fixed, $2\le k<n$, and $0<\varepsilon\le1$.
\begin{theorem}[Fixed-dimensional parameterized approximation]\label{thm:exchange-epas}
For $P\subset\R^d$, a deterministic algorithm retains $Q\subseteq P$ with
\begin{equation}\label{eq:exchange-size}
 k\le M:=|Q|\le n,\qquad
 M=O_{d,s}\!\left(\frac{k\bigl(H_{2k}^{(s/d)}\bigr)^{d/s}}{\varepsilon^d}\right),
\end{equation}
and computes its exact cardinality-$k$ optimum $V$ using
\begin{equation}\label{eq:exchange-time}
 O(nk+n\log n)+
 \left(\frac{C_{d,s}\bigl(H_{2k}^{(s/d)}\bigr)^{d/s}}{\varepsilon^d}\right)^k
\end{equation}
operations in the model of Section~\ref{sec:model}. The returned original candidates satisfy
\begin{equation}\label{eq:epas-certificate}
 V/(1+\varepsilon)\le\OPT_s(P,k)\le V.
\end{equation}
No input separation, bounded aspect ratio, or partition is required. For rational coordinates and fixed positive even $s$, each operation has polynomial bit cost.
\end{theorem}
The proof is given in Section~\ref{sec:rounding-proof}, using the exchange and rounding lemmas below.

\subsection{A finite set of feasible replacements}
If $n\le2k$, retain $P$ and enumerate directly; this fits the theorem because $\binom nk\le4^k$. Hence suppose $n>2k$. Choose $2k$ centers $C\subset P$ by farthest-first traversal~\cite{gonzalez1985}, starting arbitrarily, and let $r>0$ be the last insertion distance. Insertion radii are nonincreasing, so $C$ is $r$-separated. Before the final insertion all candidates are within $r$ of the previous centers; thus the radius-$r$ balls around $C$ cover $P$. Maintaining nearest-center distances takes $O(nk)$ operations. Squared rational distances suffice for all comparisons.

\begin{lemma}[Replacement potential and separation]\label{lem:exchange}
Let a feasible set $P$ contain the $2k$ mutually $r$-separated centres, and let $S$ be a cardinality-$k$ subset admitting no improving one-point replacement from $P$. For every $x\in S$, $\sum_{y\in S\setminus\{x\}}\|x-y\|^{-s}\le4^s H_{2k}^{(s/d)}/r^s$.
In particular, every optimum is separated by
\begin{equation}\label{eq:exchange-separation}
 \delta=\frac{r}{4\bigl(H_{2k}^{(s/d)}\bigr)^{1/s}}.
\end{equation}
\end{lemma}
\begin{proof}
Fix $x$, put $A=S\setminus\{x\}$, and let
$T_x=\{c\in C:\|c-y\|\ge r/2\text{ for all }y\in A\}$. Each $y$ excludes at most one center at distance \emph{strictly less than} $r/2$, since two such centers would be less than $r$ apart. Therefore $|T_x|\ge k+1$. Every $c\in T_x$ lies outside $A$ and is a feasible replacement; $c=x$ is harmless. With $\Psi_A(z)=\sum_{y\in A}\|z-y\|^{-s}$, replacement optimality gives $\Psi_A(x)\le\Psi_A(c)$.

For fixed $y\in A$, sort the free-center distances as $r/2\le\rho_1\le\cdots\le\rho_{|T_x|}$. The disjoint open radius-$r/2$ balls about the first $j$ centers lie in the radius-$(\rho_j+r/2)$ ball about $y$. Volume comparison yields $j\le(1+2\rho_j/r)^d\le(4\rho_j/r)^d$, hence
$\rho_j^{-s}\le4^s r^{-s}j^{-s/d}$. The point $y$ need not be a center. Averaging over the feasible replacements gives
\begin{align*}
 \Psi_A(x)&\le\frac1{|T_x|}\sum_{c\in T_x}\sum_{y\in A}\|c-y\|^{-s}\\
 &\le\frac{k-1}{|T_x|}\,4^s r^{-s}H_{2k}^{(s/d)}
 \le4^s r^{-s}H_{2k}^{(s/d)}.
\end{align*}
Every individual pair term is at most this potential, proving~\eqref{eq:exchange-separation}.
\end{proof}
The averaging removes the factor $k$ lost by bounding one pair with the entire incumbent energy. Only one-swap optimality is used: the algorithm need not run local search. The lemma bounds separation, not the energy ratio or convergence time of local search. Improving one-point replacements also underlie the practical heuristic of~\cite[Algorithm 3]{falcon2026}; that algorithm has no approximation bound.

\begin{theorem}[Sharpness of planar separation]\label{thm:sharp-planar}
For every sufficiently large $k$ there is a finite rational $P_k\subset\mathbb R^2$ such that,
for any start and tie-breaking of farthest-first traversal, every one-swap local optimum
for $E_2$, and hence every global optimum, has minimum distance
$\Theta(r_{2k}/\sqrt{\log k})$. Here $r_{2k}$ is the last insertion distance among $2k$ centers;
the constants are absolute.
\end{theorem}
\begin{proof}[Proof sketch]
First, $m\ge4$ points in a square of side $a$ have energy
$\Omega(m^2\log m/a^2)$. Indeed, a nested quadtree has at least
$N_j\ge m^2/(4\cdot4^j)$ same-cell pairs at every level
$0\le j\le J=\lfloor\log_4(m/2)\rfloor$, by Cauchy--Schwarz.
A pair whose last common cell is at level $\ell$ receives total weight
$\sum_{j=0}^{\ell}4^j/a^2\le(8/3)\|x-y\|^{-2}$. Hence
$E_2\ge(3/8)\sum_{j=0}^J4^jN_j/a^2=\Omega(m^2\log m/a^2)$.

Let $a=2\lceil\sqrt{k}\rceil$ and $G=\{0,\ldots,a-1\}^2$.
Adjoin $u=(k^2,0)$ and $v=(k^2+t,0)$, choosing dyadic
$t\in[A/\sqrt{\log k},2A/\sqrt{\log k}]$ for a fixed sufficiently large $A$.
For large $k$, $t<1$. Some unchosen lattice site stays at distance at least one
throughout the first $2k$ insertions, so $r_{2k}\ge1$. At least $2k-2$ centers
are in $G$. Its $\lceil\sqrt{k}\rceil^2<2k-2$ disjoint blocks of $2$ by $2$
sites force two centers into one block, so $r_{2k}\le\sqrt2$, regardless of
the start and tie-breaking.

A $k$-set omitting a remote point contains $m\ge k-1$ lattice sites. The square
bound, $a^2=O(k)$ and averaging give one lattice site with incident energy at
least $c_0\log k$. Moving it to an omitted remote point costs at most
$t^{-2}+O(k^{-3})\le A^{-2}\log k+O(k^{-3})$.
Choose $A^{-2}<c_0/2$; this move strictly improves the energy. Thus every
one-swap local optimum contains both remote points and has minimum distance
$t=\Theta(r_{2k}/\sqrt{\log k})$. Appendix~\ref{app:sharp-planar} supplies the
full counting details.
\end{proof}
This proves sharpness of the separation lemma, not a lower bound on
representative-set size or on the running time of other approximation schemes.

\subsection{Injective rounding and exact search}\label{sec:rounding-proof}
\begin{lemma}[Rounding an optimum]\label{lem:rounding}
Suppose every pair of an optimum has distance at least a known $\sigma>0$. Let $q\ge\max\{1,s\}$ be a fixed integer, choose $\sigma/2<\widehat\sigma\le\sigma$, and grid $\R^d$ into half-open cubes of side $h=\varepsilon\widehat\sigma/(4qd)$. Retain one feasible point from each cell intersecting the feasible set. The resulting $Q$ contains a $(1+\varepsilon)$-approximate $k$-subset. A cover by $c$ sets of diameter at most $a$ gives $|Q|\le c(2+a/h)^d$.
\end{lemma}
\begin{proof}
Cell diameter is $\sqrt d\,h<\sigma$, so snapping an optimum to retained candidates is injective. A selected distance $D\ge\sigma$ decreases by at most $2\sqrt d\,h\le\varepsilon D/(2q)$. For $s\ge1$, Bernoulli's inequality gives $(1-\varepsilon/(2q))^s\ge1-\varepsilon/2$; for $s<1$, use $(1-z)^s\ge1-z$. Since $1-\varepsilon/2\ge(1+\varepsilon)^{-1}$ for $0\le\varepsilon\le1$, every pair energy increases by at most $1+\varepsilon$. Sum over pairs. Each diameter-$a$ set has coordinate ranges at most $a$ and meets at most $(2+a/h)^d$ cells.
\end{proof}

\begin{proof}[Proof of Theorem~\ref{thm:exchange-epas}]
Apply Lemma~\ref{lem:rounding} with $\sigma=\delta$ from Lemma~\ref{lem:exchange}. The $2k$ covering balls have diameter $2r$, so
$M\le2k(2+2r/h)^d=O_{d,s}(k(H_{2k}^{(s/d)})^{d/s}/\varepsilon^d)$.
In the exact real model we may take $\widehat\sigma=\sigma$; rational implementation is described below. Sort cell keys in $O(n\log n)$ operations. Evaluate each of the $\binom Mk$ subsets in $O(k^2)$ operations and retain the best. The bound $\binom Mk\le(eM/k)^k$, with $k^2\le4^k$, proves~\eqref{eq:exchange-time}. Optional precomputation of representative pair weights is absorbed by this bound. Since $Q\subseteq P$ and the snapped optimum is feasible in $Q$, equation~\eqref{eq:epas-certificate} follows. Any additional feasible solution of energy $U$ can improve the upper endpoint to $\min\{U,V\}$.

For rational coordinates and positive even $s$, write $m=2k$, $b=\lceil\log_2m\rceil$, and use the rational upper bound
\begin{equation}\label{eq:h-surrogate}
 \overline H=
 \begin{cases}
 \dfrac{d}{d-s}\,2^{\lceil b(d-s)/d\rceil},&s<d,\\
 1+b,&s=d,\\
 \dfrac{s}{s-d},&s>d.
 \end{cases}
\end{equation}
Integral bounds give $H_m^{(s/d)}\le\overline H=O_{d,s}(H_m^{(s/d)})$. Thus $B=4^s\overline H/(r^2)^{s/2}$ is rational and $\sigma=B^{-1/s}$ is a valid separation scale with the same asymptotic bounds. Binary lengths and a constant number of rational tests find a dyadic $\widehat\sigma$ satisfying $B\widehat\sigma^s\le1<B(2\widehat\sigma)^s$. No root is computed. Grid indices use integer division. All coordinates, keys and sums of at most $\binom{k}{2}$ rational pair weights have polynomial bit length. Appendix~\ref{sec:epas-dyadic} gives the harmonic and dyadic arithmetic.
\end{proof}

\begin{corollary}[Planar inverse-square scheme]\label{thm:k-epas}
For rational planar input and $s=2$, the scheme retains $O(k\log(2k)/\varepsilon^2)$ candidates and uses $O(nk+n\log n)+(C\log(2k)/\varepsilon^2)^k$
operations of polynomial bit cost, for an absolute constant $C$.
\end{corollary}
The three regimes in Table~\ref{tab:regimes} follow from $H_m^{(p)}=O_p(m^{1-p})$ for $p<1$, $O(\log m)$ for $p=1$, and $O_p(1)$ for $p>1$. At fixed accuracy, planar $s=2$ has search bound $2^{O(k\log\log(2k))}$; $s>d$ has $2^{O_{d,s}(k)}$. These remain exponential searches, not unrestricted PTASs. The bounds on separated-set energy may be order-tight without making the representative bounds optimal: the construction still uses a uniform rounding scale.

\subsection{Comparing planar separation certificates}\label{sec:comparison}
Let $C_k$ be the first $k$ farthest-first centers, let $r_k$ be their last insertion radius, and put $U=E_2(C_k)$ and $N_k=\binom{k}{2}$. Any $B\ge U$ certifies optimum separation $B^{-1/2}$. Table~\ref{tab:comparison} lists explicit choices; $r_{2k}$ is the radius used by Lemma~\ref{lem:exchange}.
\begin{proposition}[Planar comparison of certificates]\label{prop:comparison}
The pair-count and incumbent-packing analyses give $O(k^3/\varepsilon^2)$ and $O(k^2\log(2k)/\varepsilon^2)$ representatives, respectively. The exchange analysis gives $O(k\log(2k)/\varepsilon^2)$. All preserve a $(1+\varepsilon)$ approximation to the original objective.
\end{proposition}
\begin{proof}
The $r_k$-separated incumbent satisfies $U\le N_k/r_k^2$. For a center's $j$th nearest companion at distance $d_j$, packing gives $j+1\le(1+2d_j/r_k)^2$, hence $j\le8(d_j/r_k)^2$ because $d_j\ge r_k$. Summing over companions and dividing by two yields $U\le4kH_{k-1}/r_k^2$. The $k$ covering disks and Lemma~\ref{lem:rounding} give $M=O(k r_k^2 B/\varepsilon^2)$ in either case. Lemma~\ref{lem:exchange} instead bounds one point's potential, yielding the third row. Enumeration uses $\binom Mk\le(eM/k)^k$.
\end{proof}
\begin{table}[t]\centering\small
\caption{Explicit planar certificates and bounds. The first two rows are baselines derived here from packing and proxy reasoning, not Riesz theorems attributed to prior papers. $H_j=\sum_{i=1}^j1/i$. $B$ is an upper bound on every optimum pair contribution.}\label{tab:comparison}
\begin{tabular}{@{}lll@{}}\toprule
Certificate & $B$ & Representative upper bound\\\midrule
Pair count & $N_k/r_k^2$ & $O(k^3/\varepsilon^2)$\\
Incumbent packing & $\min\{N_k,4kH_{k-1}\}/r_k^2$ & $O(k^2\log(2k)/\varepsilon^2)$\\
Finite exchange & $16H_{2k}/r_{2k}^2$ & $O(k\log(2k)/\varepsilon^2)$\\\bottomrule
\end{tabular}\end{table}
Let $B_X=16H_{2k}/r_{2k}^2$, and let $\overline U\ge U$ be any evaluated incumbent upper bound. Use $B_H=\min\{\overline U,B_X\}$ in the same dyadic rounding rule. Both bounds are valid. Since $B_H\le B_X$, the width is no smaller than the exchange width; the $2k$ covering disks give the same asymptotic guarantee. With a common physical origin, the dyadic grids are nested, so the hybrid retains at most the smaller representative count. Returned energies need not be ordered. Computing $U$ costs $O(k^2)$ operations and does not change the theorem's bound. Appendix~\ref{app:nesting} proves nesting and gives a bounded-denominator evaluation of $\overline U$.

For fixed $d,s$, the EPAS is parameterized by $k$ and $1/\varepsilon$. Sorting cell keys gives the observed count $M$ before subset search. With a budget of $T$ subset evaluations, enumerate if $\binom Mk\le T$; each sums $\binom{k}{2}$ pair weights. Pruning can make larger counts tractable, as in the Swiss example. A completed exact solve gives~\eqref{eq:epas-certificate}.

The certified cell side $h$ sets the scale at which clustering matters. If each of $c$ clusters lies in a distinct single cell, then $M=c$. When $c$ is close to $k$, this leaves few extra representatives. A cluster of diameter below $h$ can still meet up to $2^d$ cells at grid boundaries. For $n>2k$ the exchange grid occupies at least $2k$ cells because its cell diameter is smaller than the separation $r$ of its $2k$ centres. With fewer than $2k$ tight groups, traversal must enter some group twice and can produce a very fine grid. The hybrid in Section~\ref{sec:comparison} can give $M=k$, leaving a single reduced choice. In a uniformly sampled region with $G$ cells of positive sampling probability, $M\le\min\{n,G\}$. For a fixed grid, dense sampling fills an increasing fraction of those cells.

\section{Extensions of the geometric approximation scheme}\label{sec:extensions}
\begin{corollary}[Compact feasible domains]\label{cor:domains}
Let $D\subset\mathbb R^d$ be compact. Suppose one can compute $2k$ feasible
centres at mutual distance at least $r>0$ whose closed radius-$r$ balls cover
$D$, and can test a half-open grid cell for intersection with $D$ and return a
feasible witness when it intersects. In the exact-operation model, retaining
one witness per intersecting cell gives a $(1+\varepsilon)$-approximation with
the representative bound~\eqref{eq:exchange-size} (omitting $M\le n$).
The cost is $T_{\rm cert}+T_{\rm grid}+O(k^2\binom Mk)$, where $T_{\rm grid}$
includes scale computation, cell enumeration, and witness queries.
\end{corollary}
\begin{proof}
On $D^k$ extend the energy to $+\infty$ at collisions. It is lower semicontinuous
and some feasible tuple has finite energy, so its minimum is attained at distinct
points. The given centres are feasible replacements in Lemma~\ref{lem:exchange},
which uses only their separation. Lemma~\ref{lem:rounding} applies to the
witness in each optimal point's cell. Enumerate the grid keys in the axis-aligned
bounding boxes of the $2k$ covering balls, remove repeated keys, and query the
remaining cells. Each box yields at most $(2+2r/h)^d$ keys. The ratio $r/h$ is
$O_{d,s}((H_{2k}^{(s/d)})^{1/s}/\varepsilon)$, so the absolute scale cancels
from the number of cell queries. For $s=d=2$ the algorithm queries
$O(k\log(2k)/\varepsilon^2)$ cells, followed by at most
$(C\log(2k)/\varepsilon^2)^k$ operations for subset search.
\end{proof}
In a closed simple polygon with $m$ vertices, a continuous farthest-first step
with $j$ sites maximizes the empty-circle radius subject to its centre lying in
the polygon. Chew and Drysdale's Voronoi algorithm uses $O(mj+j\log(2j))$
geometric operations for this step~\cite{chew1986}. A triangulation supplies
cell witnesses by clipping triangles to the closed cell and testing the excluded
upper faces. Appendix~\ref{app:polygons} gives the boundary tests and coordinate bit costs. For
polyhedral domains in higher dimensions the corollary applies with the costs of
the supplied certificate and witness routines.

\begin{theorem}[Finite metrics of bounded doubling dimension]\label{thm:doubling}
Let $P$ be an $n$-point metric in which every closed radius-$R$ ball is
covered by at most $2^\lambda$ closed radius-$R/2$ balls with centres in $P$.
For fixed $\lambda\ge1$ and $s>0$, an exact distance oracle and the stated
power and comparison primitives suffice for a deterministic EPAS. It retains $M=O_{\lambda,s}(k(H_{2k}^{(s/\lambda)})^{\lambda/s}/\varepsilon^\lambda)$ input sites. Centre and net construction take $O(nk+nM)$ distance queries.
Enumeration takes $O(k^2\binom Mk)$ distance and arithmetic operations.
\end{theorem}
\begin{proof}
For $n\le2k$ enumerate the input directly. Otherwise use its $2k$ farthest-first
centres. For an $a$-separated set in a closed radius-$R$ ball, if $R<a/2$
there is at most one point. If $R\ge a/2$, iterate the doubling cover
$\ell=\lfloor\log_2(2R/a)\rfloor+1$ times. Each resulting ball has diameter
strictly below $a$, including when $2R/a$ is a power of two. The number of
points is at most $2^{\lambda\ell}\le(4R/a)^\lambda$.

Fix an optimal site $x$ and its $k-1$ companions. Their open radius-$r/2$ balls
exclude at most $k-1$ centres. For each companion the $j$th nearest remaining
centre has distance $\rho_j\ge r/2$, so $j\le(4\rho_j/r)^\lambda$.
Averaging the exchange inequalities over the at least $k+1$ remaining centres,
as in Lemma~\ref{lem:exchange}, bounds the incident potential by
$4^sr^{-s}H_{2k}^{(s/\lambda)}$. Thus every optimal pair has distance at least
$\delta=r/[4(H_{2k}^{(s/\lambda)})^{1/s}]$.

Put $\eta=\varepsilon\delta/(4q)$ with integer $q\ge\max(1,s)$. Scan $P$,
retaining a site when all earlier representatives are farther than $\eta$.
Every site has a representative within $\eta$. Rounding an optimal pair at
distance $b\ge\delta$ leaves distance at least $b-2\eta\ge
(1-\varepsilon/(2q))b>0$. This proves injectivity and, by the scalar inequality
in Lemma~\ref{lem:rounding}, the energy guarantee. The $\eta$-separated
representatives in each covering ball number at most $(4r/\eta)^\lambda$.
Summing over the $2k$ balls proves the bound. A scan compares each input site
with at most $M$ representatives.
\end{proof}
For $s<\lambda$ the count is $O_{\lambda,s}(k^{\lambda/s}/\varepsilon^\lambda)$.
At $s=\lambda$ it is $O_\lambda(k\log(2k)/\varepsilon^\lambda)$.
For $s>\lambda$ it is $O_{\lambda,s}(k/\varepsilon^\lambda)$.
Here $\lambda$ bounds the doubling
dimension of the metric supplied to the oracle. Section~\ref{sec:epas} instead
uses ambient Euclidean volume. The net is the familiar diversity proxy
construction~\cite[Section 5]{ceccarello2017}. The energy-specific step is the
bound on the potential of an optimal site, which determines how far a proxy
may move while preserving the entire inverse-power sum.

\section{General-metric parameterized complexity}\label{sec:metric-barrier}
\begin{theorem}[Exact classification and approximation barrier]\label{thm:metric-barrier}
At $s=2$, exact Riesz selection on metrics with nonzero distances in $\{1,2\}$ is W[1]-complete parameterized by $k$. Unless $\mathrm{FPT}=\mathsf{W[1]}$, even this restriction has no deterministic $(1+\varepsilon)$ approximation in $f(k,1/\varepsilon)L^{O(1)}$ time.
\end{theorem}
\begin{proof}
Use Atta's two-distance reduction~\cite[Section 5]{atta2026}, equivalently the complementary-clique construction in~\cite[Theorem 1]{metric2026}. For a graph $G$, assign distance $1$ on edges and $2$ on nonedges. Every positive distance is at most the sum of two positive distances, so this is a metric. For $N_k=\binom{k}{2}$ and a $k$-set $S$, $E_2(S)=N_k/4+(3/4)e_G(S)$. The objective is an affine transformation of Sparsest $k$-Subgraph, whose general-graph W[1]-hardness is documented in~\cite{bougeret2014}. Thus threshold $B=N_k/4$ is attainable exactly when $G$ has a $k$-independent set. The reduction preserves $k$, proving W[1]-hardness.

For membership at any rational threshold, enumerate the $2^{N_k}$ labeled distance patterns on $[k]$ and discard those exceeding the threshold. For each remaining pattern, form a $k$-partite graph with vertices $(i,u)$, $i\in[k]$, $u\in P$. Join $(i,u)$ and $(j,v)$ precisely when $i\ne j$, $u\ne v$, and their input distance matches the pattern on $ij$. A $k$-clique realizes the pattern with distinct candidates. The disjoint union of these graphs reduces the problem to $k$-Clique in $f(k)L^{O(1)}$ time, with unchanged parameter. For $k\ge2$ a clique lies in one component; smaller cases are direct. This proves W[1] membership.

For the approximation barrier, assume $k\ge2$ and set $\varepsilon=1/N_k$. A yes-instance has optimum $B$, so a $(1+\varepsilon)$ algorithm returns energy at most $B+1/4$. A no-instance has every feasible energy at least $B+3/4$. Comparing the returned energy with $B+1/2$ decides Independent Set. Because $1/\varepsilon=N_k$ depends only on $k$, the assumed EPAS would be an FPT decision algorithm.
\end{proof}

For arbitrary rational weights, guessing $k$ indices uses $O(k\log n)$ nondeterministic bits and permits polynomial-time exact verification. The bounded-nondeterminism characterization gives W[P] membership~\cite{cygan2015}; it does not give W[1] membership when weight precision is unrestricted. Table~\ref{tab:complexity-v2} separates these statements.

\begin{table}[t]\centering\small
\caption{Subset-size complexity at $s=2$. Approximation uses the joint parameter $k+1/\varepsilon$; exact decision uses $k$. Rational encodings are assumed.}\label{tab:complexity-v2}
\begin{tabular}{@{}p{.27\linewidth}p{.30\linewidth}p{.35\linewidth}@{}}\toprule
Setting & Exact decision & Approximation\\\midrule
Metrics with distances $1,2$ & W[1]-complete & No EPAS unless FPT = W[1]\\
General rational metrics & W[1]-hard; in W[P] & Same obstruction\\
Bounded doubling (rational distances) & In W[P] & EPAS for fixed $\lambda$, Theorem~\ref{thm:doubling}\\
Rational Euclidean plane & In W[P]; exact FPT status not settled here & EPAS, Corollary~\ref{thm:k-epas}\\\bottomrule
\end{tabular}
\end{table}

The relative gap is $3/N_k$, so this proof does not exclude constant-factor approximation or a scheme whose input-size exponent depends on accuracy. The metric of an edgeless input graph has all distances two: a radius-two ball can contain all $n$ sites, while its radius-one subballs are singletons. Thus the reduction family has unbounded doubling dimension, as required to coexist with Theorem~\ref{thm:doubling}. In particular, the reduction is not planar parameterized hardness. Conversely, an EPAS does not decide an exact planar threshold at precision depending only on $k$; rational energy gaps can depend on input bit length. The underlying reduction and graph problem are prior work; the classification and parameterized gap delimit the geometric approximation theorem.

\section{Experiments and geometric examples}\label{sec:experiments}
\label{sec:refine}
\begin{proposition}[EPAS-Refine]\label{prop:refine}
Starting at the exact reduced solution of energy $V$, visit the selected cells
in a fixed order. In each, choose an original candidate of minimum potential
against the other current sites, retaining the current site on a tie.
After $t$ complete sweeps,
$V/(1+\varepsilon)\le\OPT\le U_t\le U_{t-1}\le\cdots\le U_0=V\le(1+\varepsilon)\OPT$.
Grouped cell membership gives $O(tnk)$ additional arithmetic operations.
\end{proposition}
\begin{proof}
The $k$ selected cells are distinct. Replacing one site changes precisely its
$k-1$ incident terms, whose sum is minimized over a set containing its current
position. Feasibility and the upper bound follow by induction. Within a sweep
the selected cells are disjoint, so at most $n$ candidates each require $k-1$
pair evaluations. The lower bound follows from~\eqref{eq:epas-certificate}.
\end{proof}

We measure occupancy, exact reduced search and within-cell refinement at $d=s=2$ and $\varepsilon\in\{1/2,1\}$ (energy factors $1.5$ and $2$). The Swiss sites use whole-metre national coordinates and the synthetic benchmarks use dyadic coordinates. We count integer cell keys before allocating pair weights and use rational branch and bound. Swiss refinement compares exact fractions. On synthetic benchmarks, outward-rounded intervals precede exact comparisons. Appendices~\ref{app:exact-search} and~\ref{sec:paired-protocol} give arithmetic details and 157 independent checks.

\begin{figure}[tbp]
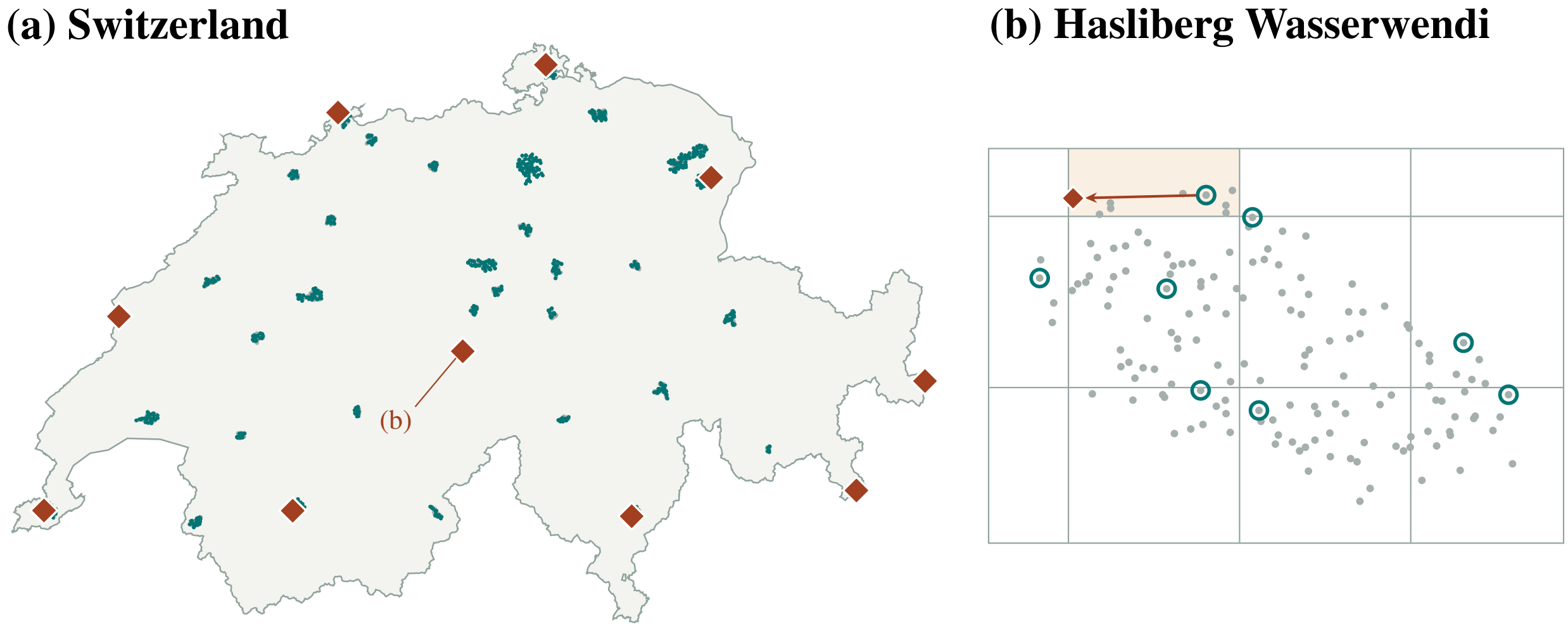
\centering
\resizebox{\linewidth}{!}{
}

\caption{Ten facilities selected from 5,403 synthetic Swiss sites ($s=2$, $\varepsilon=1$).
(a) Grey points are candidates, teal points are the 706 hybrid grid representatives,
and rust diamonds are the refined facilities, verified optimal on the full input.
(b) At Hasliberg Wasserwendi, 150 candidates occupy eight cells of side 1,024~m;
rings mark their representatives. The arrow shows the replacement within the shaded
cell. Settlement geometry and outline: swisstopo~\cite{swissaddresses,swissboundary}.}\label{fig:city-v2}
\end{figure}

Figure~\ref{fig:city-v2} selects $k=10$ facilities from 5,403 synthetic sites.
The clouds follow 26 cantonal seats and ten rural settlements; a singleton near
the eastern border represents Müstair. Candidates were made denser within
settlements before constructing either grid. At $\varepsilon=1$, the exchange
grid retains 3,365 sites (cell side 256~m). The hybrid instead retains one
representative per occupied 1,024~m cell, leaving 706 sites ($86.93\%$ removed).
Among $\binom{706}{10}\approx7.95\times10^{21}$ selections, diameter bounds exclude two sites in one locality. Interaction lower bounds leave only 69 sets of ten localities for exact search.

The hybrid representative optimum is $0.32553\%$ above the full optimum.
Eight within-cell replacements recover the latter exactly, at energy
$2.81234821\times10^{-9}\,\mathrm{m}^{-2}$. One move selects a site
in Hasliberg Wasserwendi, 5.87~km from the national area centroid, without a
central-coverage constraint. The EPAS guarantees a factor of two here; exact
recovery is an observation proved by a separate full-input search. Appendix~\ref{sec:swiss-density} gives the construction, exact audit, and executable experiment.

Could the small Swiss gap be certified without solving the full instance? One route is to compute a lower bound on the full optimum. For $s=2$, a cutoff at $R>0$ drops the cost of distant pairs. Its optimum is a lower bound with additive error at most $\binom{k}{2}/R^2$: at most $\binom{k}{2}$ omitted pairs each contribute at most $R^{-2}$. A stepwise lower envelope instead rounds distances up to bin endpoints. With a diameter upper bound $D$ as the last endpoint, every selection contributes the same baseline $\binom{k}{2}D^{-s}$; only pairs with a positive residual need edges. Given a width-$w$ decomposition of this graph, the dynamic program in Appendix~\ref{sec:structure-short} computes the proxy optimum $B$, including the baseline, in $2^{O(w)}n(k+1)^2$ operations after preprocessing. To see this, use a nice
decomposition and store the selected bag vertices and number of selected forgotten
vertices. Charge each residual edge once at its introduce-edge node; at joins,
convolve the forgotten counts for the same bag selection. The empty-root state
of count $k$ gives $B$ after restoring the baseline. For reduced optimum $V$ and any feasible energy $U$, the bound $L_*=\max\{B,V/(1+\varepsilon)\}$ certifies $L_*\le\OPT\le U$. Thus $U\le1.01L_*$ would certify one per cent accuracy on that instance. Choosing bins that give a useful bound at small width remains an open problem.

With $n=10^6$, $k=5$ and $\varepsilon=1/2$, the exchange grid had side $h=1/256$ on all six inputs. Each input with ten tight lattice clusters gave $M=10$ and $\binom{10}{5}=252$ subsets. Median preprocessing and exact search took $0.377$~s and $0.00326$~s. Uniform inputs in $[0,4)^2$ occupied $644{,}670$ to $645{,}259$ cells, leaving about $9.3\times10^{26}$ subsets. With $10^5$ sites in just five tight groups, the hybrid reached $M=k=5$ on all three seeds when the groups lay inside cells. The exchange grid retained over $89{,}000$ sites on the same inputs. Appendix~\ref{app:occupancy} and Table~\ref{tab:occupancy-budget} give the widths and the effect of translating their common origin.

We compare pair count (PC), packing (PK), an evaluated incumbent (I), exchange (X), and the hybrid (H) at $\varepsilon=1/2$. The 114 paired configurations give 570 method runs on 99 distinct input arrays. All five use identical inputs, traversal, physical grid origin, first-index representatives and search limits. PC and PK use Table~\ref{tab:comparison}. I rounds the incumbent energy upwards with error below $\binom{k}{2}2^{-48}$. H uses the tighter I/X bound. We report medians and ranges over seeds $17,29,43$.

The $n=2000$, $k=5$ inputs vary group shape, location, population, width and grid origin, and include uniform controls. Nine $n=24$, $k=4$ inputs also permit exhaustive checks of the original optimum. Across the 72 search-eligible cases per method, we attempt search when $M\le160$ and $\binom Mk\le10^7$, with a five-second cap. PC/PK/I/X/H completed 60/60/60/48/60 searches, respectively. The other 72 runs exceeded a size limit.

At small $k$, the evaluated incumbent often gives a coarser grid than exchange. The hybrid preserves that benefit and the exchange theorem's worst-case bound.
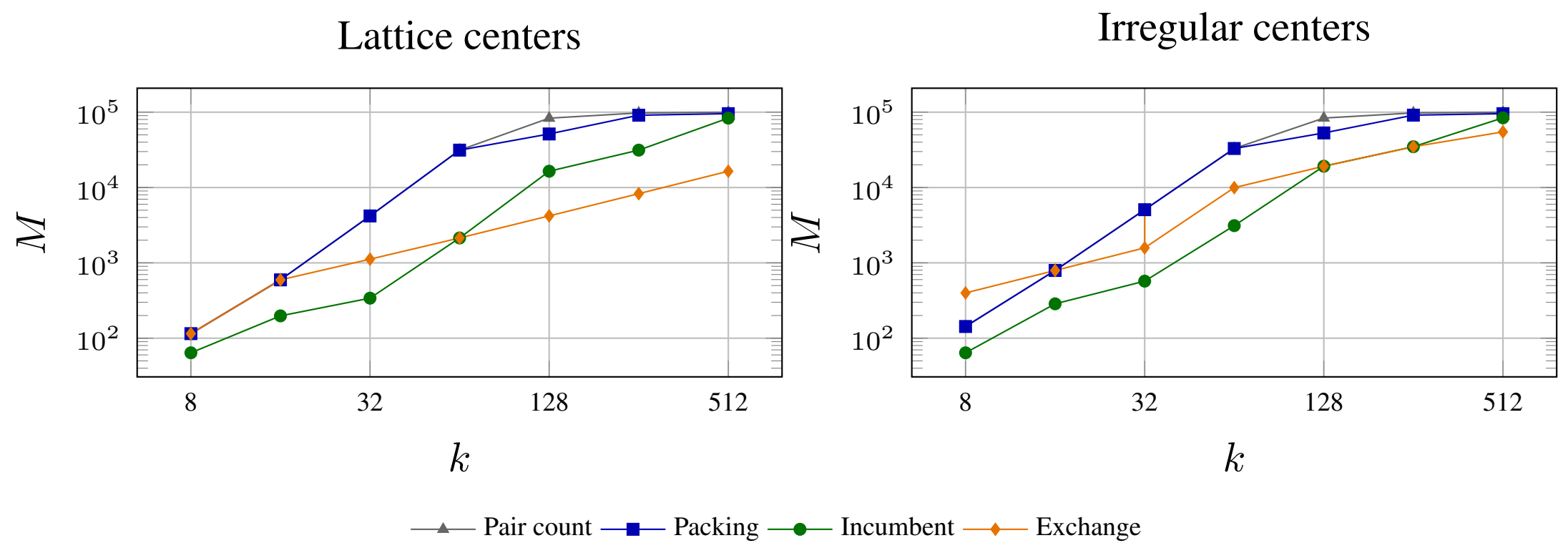
\begin{figure}[t]\centering
\begin{tikzpicture}
\begin{groupplot}[group style={group size=2 by 1,horizontal sep=1.2cm},width=.46\linewidth,height=4.25cm,xmode=log,ymode=log,log basis x=2,xlabel={$k$},ylabel={$M$},xtick={8,32,128,512},xticklabels={8,32,128,512},tick label style={font=\scriptsize},label style={font=\small},legend style={font=\scriptsize,draw=none},legend columns=4,grid=major]
\nextgroupplot[title={Lattice centers},title style={font=\small},legend to name=paired-v2-legend]
\addplot[forget plot,black!60,thin] coordinates {(8,114) (8,117)};
\addplot[forget plot,black!60,thin] coordinates {(16,593) (16,602)};
\addplot[forget plot,black!60,thin] coordinates {(32,4185) (32,4207)};
\addplot[forget plot,black!60,thin] coordinates {(64,31331) (64,31345)};
\addplot[forget plot,black!60,thin] coordinates {(128,83133) (128,83314)};
\addplot[forget plot,black!60,thin] coordinates {(256,97655) (256,97720)};
\addplot[forget plot,black!60,thin] coordinates {(512,99703) (512,99741)};
\addplot[black!60,mark=triangle*,mark size=1.5pt] coordinates {(8,115) (16,595) (32,4194) (64,31340) (128,83251) (256,97657) (512,99704)};
\addlegendentry{Pair count}
\addplot[forget plot,blue!70!black,thin] coordinates {(8,114) (8,117)};
\addplot[forget plot,blue!70!black,thin] coordinates {(16,593) (16,602)};
\addplot[forget plot,blue!70!black,thin] coordinates {(32,4185) (32,4207)};
\addplot[forget plot,blue!70!black,thin] coordinates {(64,31331) (64,31345)};
\addplot[forget plot,blue!70!black,thin] coordinates {(128,51345) (128,51608)};
\addplot[forget plot,blue!70!black,thin] coordinates {(256,91109) (256,91293)};
\addplot[forget plot,blue!70!black,thin] coordinates {(512,95510) (512,95545)};
\addplot[blue!70!black,mark=square*,mark size=1.5pt] coordinates {(8,115) (16,595) (32,4194) (64,31340) (128,51398) (256,91133) (512,95533)};
\addlegendentry{Packing}
\addplot[forget plot,green!45!black,thin] coordinates {(8,64) (8,64)};
\addplot[forget plot,green!45!black,thin] coordinates {(16,195) (16,206)};
\addplot[forget plot,green!45!black,thin] coordinates {(32,332) (32,353)};
\addplot[forget plot,green!45!black,thin] coordinates {(64,2138) (64,2142)};
\addplot[forget plot,green!45!black,thin] coordinates {(128,16440) (128,16461)};
\addplot[forget plot,green!45!black,thin] coordinates {(256,31298) (256,31357)};
\addplot[forget plot,green!45!black,thin] coordinates {(512,83385) (512,83436)};
\addplot[green!45!black,mark=*,mark size=1.5pt] coordinates {(8,64) (16,198) (32,340) (64,2140) (128,16452) (256,31349) (512,83390)};
\addlegendentry{Incumbent}
\addplot[forget plot,orange!90!black,thin] coordinates {(8,114) (8,117)};
\addplot[forget plot,orange!90!black,thin] coordinates {(16,593) (16,602)};
\addplot[forget plot,orange!90!black,thin] coordinates {(32,1110) (32,1129)};
\addplot[forget plot,orange!90!black,thin] coordinates {(64,2138) (64,2142)};
\addplot[forget plot,orange!90!black,thin] coordinates {(128,4190) (128,4213)};
\addplot[forget plot,orange!90!black,thin] coordinates {(256,8283) (256,8317)};
\addplot[forget plot,orange!90!black,thin] coordinates {(512,16440) (512,16468)};
\addplot[orange!90!black,mark=diamond*,mark size=1.5pt] coordinates {(8,115) (16,595) (32,1119) (64,2140) (128,4201) (256,8288) (512,16458)};
\addlegendentry{Exchange}
\nextgroupplot[title={Irregular centers},title style={font=\small}]
\addplot[forget plot,black!60,thin] coordinates {(8,143) (8,144)};
\addplot[forget plot,black!60,thin] coordinates {(16,792) (16,798)};
\addplot[forget plot,black!60,thin] coordinates {(32,5067) (32,5086)};
\addplot[forget plot,black!60,thin] coordinates {(64,32856) (64,33064)};
\addplot[forget plot,black!60,thin] coordinates {(128,83492) (128,83746)};
\addplot[forget plot,black!60,thin] coordinates {(256,97664) (256,97728)};
\addplot[forget plot,black!60,thin] coordinates {(512,99661) (512,99720)};
\addplot[black!60,mark=triangle*,mark size=1.5pt] coordinates {(8,143) (16,793) (32,5077) (64,32955) (128,83527) (256,97692) (512,99701)};
\addplot[forget plot,blue!70!black,thin] coordinates {(8,143) (8,144)};
\addplot[forget plot,blue!70!black,thin] coordinates {(16,792) (16,798)};
\addplot[forget plot,blue!70!black,thin] coordinates {(32,5067) (32,5086)};
\addplot[forget plot,blue!70!black,thin] coordinates {(64,32856) (64,33064)};
\addplot[forget plot,blue!70!black,thin] coordinates {(128,52907) (128,53136)};
\addplot[forget plot,blue!70!black,thin] coordinates {(256,91283) (256,91297)};
\addplot[forget plot,blue!70!black,thin] coordinates {(512,95449) (512,95594)};
\addplot[blue!70!black,mark=square*,mark size=1.5pt] coordinates {(8,143) (16,793) (32,5077) (64,32955) (128,52989) (256,91294) (512,95511)};
\addplot[forget plot,green!45!black,thin] coordinates {(8,64) (8,64)};
\addplot[forget plot,green!45!black,thin] coordinates {(16,285) (16,288)};
\addplot[forget plot,green!45!black,thin] coordinates {(32,566) (32,572)};
\addplot[forget plot,green!45!black,thin] coordinates {(64,3110) (64,3125)};
\addplot[forget plot,green!45!black,thin] coordinates {(128,19142) (128,19195)};
\addplot[forget plot,green!45!black,thin] coordinates {(256,34726) (256,34905)};
\addplot[forget plot,green!45!black,thin] coordinates {(512,83979) (512,84108)};
\addplot[green!45!black,mark=*,mark size=1.5pt] coordinates {(8,64) (16,286) (32,572) (64,3118) (128,19166) (256,34796) (512,84091)};
\addplot[forget plot,orange!90!black,thin] coordinates {(8,398) (8,400)};
\addplot[forget plot,orange!90!black,thin] coordinates {(16,792) (16,798)};
\addplot[forget plot,orange!90!black,thin] coordinates {(32,1576) (32,5086)};
\addplot[forget plot,orange!90!black,thin] coordinates {(64,9940) (64,10002)};
\addplot[forget plot,orange!90!black,thin] coordinates {(128,19142) (128,19195)};
\addplot[forget plot,orange!90!black,thin] coordinates {(256,34726) (256,34905)};
\addplot[forget plot,orange!90!black,thin] coordinates {(512,54718) (512,54799)};
\addplot[orange!90!black,mark=diamond*,mark size=1.5pt] coordinates {(8,399) (16,793) (32,1582) (64,9964) (128,19166) (256,34796) (512,54751)};
\end{groupplot}\end{tikzpicture}\par\smallskip\pgfplotslegendfromname{paired-v2-legend}\par
\caption{Representatives retained at $\varepsilon=1/2$ on $n=10^5$ sites in $2k$ groups of half-width $1/256$. Curves show medians over three seeds; vertical segments span the full range. All four certificates use the same points and grid origin. The hybrid count is the smaller of incumbent and exchange. No subset search was attempted on these inputs.}\label{fig:paired-v2}
\end{figure}
For $n=10^5$, Figure~\ref{fig:paired-v2} varies $k$ and the number of groups $2k$ together. At $k=512$, lattice medians are $M_I=83{,}390$ and $M_X=16{,}458$; irregular-center medians are $84{,}091$ and $54{,}751$. These reduced sets remain too large to enumerate. Among the 72 search-protocol cases, X retained fewer/equal/more sites than I in $0/13/59$ cases. The 60 searches completed by both H and I had identical representative lists and energies. All twelve strict exchange improvements occurred among the 42 occupancy-only cases.

To separate cardinality from changing data, we also varied $k$ on six frozen arrays (Appendix~\ref{app:fixed-input}). On the three 256-group arrays at $k=128$, median $M$ fell from 1,004 for I to 933 for X. All nine I/X/H searches timed out after five seconds when launched without size gates. Thus exchange sometimes shrinks $M$ on fixed points, but no reported gain produced a newly completed solve.

Of 84 completed searches in the 156-run scale study, the first sweep attained the
final recorded energy in 78 runs and the second in six. The median improvement
was $0.05231\%$ (maximum $0.35347\%$); refinement recovered the full optimum in
seven of nine small audited cases. The 72 timeouts or budget exclusions yielded no refinement observations.
Appendix~\ref{sec:sweeps} gives timings and a global coordinate baseline.

\begin{figure}[tbp]
\centering
\includegraphics[width=\linewidth]{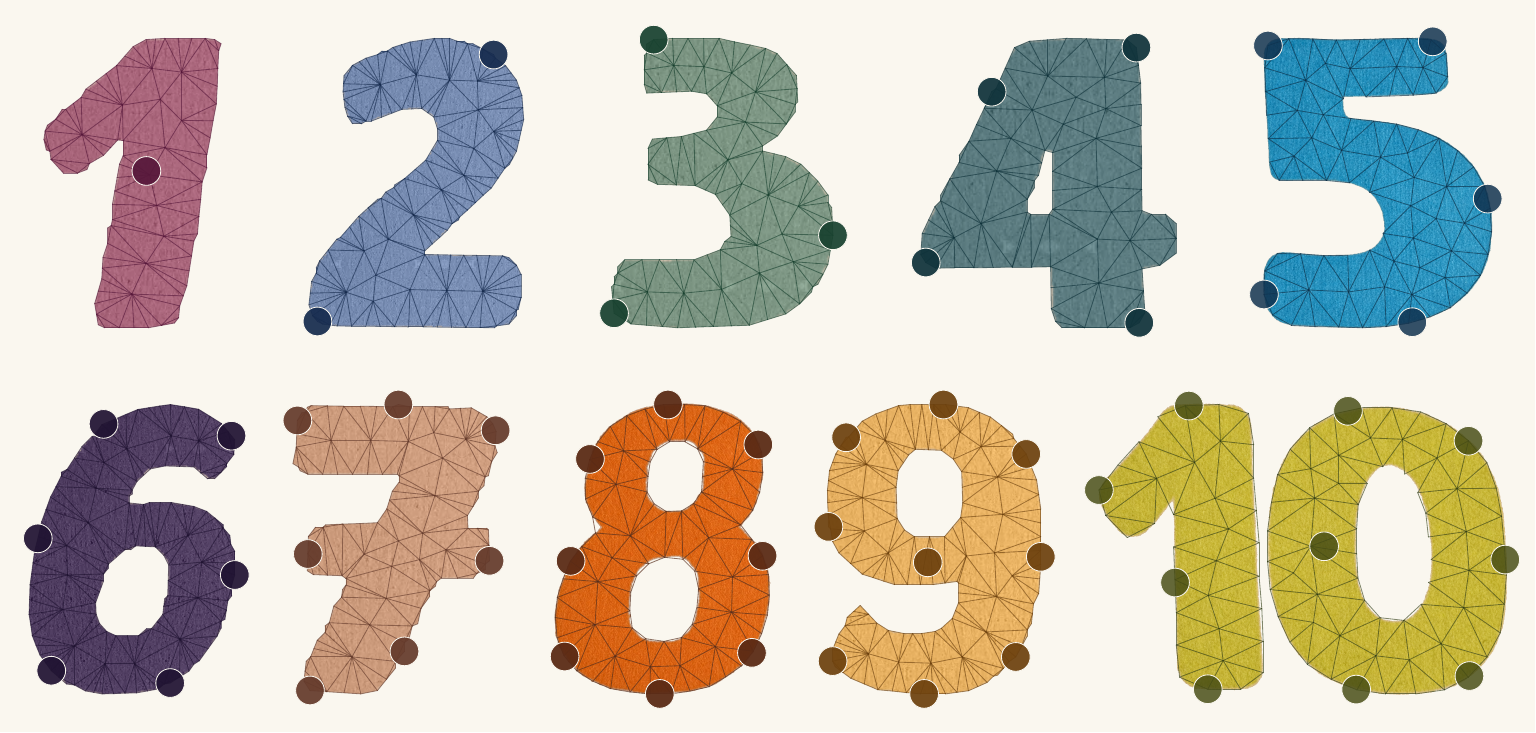}
\caption{Continuous inverse-square placements on the ten filled numeral shaped domains.
Each disk centre is a feasible site; triangles define the domain, including holes.
The numeral $k$ contains $k$ sites, giving 55 sites in total. The six uses a
completed outline. Display scales differ between panels.}
\label{fig:numerals}
\end{figure}

We also applied the EPAS with refinement in a playful example on triangulated
continuous domains (Figure~\ref{fig:numerals}), placing on each numeral as many
points as it represents. With $s=2$ and $\varepsilon=1$, we solved the full
reference grids for $2$--$8$; for $9$ and $10$, we searched smaller witness sets
and used independent continuous lower bounds. Refinement improved energy by up
to $1.078\%$. Each complete numeral run took at most about five minutes on
a Ryzen~7~5800H laptop, using the supplied Python drivers and native C++ search.
Certified relative energy excesses are:
$1$~(exact), $2$~(exact), $3$~(exact), $4$~($\leq0.000101\%$),
$5$~($\leq10.02\%$), $6$~($\leq5.97\%$), $7$~($\leq11.78\%$),
$8$~($\leq10.24\%$), $9$~($\leq9.31\%$), and $10$~($\leq11.31\%$).
The bounds concern energy on the polygons, not distance to an optimal
placement or tracing accuracy (Appendix~\ref{app:numerals}).

\section{Discussion and conclusion}\label{sec:conclusion}
For fixed $d,s$ and accuracy, our EPAS confines dependence on $n$ to preprocessing and reduces Riesz-energy selection to a finite search. Its strongest bound comes from averaging incident energy over feasible replacements. In the critical planar case the separation scale is sharp and yields $O(k\log(2k)/\varepsilon^2)$ representatives. For $s>d$ the bound is linear in $k$. Nets extend the argument to bounded doubling, and effective cell witnesses extend it to filled polygons with infinitely many feasible locations.

The two-distance barrier excludes a deterministic EPAS in unrestricted metrics unless $\mathrm{FPT}=\mathsf{W[1]}$; its shrinking gap does not rule out fixed-factor approximation or settle exact planar FPT. Adaptive rounding, exact planar FPT, and a planar PTAS for unrestricted $k$ remain open here. Ordered inputs also suggest studying a few exceptions to a line or Manhattan staircase, or projections of narrow strips~\cite{line2026}.

Practical use starts by measuring $M$ and comparing $\binom Mk$ with the search budget. A million clustered sites left 252 subsets, whereas uniform inputs left about $9.3\times10^{26}$. Exchange improves the worst-case bound but did not produce a newly completed solve in our experiments. The incumbent/hybrid Swiss search illustrates the potential of geometric pruning and refinement: eight within-cell moves recovered the independently verified full optimum despite a factor-two general certificate. Useful lower bounds on individual instances could certify this accuracy without a full-input solve.

\bibliographystyle{plainurl}
\bibliography{references}

\begin{thebibliography}{10}

\bibitem{atta2026}
Soumen Atta.
\newblock {Riesz} energy minimization facility location problem in the plane:
  Complexity and a polynomial-time approximation scheme.
\newblock {\em Theoretical Computer Science}, 1070:115833, 2026.
\newblock \href {https://doi.org/10.1016/j.tcs.2026.115833}
  {\path{doi:10.1016/j.tcs.2026.115833}}.

\bibitem{attaEmmerich2026}
Soumen Atta and Michael Emmerich.
\newblock {Riesz} {$s$}-energy regularized bi-objective maximal covering
  location problem and its solution via a memetic algorithm.
\newblock {\em Memetic Computing}, 18:47, 2026.
\newblock \href {https://doi.org/10.1007/s12293-026-00531-4}
  {\path{doi:10.1007/s12293-026-00531-4}}.

\bibitem{borodachov2019}
Sergiy~V. Borodachov, Douglas~P. Hardin, and Edward~B. Saff.
\newblock {\em Discrete Energy on Rectifiable Sets}.
\newblock Springer Monographs in Mathematics. Springer, New York, 2019.
\newblock \href {https://doi.org/10.1007/978-0-387-84808-2}
  {\path{doi:10.1007/978-0-387-84808-2}}.

\bibitem{bougeret2014}
Marin Bougeret, Nicolas Bousquet, Rodolphe Giroudeau, and R{\'e}mi Watrigant.
\newblock Parameterized complexity of the sparsest {$k$}-subgraph problem in
  chordal graphs.
\newblock In {\em {SOFSEM} 2014: Theory and Practice of Computer Science},
  volume 8327 of {\em Lecture Notes in Computer Science}, pages 150--161.
  Springer, 2014.
\newblock \href {https://doi.org/10.1007/978-3-319-04298-5_14}
  {\path{doi:10.1007/978-3-319-04298-5_14}}.

\bibitem{ceccarello2017}
Matteo Ceccarello, Andrea Pietracaprina, Geppino Pucci, and Eli Upfal.
\newblock {MapReduce} and streaming algorithms for diversity maximization in
  metric spaces of bounded doubling dimension.
\newblock {\em Proceedings of the VLDB Endowment}, 10(5):469--480, 2017.
\newblock \href {https://doi.org/10.14778/3055540.3055541}
  {\path{doi:10.14778/3055540.3055541}}.

\bibitem{cevallos2018}
Alfonso Cevallos, Friedrich Eisenbrand, and Sarah Morell.
\newblock Diversity maximization in doubling metrics.
\newblock In {\em 29th International Symposium on Algorithms and Computation
  ({ISAAC} 2018)}, volume 123 of {\em Leibniz International Proceedings in
  Informatics (LIPIcs)}, pages 33:1--33:12. Schloss Dagstuhl--Leibniz-Zentrum
  f{\"u}r Informatik, 2018.
\newblock \href {https://doi.org/10.4230/LIPIcs.ISAAC.2018.33}
  {\path{doi:10.4230/LIPIcs.ISAAC.2018.33}}.

\bibitem{chew1986}
L.~Paul Chew and Robert L.~Scot Drysdale.
\newblock Finding largest empty circles with location constraints.
\newblock Computer Science Technical Report PCS-TR86-130, Dartmouth College,
  1986.
\newblock URL: \url{https://digitalcommons.dartmouth.edu/cs_tr/29/}.

\bibitem{cygan2015}
Marek Cygan, Fedor~V. Fomin, {\L}ukasz Kowalik, Daniel Lokshtanov, D{\'a}niel
  Marx, Marcin Pilipczuk, Micha{\l} Pilipczuk, and Saket Saurabh.
\newblock {\em Parameterized Algorithms}.
\newblock Springer, 2015.
\newblock \href {https://doi.org/10.1007/978-3-319-21275-3}
  {\path{doi:10.1007/978-3-319-21275-3}}.

\bibitem{plane2026}
Michael Emmerich.
\newblock {Riesz} energy subset selection in the {Euclidean} plane is
  {NP}-hard: A reduction from the {Ising} model on planar cubic graphs, 2026.
\newblock Preprint, arXiv:2608.23506v3.
\newblock URL: \url{https://arxiv.org/abs/2608.23506v3}.

\bibitem{line2026}
Michael T.~M. Emmerich.
\newblock Polynomial-time {Riesz}-energy subset selection for ordered point
  sets on lines and {$\ell_1$}-staircases, 2026.
\newblock Preprint, arXiv:2606.16946v5.
\newblock URL: \url{https://arxiv.org/abs/2606.16946v5}.

\bibitem{metric2026}
Michael T.~M. Emmerich, Ksenia Pereverdieva, and Andr{\'e} Deutz.
\newblock On the complexity of minimum {Riesz} {$s$}-energy subset selection in
  {Euclidean} and ultrametric spaces, 2026.
\newblock Preprint, arXiv:2605.04715v1.
\newblock URL: \url{https://arxiv.org/abs/2605.04715v1}.

\bibitem{falcon2026}
Jes{\'u}s~Guillermo Falc{\'o}n-Cardona, Julio Ju{\'a}rez, Luis~A.
  M{\'a}rquez-Vega, and Michael T.~M. Emmerich.
\newblock Fast high-diversity subset selection for multiobjective optimization
  by {Riesz} {$s$}-energy.
\newblock {\em IEEE Transactions on Evolutionary Computation}, 30(2):747--761,
  2026.
\newblock Early access 2025.
\newblock \href {https://doi.org/10.1109/TEVC.2025.3570938}
  {\path{doi:10.1109/TEVC.2025.3570938}}.

\bibitem{falconDiversity2025}
Jes{\'u}s~Guillermo Falc{\'o}n-Cardona, Lourdes Uribe, and Pablo Rosas.
\newblock {Riesz} {$s$}-energy as a diversity indicator in evolutionary
  multiobjective optimization.
\newblock {\em IEEE Transactions on Evolutionary Computation},
  29(4):1168--1182, 2025.
\newblock \href {https://doi.org/10.1109/TEVC.2024.3405197}
  {\path{doi:10.1109/TEVC.2024.3405197}}.

\bibitem{swissaddresses}
{Federal Office of Topography swisstopo}.
\newblock Official directory of building addresses, 2026.
\newblock National LV95 CSV, edition of 5 November 2026.
\newblock URL:
  \url{https://www.swisstopo.admin.ch/en/official-directory-of-building-addresses}.

\bibitem{swissboundary}
{Federal Office of Topography swisstopo}.
\newblock {swissBOUNDARIES3D}, 2026.
\newblock January 2026, national territory in LV95.
\newblock URL:
  \url{https://data.geo.admin.ch/api/stac/v1/collections/ch.swisstopo.swissboundaries3d/items/swissboundaries3d_2026-01}.

\bibitem{gonzalez1985}
Teofilo~F. Gonzalez.
\newblock Clustering to minimize the maximum intercluster distance.
\newblock {\em Theoretical Computer Science}, 38:293--306, 1985.
\newblock \href {https://doi.org/10.1016/0304-3975(85)90224-5}
  {\path{doi:10.1016/0304-3975(85)90224-5}}.

\bibitem{hardinEtAl2019}
Douglas~P. Hardin, Alexander Reznikov, Edward~B. Saff, and Alexander Volberg.
\newblock Local properties of {Riesz} minimal energy configurations and
  equilibrium measures.
\newblock {\em International Mathematics Research Notices},
  2019(16):5066--5086, 2019.
\newblock \href {https://doi.org/10.1093/imrn/rnx262}
  {\path{doi:10.1093/imrn/rnx262}}.

\bibitem{hardinSaff2005}
Douglas~P. Hardin and Edward~B. Saff.
\newblock Minimal {Riesz} energy point configurations for rectifiable
  {$d$}-dimensional manifolds.
\newblock {\em Advances in Mathematics}, 193(1):174--204, 2005.
\newblock \href {https://doi.org/10.1016/j.aim.2004.05.006}
  {\path{doi:10.1016/j.aim.2004.05.006}}.

\bibitem{navarroEmmerich2026}
Paco Navarro and Michael Emmerich.
\newblock Computational complexity and energy landscapes of {Coulomb} crystals,
  2026.
\newblock Preprint, arXiv:2611.30657v1.
\newblock URL: \url{https://arxiv.org/abs/2611.30657v1}.

\bibitem{pereverdieva2025}
Ksenia Pereverdieva, Andr{\'e} Deutz, Tessa Ezendam, Thomas B{\"a}ck, H{\`e}rm
  Hofmeyer, and Michael Emmerich.
\newblock Comparative analysis of indicators for multi-objective diversity
  optimization.
\newblock In {\em Evolutionary Multi-Criterion Optimization ({EMO} 2025), Part
  II}, volume 15513 of {\em Lecture Notes in Computer Science}, pages 58--71.
  Springer, 2025.
\newblock \href {https://doi.org/10.1007/978-981-96-3538-2_5}
  {\path{doi:10.1007/978-981-96-3538-2_5}}.

\end{thebibliography}

\section*{AI Statement}
OpenAI ChatGPT/Codex Sol 5.6 and Lumo.ai (Proton) assisted with co-developing algorithmic ideas, developing and checking proof drafts, implementing and testing the algorithms, designing and running experiments, and preparing text and figures. The authors directed all experiments, proof development, and text preparation. They take responsibility for all statements, mathematical claims, references, software and experimental interpretation.

\appendix
\setcounter{secnumdepth}{2}
\section{Table of symbols}\label{app:symbols}
The table collects the notation used throughout the main argument. Local variables
in individual proofs and in the exact-search implementation are defined where used.
For the proofs and code, Appendix~\ref{sec:supp-extra} gives the arithmetic,
Appendices~\ref{app:walk}--\ref{app:numerals} give the worked examples
and experiments, and Appendix~\ref{app:reproduce} gives replay commands. Appendix~\ref{sec:structure-short} gives the treewidth algorithm for proxy lower bounds.

\begingroup\small
\setlength{\tabcolsep}{5pt}\renewcommand{\arraystretch}{1.07}
\begin{longtable}{@{}p{.25\linewidth}p{.71\linewidth}@{}}
\caption{Notation for selection, geometric compression and certification.}\label{tab:symbols}\\
\toprule Symbol & Meaning\\\midrule\endfirsthead
\toprule Symbol & Meaning (continued)\\\midrule\endhead
\bottomrule\endfoot
\multicolumn{2}{@{}l}{\textbf{Problem and computational model}}\\
$P,\ n$ & Feasible input sites and their number; in the continuous extension the feasible set is a compact domain $D$.\\
$S,\ k,\ S^\star$ & A selection, its prescribed cardinality and an optimal selection.\\
$d,\ s$ & Euclidean dimension and positive Riesz exponent. In metric notation $d(x,y)$ denotes distance.\\
$E_s(S),\ \OPT_s(P,k)$ & Energy summed over unordered pairs and the minimum energy of a feasible $k$-set.\\
$N_k,\ \Delta(S)$ & Number $\binom{k}{2}$ of selected pairs and shortest distance in $S$.\\
$\varepsilon,\ L$ & Relative tolerance ($0<\varepsilon\le1$) and total binary input length, including rational accuracy.\\[3pt]
\multicolumn{2}{@{}l}{\textbf{Exchange and geometric rounding}}\\
$C_j,\ r_j$ & First $j$ farthest-first centers and distance of center $j$ to its nearest predecessor ($j\ge2$). The proof uses $C=C_{2k}$ and $r=r_{2k}$.\\
$A,\ T_x,\ \Psi_A(z)$ & Companions $S\setminus\{x\}$, centers at distance at least $r/2$ from all of $A$, and the interaction of $z$ with $A$.\\
$H_m^{(p)},\ H_m$ & Generalized harmonic sum $\sum_{j=1}^m j^{-p}$ and its value at $p=1$. $\overline H$ denotes a computable upper bound.\\
$\delta,\ \sigma,\ \widehat\sigma$ & Separation proved by exchange, any valid lower bound on optimal pair distances, and a rounding scale in $(\sigma/2,\sigma]$.\\
$q,\ h,\ o$ & Fixed integer $q\ge\max\{1,s\}$, grid side $\varepsilon\widehat\sigma/(4qd)$, and common physical grid origin.\\
$Q,\ M,\ T$ & Representatives, their number $M=|Q|$, and a budget of complete subset evaluations, to compare with $\binom Mk$.\\
$\lambda,\ \eta$ & Upper bound on doubling dimension and covering radius of the greedy representative net.\\[3pt]
\multicolumn{2}{@{}l}{\textbf{Bounds and experiments}}\\
$U,\ \overline U$ & In the certificate comparison: energy of the $k$-center incumbent and an upper bound on that energy.\\
$B_X,\ B_H$ & Upper bounds on a single optimal pair contribution: exchange and hybrid, with $B_H=\min\{\overline U,B_X\}$ in the plane.\\
$V,\ U_t$ & Exact reduced optimum and original energy after $t$ within-cell refinement sweeps ($U_0=V$). The certified interval is $[V/(1+\varepsilon),U_t]$.\\
$c,\ G$ & Number of clusters and number of positive-probability cells in the fixed-grid occupancy model.\\
PC, PK, I, X, H & Pair-count, incumbent-packing, evaluated-incumbent, exchange and hybrid certificates. $M_I,M_X,M_H$ are the corresponding grid occupancies for I, X and H; $T$ is the enumeration budget.\\
\end{longtable}
\endgroup

\section{Complete proofs and implementation}\label{sec:supp-extra}
This appendix collects full proofs of the main results, including those presented in the main text, followed by the arithmetic and implementation details. All approximation statements concern the original full Riesz energy.

\subsection{Exchange potential and separation}
\begin{proof}[Proof of Lemma~\ref{lem:exchange}]
Fix $x$, put $A=S\setminus\{x\}$, and let
\[
 T_x=\{c\in C:\|c-y\|\ge r/2\text{ for all }y\in A\}.
\]
Each $y$ excludes at most one center at distance \emph{strictly less than} $r/2$, since two such
centers would be less than $r$ apart. Therefore $|T_x|\ge k+1$. Every $c\in T_x$ lies outside $A$
and is a feasible replacement; $c=x$ is harmless. With $\Psi_A(z)=\sum_{y\in A}\|z-y\|^{-s}$,
replacement optimality gives $\Psi_A(x)\le\Psi_A(c)$.

For fixed $y\in A$, sort the free-center distances as $r/2\le\rho_1\le\cdots\le\rho_{|T_x|}$. The
disjoint open radius-$r/2$ balls about the first $j$ centers lie in the radius-$(\rho_j+r/2)$ ball
about $y$. Volume comparison yields $j\le(1+2\rho_j/r)^d\le(4\rho_j/r)^d$, hence $\rho_j^{-s}\le4^s r^{-s}j^{-s/d}$. The point $y$ need not be a center. Averaging over the feasible replacements gives
\begin{align*}
 \Psi_A(x)&\le\frac1{|T_x|}\sum_{c\in T_x}\sum_{y\in A}\|c-y\|^{-s}\\
 &\le\frac{k-1}{|T_x|}\,4^s r^{-s}H_{2k}^{(s/d)}
 \le4^s r^{-s}H_{2k}^{(s/d)}.
\end{align*}
Every individual pair term is at most this potential, proving~\eqref{eq:exchange-separation}.
\end{proof}

\subsection{Injective rounding}
\begin{proof}[Proof of Lemma~\ref{lem:rounding}]
Cell diameter is $\sqrt d\,h<\sigma$, so snapping an optimum to retained candidates is injective.
A selected distance $D\ge\sigma$ decreases by at most $2\sqrt d\,h\le\varepsilon D/(2q)$. For
$s\ge1$, Bernoulli's inequality gives $(1-\varepsilon/(2q))^s\ge1-\varepsilon/2$; for $s<1$, use
$(1-z)^s\ge1-z$. Since $1-\varepsilon/2\ge(1+\varepsilon)^{-1}$ for $0\le\varepsilon\le1$, every
pair energy increases by at most $1+\varepsilon$. Sum over pairs. Each diameter-$a$ set has
coordinate ranges at most $a$ and meets at most $(2+a/h)^d$ cells.
\end{proof}

\subsection{The finite Euclidean EPAS}
\begin{proof}[Proof of Theorem~\ref{thm:exchange-epas}]
If $n\le2k$, retain the input and enumerate at most $4^k$ subsets.
Otherwise choose $2k$ farthest-first centers in $O(nk)$ operations, starting arbitrarily.
Their last insertion distance $r>0$ is a lower bound on all inter-center distances.
Every input point is within $r$ of the centers already chosen before the final insertion;
hence the $2k$ closed radius-$r$ balls cover the input. Apply Lemma~\ref{lem:rounding} with $\sigma=\delta$ from Lemma~\ref{lem:exchange}. The $2k$
covering balls have diameter $2r$, so $M\le2k(2+2r/h)^d=O_{d,s}(k(H_{2k}^{(s/d)})^{d/s}/\varepsilon^d)$. In the exact real model we may take $\widehat\sigma=\sigma$; rational implementation is described
below. Sort cell keys in $O(n\log n)$ operations. Evaluate each of the $\binom Mk$ subsets in
$O(k^2)$ operations and retain the best. The bound $\binom Mk\le(eM/k)^k$, with $k^2\le4^k$,
proves~\eqref{eq:exchange-time}. Optional precomputation of representative pair weights is
absorbed by this bound. Since $Q\subseteq P$ and the snapped optimum is feasible in $Q$,
equation~\eqref{eq:epas-certificate} follows. Any additional feasible solution of energy $U$ can
improve the upper endpoint to $\min\{U,V\}$.

For rational coordinates and positive even $s$, write $m=2k$, $b=\lceil\log_2m\rceil$, and use the rational upper bound
\begin{equation}
 \overline H=
 \begin{cases}
 \dfrac{d}{d-s}\,2^{\lceil b(d-s)/d\rceil},&s<d,\\
 1+b,&s=d,\\
 \dfrac{s}{s-d},&s>d.
 \end{cases}
\end{equation}
Integral bounds give $H_m^{(s/d)}\le\overline H=O_{d,s}(H_m^{(s/d)})$. Thus $B=4^s\overline
H/(r^2)^{s/2}$ is rational and $\sigma=B^{-1/s}$ is a valid separation scale with the same
asymptotic bounds. Binary lengths and a constant number of rational tests find a dyadic
$\widehat\sigma$ satisfying $B\widehat\sigma^s\le1<B(2\widehat\sigma)^s$. No root is computed.
Grid indices use integer division. All coordinates, keys and sums of at most $\binom{k}{2}$
rational pair weights have polynomial bit length. Appendix~\ref{sec:epas-dyadic} gives the
arithmetic details.
\end{proof}

The planar corollary follows by setting $d=s=2$ and using $H_{2k}=O(\log(2k))$. The other regimes follow from the elementary integral bounds on generalized harmonic sums given below.

\subsection{Comparison of planar certificates}
\begin{proof}[Proof of Proposition~\ref{prop:comparison}]
The $r_k$-separated incumbent satisfies $U\le N_k/r_k^2$. For a center's $j$th nearest companion
at distance $d_j$, packing gives $j+1\le(1+2d_j/r_k)^2$, hence $j\le8(d_j/r_k)^2$ because $d_j\ge
r_k$. Summing over companions and dividing by two yields $U\le4kH_{k-1}/r_k^2$. The $k$ covering
disks and Lemma~\ref{lem:rounding} give $M=O(k r_k^2 B/\varepsilon^2)$ in either case.
For exchange, the $2k$ covering disks have radius $r_{2k}$, and
Lemma~\ref{lem:exchange} bounds each selected site's incident energy by $16H_{2k}/r_{2k}^2$.
That radius again cancels from the count, yielding the third row. Enumeration
uses $\binom Mk\le(eM/k)^k$.
\end{proof}

\subsection{Compact feasible domains}
\begin{proof}[Proof of Corollary~\ref{cor:domains}]
On $D^k$ extend the energy to $+\infty$ at collisions. It is lower semicontinuous
and some feasible tuple has finite energy, so its minimum is attained at distinct
points. The given centers are feasible replacements in Lemma~\ref{lem:exchange},
which uses only their separation. Lemma~\ref{lem:rounding} applies to the
witness in each optimal point's cell. Enumerate the grid keys in the axis-aligned
bounding boxes of the $2k$ covering balls, remove repeated keys, and query the
remaining cells. Each box yields at most $(2+2r/h)^d$ keys. The ratio $r/h$ is
$O_{d,s}((H_{2k}^{(s/d)})^{1/s}/\varepsilon)$, so the absolute scale cancels
from the number of cell queries. For $s=d=2$ the algorithm queries
$O(k\log(2k)/\varepsilon^2)$ cells, followed by at most
$(C\log(2k)/\varepsilon^2)^k$ operations for subset search.
\end{proof}

\subsection{Finite metrics of bounded doubling dimension}
\begin{proof}[Proof of Theorem~\ref{thm:doubling}]
For $n\le2k$ enumerate the input directly. Otherwise use its $2k$ farthest-first
centers. For an $a$-separated set in a closed radius-$R$ ball, if $R<a/2$
there is at most one point. If $R\ge a/2$, iterate the doubling cover
$\ell=\lfloor\log_2(2R/a)\rfloor+1$ times. Each resulting ball has diameter
strictly below $a$, including when $2R/a$ is a power of two. The number of
points is at most $2^{\lambda\ell}\le(4R/a)^\lambda$.

Fix an optimal site $x$ and its $k-1$ companions. Their open radius-$r/2$ balls
exclude at most $k-1$ centers. For each companion the $j$th nearest remaining
center has distance $\rho_j\ge r/2$, so $j\le(4\rho_j/r)^\lambda$.
Averaging the exchange inequalities over the at least $k+1$ remaining centers,
as in Lemma~\ref{lem:exchange}, bounds the incident potential by
$4^sr^{-s}H_{2k}^{(s/\lambda)}$. Thus every optimal pair has distance at least
$\delta=r/[4(H_{2k}^{(s/\lambda)})^{1/s}]$.

Put $\eta=\varepsilon\delta/(4q)$ with integer $q\ge\max(1,s)$. Scan $P$,
retaining a site when all earlier representatives are farther than $\eta$.
Every site has a representative within $\eta$. Rounding an optimal pair at
distance $b\ge\delta$ leaves distance at least $b-2\eta\ge
(1-\varepsilon/(2q))b>0$. This proves injectivity and, by the scalar inequality
in Lemma~\ref{lem:rounding}, the energy guarantee. The $\eta$-separated
representatives in each covering ball number at most $(4r/\eta)^\lambda$.
Summing over the $2k$ balls proves the bound. A scan compares each input site
with at most $M$ representatives.
\end{proof}

\subsection{Two-distance classification and approximation barrier}
\begin{proof}[Proof of Theorem~\ref{thm:metric-barrier}]
Use Atta's two-distance reduction~\cite[Section 5]{atta2026}, equivalently the
complementary-clique construction in~\cite[Theorem 1]{metric2026}. For a graph $G$, assign
distance $1$ on edges and $2$ on nonedges. Every positive distance is at most the sum of two
positive distances, so this is a metric. For $N_k=\binom{k}{2}$ and a $k$-set $S$, $E_2(S)=N_k/4+(3/4)e_G(S)$. The objective is an affine transformation of Sparsest $k$-Subgraph, whose general-graph
W[1]-hardness is documented in~\cite{bougeret2014}. Thus threshold $B=N_k/4$ is attainable exactly
when $G$ has a $k$-independent set. The reduction preserves $k$, proving W[1]-hardness.

For membership at any rational threshold, enumerate the $2^{N_k}$ labeled distance patterns on
$[k]$ and discard those exceeding the threshold. For each remaining pattern, form a $k$-partite
graph with vertices $(i,u)$, $i\in[k]$, $u\in P$. Join $(i,u)$ and $(j,v)$ precisely when $i\ne
j$, $u\ne v$, and their input distance matches the pattern on $ij$. A $k$-clique realizes the
pattern with distinct candidates. The disjoint union of these graphs reduces the problem to
$k$-Clique in $f(k)L^{O(1)}$ time, with unchanged parameter. For $k\ge2$ a clique lies in one
component; smaller cases are direct. This proves W[1] membership.

For the approximation barrier, assume $k\ge2$ and set $\varepsilon=1/N_k$. A yes-instance has
optimum $B$, so a $(1+\varepsilon)$ algorithm returns energy at most $B+1/4$. A no-instance has
every feasible energy at least $B+3/4$. Comparing the returned energy with $B+1/2$ decides
Independent Set. The no/yes optimum ratio is at least $1+3/N_k$; taking
$1/\varepsilon=N_k$ keeps the runtime parameter a function of $k$ alone. Thus the assumed
scheme would decide Independent Set in FPT time.
\end{proof}

\subsection{Certified within-cell refinement}
\begin{proof}[Proof of Proposition~\ref{prop:refine}]
The $k$ selected cells are distinct. Replacing one site changes precisely its
$k-1$ incident terms, whose sum is minimized over a set containing its current
position. Feasibility and the upper bound follow by induction. Within a sweep
the selected cells are disjoint, so at most $n$ candidates each require $k-1$
pair evaluations. The lower bound follows from~\eqref{eq:epas-certificate}.
\end{proof}

\subsection{Why the planar separation scale is sharp}\label{app:sharp-planar}
The critical energy in a square is of order $m^2\log m$ on the unit scale; compare the
continuous asymptotic of Hardin and Saff~\cite[Theorem 2.1, (8)]{hardinSaff2005}.
For the finite exchange construction we need just the following uniform lower bound, which
has an elementary proof. All pair sums here are unordered.

\begin{lemma}[Energy forced inside a square]\label{lem:square-bulk}
There is an absolute $c>0$ such that every $m\ge4$ distinct points in a square of side
$a>0$ have inverse-square energy at least $c\,m^2\log(m)/a^2$.
\end{lemma}
\begin{proof}
At level $j$, subdivide the square into $4^j$ congruent subsquares, assigning boundary points
consistently so that the partitions are nested. If $m_C$ points are in cell $C$, the number of
same-cell pairs satisfies
\[
 N_j=\tfrac12\bigl(\textstyle\sum_C m_C^2-m\bigr)
 \ \ge\ \tfrac12(m^2/4^j-m)
 \ \ge\ m^2/(4\cdot4^j),\qquad 0\le j\le J:=\lfloor\log_4(m/2)\rfloor .
\]
For any pair $p,q$, let $\ell$ be its last same-cell level through $J$. Its squared distance is
at most $2a^2/4^\ell$. Thus its total weight in the sum of the level counts is
$\sum_{j=0}^{\ell}4^j/a^2\le(4/3)4^\ell/a^2\le(8/3)\|p-q\|^{-2}$. Summing over pairs
yields $E_2\ge(3/8)\sum_{j=0}^J4^jN_j/a^2\ge3(J+1)m^2/(32a^2)$.
\end{proof}

\begin{proof}[Proof of Theorem~\ref{thm:sharp-planar}]
Let $a=2\lceil\sqrt{k}\rceil$, and take the lattice square $G=\{0,\ldots,a-1\}^2$.
Adjoin $u=(k^2,0)$ and $v=(k^2+t,0)$, where $t$ is a dyadic rational in
$[A/\sqrt{\log k},2A/\sqrt{\log k}]$. The absolute constant $A$ is chosen below. For all
sufficiently large $k$, $t<1$ and the remote sites are at distance $\Omega(k^2)$ from $G$.
There are at least $4k$ lattice sites. Until $2k$ centers have been chosen some untouched lattice
site is at distance at least one from every center; hence $r_{2k}\ge1$. At least $2k-2$ centers
are in $G$. Partition the lattice sites into $\lceil\sqrt{k}\rceil^2$ disjoint blocks of
$2$ by $2$ sites. For all sufficiently large $k$ there are fewer blocks than lattice centers.
Two of these $r_{2k}$-separated centers lie in the same block, so $r_{2k}\le\sqrt2$.
Neither argument uses the initial site or tie-breaking.

Consider a $k$-set omitting at least one of $u,v$. It selects $m\ge k-1$ lattice sites.
Their mean incident energy, counting only their mutual interactions, is $2E_2/m$.
Lemma~\ref{lem:square-bulk} and $a^2=O(k)$ give a lattice site with incident
energy at least $c_0\log k$ for an absolute $c_0>0$. Replacing it with an omitted remote
point costs at most $t^{-2}+O(k^{-3})$: at most one interaction is across the remote pair,
and each of at most $k-1$ lattice interactions costs $O(k^{-4})$. Choose $A$ so that
$A^{-2}<c_0/2$. The replacement is strictly improving for every sufficiently large $k$.
Every one-swap local optimum must therefore contain both remote points. All its other
distances are at least one, so its minimum distance is exactly $t$.
Since $1\le r_{2k}\le\sqrt2$, the claimed order follows. The remote-pair term alone also
gives incident energy $\Omega(r_{2k}^{-2}\log k)$ at each remote site, matching the order of
Lemma~\ref{lem:exchange} there.
\end{proof}

\subsection{A rational harmonic surrogate and dyadic scale}\label{sec:epas-dyadic}
For fixed even integer $s>0$ and fixed integer dimension $d$, the harmonic sum $H_{2k}^{(s/d)}$
need not be rational. Put $m=2k$ and $b=\lceil\log_2m\rceil$, computed from integer bit lengths.
Define the rational upper surrogate
\begin{equation}\label{eq:exchange-harmonic-surrogate}
 \overline H=
 \begin{cases}
 \displaystyle\frac{d}{d-s}\,2^{\lceil b(d-s)/d\rceil},&s<d,\\[5pt]
 1+b,&s=d,\\[2pt]
 \displaystyle\frac{s}{s-d},&s>d.
 \end{cases}
\end{equation}
It satisfies $H_m^{(s/d)}\le\overline H=O_{d,s}(H_m^{(s/d)})$. For $p=s/d<1$, integral comparison gives $H_m^{(p)}\le m^{1-p}/(1-p)\le2^{\lceil b(1-p)\rceil}/(1-p)$. Conversely $H_m^{(p)}\ge m^{1-p}$, and $2^b<2m$ bounds the ratio by a constant depending only on
$p$. For $p=1$, $H_m\le1+\ln m\le1+b$, while $H_m\ge\ln(m+1)$ gives the constant-factor converse.
For $p>1$, $H_m^{(p)}\le1+\int_1^\infty x^{-p}\,dx=p/(p-1)$ and $H_m^{(p)}\ge1$. This proves all
three cases. In the planar critical case, the exact rational $H_{2k}$ may be used instead.

Set $B=4^s\overline H/(r^2)^{s/2}$ and $\delta_0=B^{-1/s}$. Then $\delta_0$ is a valid optimum-separation lower bound, within a fixed factor
of~\eqref{eq:exchange-separation}. Since $s$ is fixed and even, $B$ is positive rational. Write
$B=A/D$ with positive integers $A,D$, and let $\operatorname{bits}(z)=\lfloor\log_2z\rfloor+1$ for
a positive integer. Compute $e=\lfloor(\operatorname{bits}(D)-\operatorname{bits}(A))/s\rfloor$. If $B2^{es}>1$, decrement $e$ once. The resulting $\widehat\delta=2^e$ obeys $B\widehat\delta^s\le1<B(2\widehat\delta)^s$, hence $\delta_0/2<\widehat\delta\le\delta_0$. For completeness, put $\ell=\operatorname{bits}(D)-\operatorname{bits}(A)$ and write $\ell=es+t$,
$0\le t<s$, for the initial exponent. Bit lengths imply $2^{-\ell-1}<B<2^{-\ell+1}$ and $2^{-t-1}<B2^{es}<2^{1-t}$. Thus the initial scale is either valid or at most one dyadic step too large; its next larger step
is strictly invalid. After a decrement, the previous value $B2^{es}>1$ supplies the required
strict upper comparison. This proves the claimed constant-number comparison procedure.

All integers in this construction, including the grid indices, have polynomial bit length in $L$.
The magnitude of $e$ is polynomially bounded in the encoding length of $B$; shifts implement
powers of two exactly. The surrogate is computed using fixed rational constants, integer ceilings,
and shifts. Consequently preprocessing and enumeration have polynomial bit cost per stated
operation, and no algorithm step treats an irrational boundary or an arbitrary-real floor as a
bit-model primitive.

\subsection{Nested grids and incumbent bounds}\label{app:nesting}
For each positive certificate $B$, let $\widehat\sigma(B)$ be the largest power of two satisfying
$B\widehat\sigma(B)^2\le1$. If $B_1\le B_2$, then $\widehat\sigma(B_1)/\widehat\sigma(B_2)$ is an
integral power of two. With common accuracy and physical origin $o$, the cell width ratio is
therefore an integer and each coarse half-open cell is a union of fine cells. Every occupied
coarse cell contains an occupied fine cell, proving $M(B_1)\le M(B_2)$. It follows that the hybrid
has $M_H=\min(M_I,M_X)$. Representative identities can change, so this assertion is not an
ordering of the returned energies. Origins chosen as different fractions of each method's width
need not be common physical origins and do not support this nesting proof.

For integer coordinate numerators with common denominator $D_0$, write $q_{ij}=\|X_i-X_j\|^2$ and
set $A=2^{48}$. The integer sum $\widetilde U=A^{-1}\sum_{\{i,j\}\subset C_k}\lceil A D_0^2/q_{ij}\rceil$ satisfies $U\le\widetilde U<U+\binom{k}{2}/A$. Taking $\overline U=\min\{\widetilde U,B_{\rm
PK}\}$ remains valid because both quantities bound $U$. This avoids summing fractions with large
unrelated denominators in the larger-$k$ occupancy study. All ceiling operations use integers. The
small-instance tests independently verify that this bound dominates exact incumbent energy.

There is a useful endpoint check when comparing these certificates under $T=10^7$.
For a planar incumbent $C_k$ at mutual distances at least $r_k$, draw an edge for each distance
strictly below $\sqrt2\,r_k$. The straight-line edges cannot cross: in a convex quadrilateral
some angle is at least $90^\circ$, and the diagonal opposite that angle has length at least
$\sqrt2\,r_k$ by the cosine rule. Two crossing edges would be its diagonals. A short edge also
cannot pass through a third site, since its two subsegments would each have length at least
$r_k$. Thus there are at most $3k-6$ short pairs for $k\ge3$. Each short pair contributes at
most $1/r_k^2$, and each other pair at most $1/(2r_k^2)$. Consequently $U\le(N_k+3k-6)/(2r_k^2)$. At $k=12$ the numerator after division by two is $48<16H_{24}$. Together with the
$k\le11$ comparison and $M_X\ge2k$ for larger $k$, this shows that, with the exact incumbent,
exchange alone cannot newly admit a raw enumeration at this budget for any $k$ on the
$n>2k$ branch. One may also cap a rounded incumbent by this planar bound; the reported
benchmarks use the PC/PK cap described above and do not rely on this extra cap.

\subsection{Cluster size, grid placement and the search budget}\label{app:occupancy}

If each cluster is contained in a ball of radius $\rho$, its coordinate range on each axis is at
most $2\rho$. The extreme floor indices differ by at most $\lceil2\rho/h\rceil$, so $M\le
c(1+\lceil2\rho/h\rceil)^d$. When there are $M=k+t$ representatives, the enumeration count is
$\binom{k+t}{t}$. For $t=0$ this is one and for $t=1$ it is $k+1$. Even $M=2k$ leaves
$\binom{2k}{k}$ subsets, which grows exponentially in $k$. The implementation evaluates
\texttt{math.comb(M,k)} before allocating reduced pair weights.

For a fixed grid and $n$ independent draws, let $p_j>0$ be the probability of a site falling in
cell $j$, for $1\le j\le G$. Cell $j$ is empty with probability $(1-p_j)^n$. Summing the
probabilities that the cells are occupied gives $\mathbb E[M]=\sum_{j=1}^{G}[1-(1-p_j)^n]$. For equal-probability cells this becomes $G[1-(1-1/G)^n]$. At low density most sites have their
own cell. As $n$ increases at fixed $G$ the expected count approaches $G$. The experiments compute
occupancy directly because the certified width depends on the input and the dyadic uniform
generator samples without replacement.

\subsection{Exact reduced search and numerical safeguards}\label{app:exact-search}
The package contains \path{anc/benchmarks/code/planar_riesz_epas_v2.py}, the standard-library
rational reference, and \path{anc/benchmarks/code/epas_refine_v2.py}, an accelerated planar
implementation for the bounded dyadic data used here. The reference handles arbitrary rational
coordinates and small-cardinality branches. The vectorized path is intended for validated planar
arrays and the stated $2k$-cover regime; the small walkthrough additionally uses the proved $2k-2$
variant. Both use the last insertion distance to set the grid scale.
Original points are grouped by cell and remain available after reduction.

\paragraph{Exact integer geometry.}
Write a candidate as $X_i/D$, where $X_i\in\mathbb Z^2$ and $D=2^{20}$. The code first checks
distinctness and that the sum of squared coordinate spans is at most $2^{53}$. Squared pair
distances in integer units can then be evaluated exactly in signed 64-bit arithmetic and converted
exactly to binary64 when needed. They are positive for distinct candidates. If $hD=a/b$ in lowest
terms and the grid origin is $(-h/2,-h/2)$, the cell key is
\[
 \left(\left\lfloor\frac{2bX_{i,1}+a}{2a}\right\rfloor,
       \left\lfloor\frac{2bX_{i,2}+a}{2a}\right\rfloor\right).
\]
The code verifies the integer expression does not overflow; floor division handles negative
coordinates. Sorting by the two keys and then input index retains exactly one original candidate
per occupied cell. No rounded distance determines a center, cell, or certificate.

\paragraph{Exact reduced search.}
Let $w_{ij}\in\mathbb Q_{>0}$ be a representative pair weight and $A=2^{48}$. The search uses
integer lower weights $\ell_{ij}=\lfloor Aw_{ij}\rfloor$. If $S_{\rm inc}$ is the best
feasible set found so far, the pruning threshold is $U=\lceil A E_2(S_{\rm inc})\rceil$.
A deterministic greedy completion gives an initial feasible set. For a partial selection with lower cost $c$, let $b_i$ be the lower
attachment cost of available candidate $i$, and let $q$ positions remain. If $q\ge2$, assign
\[
 a_i=2b_i+\text{the sum of the }q-1\text{ smallest }\ell_{ij}
 \text{ over other available }j.
\]
For any completion $T$, $|T|=q$, the sum of its charges is at most twice its remaining lower cost:
each internal edge is counted twice and the attachment costs are doubled. Hence the sum of the $q$
smallest charges over all candidates is also a lower bound on twice the remaining cost. Prune only
when this bound plus $2c$ exceeds $2U$. For $q=1$, use $c+\min_i b_i>U$. Positivity also permits
discarding a candidate whose already forced lower cost exceeds $U$. Every surviving leaf is
compared using exact rational weights; equality is resolved by the lexicographic input-index
tuple. Conservative floors can weaken pruning but cannot remove a better or tied optimum. A
timed-out traversal is never reported as a completed EPAS.

\paragraph{Certified potential filtering in refinement.}
One candidate potential is $D^2\sum_j1/q_j$, with exact positive integers $q_j\le2^{53}$. The
common factor $D^2$ does not affect minimization. Under IEEE-754 binary64 round-to-nearest
arithmetic, each division $1/q_j$ is enclosed by the adjacent representable numbers obtained with
\texttt{nextafter}. After every positive addition, outward expansion encloses the accumulated
exact sum. Values stay far from underflow. Let $[\ell_p,u_p]$ be these enclosures and
$u_{\min}=\min_p u_p$. A true minimizer $p^\star$ satisfies $\ell_{p^\star}\le u_{\min}$: its
exact potential is no larger than that of a candidate attaining the smallest upper endpoint. Thus
candidates with $\ell_p>u_{\min}$ can be discarded safely. All others are evaluated as exact
fractions, together with the incumbent. Ties retain the incumbent; otherwise the smallest input
index among strict minimizers is used. This is a filtering implementation of exact minimization,
not floating-point acceptance with a tolerance.

\paragraph{Refinement cost and certificates.}
The selected cells and their visit order remain fixed. An update changes only the incident
potential and updates the exact total by that difference. The code checks the exact total again
after each sweep, asserts nonincrease after every update, and verifies the selected cell
identifiers and distinctness. A no-improvement sweep is a fixed point for this deterministic
within-cell rule. Requested later prefixes reuse the result without additional work. Such early
stopping is valid for fixed-$t$ output and does not establish a worst-case bound for unrestricted
convergence. The posterior interval retains the base lower endpoint $V/(1+\varepsilon)$ while
replacing the upper endpoint by the refined energy.

\subsection{Polygon witnesses and their coordinates}\label{app:polygons}
Corollary~\ref{cor:domains} assumes that both geometric primitives can be
computed. On any compact feasible set with at least $2k$ distinct points,
continuous farthest-first traversal exists: the distance to a nonempty finite
set is continuous and attains its maximum. The maximum stays positive through
the $2k$th insertion, and the maxima are nonincreasing. Hence the last insertion
distance gives separation and coverage. An existence argument alone gives no
procedure for locating these maximizers on an arbitrarily represented set.

For a filled simple polygon with $m$ vertices, start at a polygon vertex.
With $j$ sites Chew and Drysdale give $O(mj+j\log(2j))$ geometric operations,
or $O(mj)$ with the nearest-site Voronoi diagram supplied~\cite{chew1986}.
Summing gives $O(mk^2+k^2\log(2k))$ operations for the certificate.
The feasible location is the center of the empty circle. A boundary site is
eligible even when its circle extends outside the polygon. These bounds are
for a simple polygon with its boundary included.

Write $G$ for the number of cell keys visited in the cover boxes, counting
duplicates. For a box $[l_i,u_i]$ on axis $i$ and grid origin $o$, enumerate
the integers from $\lfloor(l_i-o_i)/h\rfloor$ through
$\lfloor(u_i-o_i)/h\rfloor$, inclusively. Take their Cartesian product,
sort and deduplicate, then test each key. With exact-real box arithmetic,
$G\le 2k(2+2r/h)^d$. In a rational implementation one may enclose $r$
by a dyadic $r\le r'<2r$ and use $r'$ just for the boxes. This changes the
count by at most a constant factor at fixed dimension.

Triangulate the polygon once, for example by ear clipping in $O(m^3)$
elementary operations after removing redundant collinear boundary vertices.
Clip each closed triangle against the closure of a queried square. If the result
is nonempty, average its distinct vertices, including the cases of a single
point or segment. This point lies strictly below each upper coordinate bound
unless the whole clipped set lies in the corresponding upper face. Thus it
belongs to the half-open square exactly when a witness exists in that triangle.
The lower faces are included. Stop at the first witness across all triangles.
For example the triangle with vertices $(0,0),(1,0),(0,1)$ intersects
$[1,2)\times[0,1)$ at $(1,0)$ and has empty intersection with
$[-1,0)\times[0,1)$. This direct procedure uses
$O(m^3+G\log(2G)+mG)$ operations including triangulation.

For rational vertices and planar $s=2$, use the dyadic scale of
Theorem~\ref{thm:exchange-epas}. A farthest point can be chosen among polygon
vertices, bisector--boundary intersections and Voronoi vertices: on a clipped
nearest-site cell the squared distance to that site is convex and attains a
maximum at a vertex after triangulating the clipped region. Degenerate ties
admit the same choice of a vertex, including endpoints of overlapping edges.
Every selected coordinate therefore solves a rational linear system of order
at most two using original vertices and at most three existing sites. With
$B$ bits per current coordinate, a constant number of rational arithmetic
operations gives $O(B+L)$ bits for the new coordinate, where $L$ is polygon
and accuracy encoding length. Iterating for $2k$ steps permits
$C^{O(k)}(L+1)$ bits, sufficient for polynomial bit cost in $L$ at fixed
parameters. For an elementary implementation of the certificate with this
bit accounting, enumerate the $O(j^3+mj^2+m)$ candidate intersections from
the original features of the current diagram and boundary; test polygon
membership and score nearest-site squared distance in $O(m+j)$ operations
per candidate. Extraneous feasible intersections cannot exceed the true
maximum. This slower construction avoids assumptions about bit growth inside
a particular Voronoi implementation. Dyadic cell keys, triangle clipping and
the rational energy evaluations obey the same parameter-dependent bit bound.
No polygon complexity enters the exponential subset-enumeration count.

\paragraph{Metric arithmetic.}
In Theorem~\ref{thm:doubling} an arbitrary fixed real $\lambda$ is an upper
bound and the formulas use exact powers. For an integer dimension bound and
rational distance-oracle outputs, a fixed positive integer $s$ permits the
rational harmonic surrogate in~\eqref{eq:h-surrogate} with $d$ replaced by
that bound. Rounding the resulting separation scale down to a dyadic number
loses at most another factor of two in the net radius. Bit costs then depend
on the oracle output lengths and the cost of answering a query.

\section{Worked examples: seeing the grid at three densities}\label{app:walk}
The ten-site instance lets us identify every feasible replacement. The two instances with
$n=100$ use the same initial sites and illustrate how the placement of the other candidates
changes grid occupancy. All three use $k=5$.

\subsection{Five of ten: from cover to selection}\label{sec:epas-gallery}
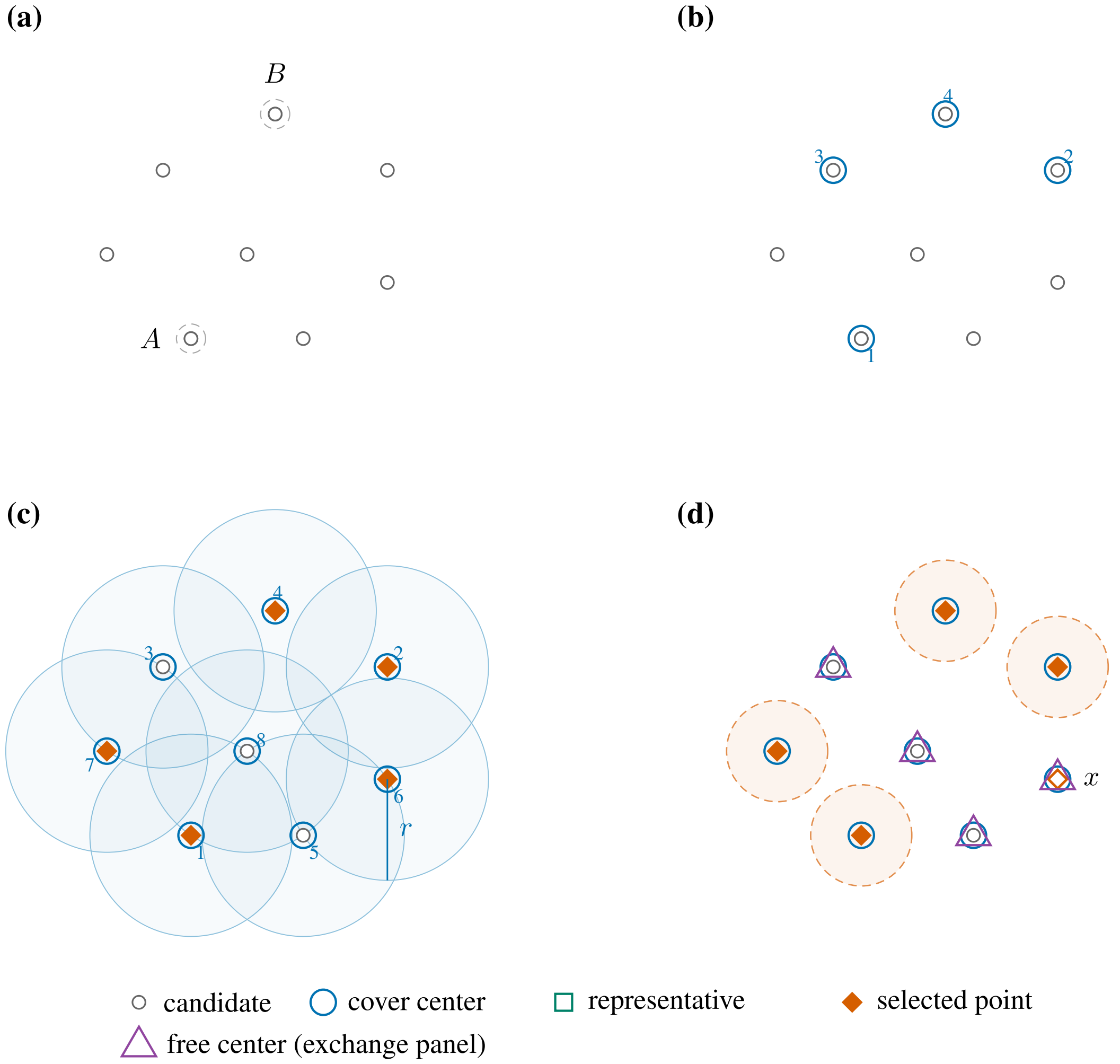
\begin{figure}[!htbp]\centering
\definecolor{epasblue}{RGB}{0,114,178}
\definecolor{epasgreen}{RGB}{0,130,105}
\definecolor{epasorange}{RGB}{213,94,0}
\definecolor{epaspurple}{RGB}{145,70,160}
\tikzset{
ginput/.style={circle,draw=black!60,fill=white,line width=.6pt,inner sep=0pt,minimum size=4.6pt},
gcenter/.style={circle,draw=epasblue,line width=.8pt,inner sep=0pt,minimum size=9pt},
gfree/.style={regular polygon,regular polygon sides=3,draw=epaspurple,line width=.8pt,inner sep=0pt,minimum size=14pt},
gkeep/.style={rectangle,draw=epasgreen,fill=white,line width=.8pt,inner sep=0pt,minimum size=6pt},
gchosen/.style={diamond,draw=epasorange,fill=epasorange,line width=.5pt,inner sep=0pt,minimum size=7pt},
gmove/.style={-{Stealth[length=1.6mm]},epasorange,line width=.8pt},
gcover/.style={draw=epasblue!45,line width=.35pt,fill=epasblue,fill opacity=.035},
gexcluded/.style={draw=epasorange!70,densely dashed,line width=.5pt,fill=epasorange,fill opacity=.07}
}
\begin{minipage}[t]{.49\textwidth}\centering
\begin{tikzpicture}[x=.35cm,y=.35cm]
\path[use as bounding box] (-7,-4.7) rectangle (11,12.3);
\node[anchor=north west,font=\small\bfseries] at (-7,12.3) {(a)};
\node[ginput] at (0.00000000,0.00000000) {};
\node[ginput] at (4.00000000,0.00000000) {};
\node[ginput] at (7.00000000,2.00000000) {};
\node[ginput] at (7.00000000,6.00000000) {};
\node[ginput] at (3.00000000,8.00000000) {};
\node[ginput] at (-1.00000000,6.00000000) {};
\node[ginput] at (-3.00000000,3.00000000) {};
\node[ginput] at (2.00000000,3.00000000) {};
\node[ginput] at (-0.00195312,-0.00195312) {};
\node[ginput] at (3.00000000,8.00195312) {};
\draw[black!35,densely dashed] (0,0) circle[radius=.52];
\node[anchor=east,font=\footnotesize] at (-.65,0) {$A$};
\draw[black!35,densely dashed] (3,8) circle[radius=.52];
\node[anchor=south,font=\footnotesize] at (3,8.7) {$B$};
\end{tikzpicture}
\end{minipage}\hfill\begin{minipage}[t]{.49\textwidth}\centering
\begin{tikzpicture}[x=.35cm,y=.35cm]
\path[use as bounding box] (-7,-4.7) rectangle (11,12.3);
\node[anchor=north west,font=\small\bfseries] at (-7,12.3) {(b)};
\node[ginput] at (0.00000000,0.00000000) {};
\node[ginput] at (4.00000000,0.00000000) {};
\node[ginput] at (7.00000000,2.00000000) {};
\node[ginput] at (7.00000000,6.00000000) {};
\node[ginput] at (3.00000000,8.00000000) {};
\node[ginput] at (-1.00000000,6.00000000) {};
\node[ginput] at (-3.00000000,3.00000000) {};
\node[ginput] at (2.00000000,3.00000000) {};
\node[ginput] at (-0.00195312,-0.00195312) {};
\node[ginput] at (3.00000000,8.00195312) {};
\node[gcenter] at (0.00000000,0.00000000) {};\node[text=epasblue,font=\scriptsize] at (0.35000000,-0.60000000) {1};
\node[gcenter] at (7.00000000,6.00000000) {};\node[text=epasblue,font=\scriptsize] at (7.40000000,6.50000000) {2};
\node[gcenter] at (-1.00000000,6.00000000) {};\node[text=epasblue,font=\scriptsize] at (-1.50000000,6.50000000) {3};
\node[gcenter] at (3.00000000,8.00195312) {};\node[text=epasblue,font=\scriptsize] at (3.10000000,8.65195313) {4};
\end{tikzpicture}
\end{minipage}\par\medskip\begin{minipage}[t]{.49\textwidth}\centering
\begin{tikzpicture}[x=.35cm,y=.35cm]
\path[use as bounding box] (-7,-4.7) rectangle (11,12.3);
\node[anchor=north west,font=\small\bfseries] at (-7,12.3) {(c)};
\draw[gcover] (0.00000000,0.00000000) circle[radius=3.60555128];
\draw[gcover] (7.00000000,6.00000000) circle[radius=3.60555128];
\draw[gcover] (-1.00000000,6.00000000) circle[radius=3.60555128];
\draw[gcover] (3.00000000,8.00195312) circle[radius=3.60555128];
\draw[gcover] (4.00000000,0.00000000) circle[radius=3.60555128];
\draw[gcover] (7.00000000,2.00000000) circle[radius=3.60555128];
\draw[gcover] (-3.00000000,3.00000000) circle[radius=3.60555128];
\draw[gcover] (2.00000000,3.00000000) circle[radius=3.60555128];
\node[ginput] at (0.00000000,0.00000000) {};
\node[ginput] at (4.00000000,0.00000000) {};
\node[ginput] at (7.00000000,2.00000000) {};
\node[ginput] at (7.00000000,6.00000000) {};
\node[ginput] at (3.00000000,8.00000000) {};
\node[ginput] at (-1.00000000,6.00000000) {};
\node[ginput] at (-3.00000000,3.00000000) {};
\node[ginput] at (2.00000000,3.00000000) {};
\node[ginput] at (-0.00195312,-0.00195312) {};
\node[ginput] at (3.00000000,8.00195312) {};
\node[gcenter] at (0.00000000,0.00000000) {};\node[text=epasblue,font=\scriptsize] at (0.35000000,-0.60000000) {1};
\node[gcenter] at (7.00000000,6.00000000) {};\node[text=epasblue,font=\scriptsize] at (7.40000000,6.50000000) {2};
\node[gcenter] at (-1.00000000,6.00000000) {};\node[text=epasblue,font=\scriptsize] at (-1.50000000,6.50000000) {3};
\node[gcenter] at (3.00000000,8.00195312) {};\node[text=epasblue,font=\scriptsize] at (3.10000000,8.65195313) {4};
\node[gcenter] at (4.00000000,0.00000000) {};\node[text=epasblue,font=\scriptsize] at (4.40000000,-0.60000000) {5};
\node[gcenter] at (7.00000000,2.00000000) {};\node[text=epasblue,font=\scriptsize] at (7.40000000,1.40000000) {6};
\node[gcenter] at (-3.00000000,3.00000000) {};\node[text=epasblue,font=\scriptsize] at (-3.60000000,2.50000000) {7};
\node[gcenter] at (2.00000000,3.00000000) {};\node[text=epasblue,font=\scriptsize] at (2.50000000,3.40000000) {8};
\node[gchosen] at (7.00000000,2.00000000) {};
\node[gchosen] at (7.00000000,6.00000000) {};
\node[gchosen] at (-3.00000000,3.00000000) {};
\node[gchosen] at (-0.00195312,-0.00195312) {};
\node[gchosen] at (3.00000000,8.00195312) {};\draw[epasblue,line width=.55pt] (7.00000000,2.00000000) -- ++(0,-3.60555128) node[midway,right,font=\footnotesize] {$r$};
\end{tikzpicture}
\end{minipage}\hfill\begin{minipage}[t]{.49\textwidth}\centering
\begin{tikzpicture}[x=.35cm,y=.35cm]
\path[use as bounding box] (-7,-4.7) rectangle (11,12.3);
\node[anchor=north west,font=\small\bfseries] at (-7,12.3) {(d)};
\draw[gexcluded] (7.00000000,6.00000000) circle[radius=1.80277564];
\draw[gexcluded] (-3.00000000,3.00000000) circle[radius=1.80277564];
\draw[gexcluded] (-0.00195312,-0.00195312) circle[radius=1.80277564];
\draw[gexcluded] (3.00000000,8.00195312) circle[radius=1.80277564];
\node[ginput] at (0.00000000,0.00000000) {};
\node[ginput] at (4.00000000,0.00000000) {};
\node[ginput] at (7.00000000,2.00000000) {};
\node[ginput] at (7.00000000,6.00000000) {};
\node[ginput] at (3.00000000,8.00000000) {};
\node[ginput] at (-1.00000000,6.00000000) {};
\node[ginput] at (-3.00000000,3.00000000) {};
\node[ginput] at (2.00000000,3.00000000) {};
\node[ginput] at (-0.00195312,-0.00195312) {};
\node[ginput] at (3.00000000,8.00195312) {};
\node[gcenter] at (0.00000000,0.00000000) {};
\node[gcenter] at (7.00000000,6.00000000) {};
\node[gcenter] at (-1.00000000,6.00000000) {};
\node[gcenter] at (3.00000000,8.00195312) {};
\node[gcenter] at (4.00000000,0.00000000) {};
\node[gcenter] at (7.00000000,2.00000000) {};
\node[gcenter] at (-3.00000000,3.00000000) {};
\node[gcenter] at (2.00000000,3.00000000) {};
\node[gchosen] at (7.00000000,6.00000000) {};
\node[gchosen] at (-3.00000000,3.00000000) {};
\node[gchosen] at (-0.00195312,-0.00195312) {};
\node[gchosen] at (3.00000000,8.00195312) {};\node[diamond,draw=epasorange,fill=white,minimum size=7pt,inner sep=0pt,line width=.9pt] at (7.00000000,2.00000000) {};\node[anchor=west,font=\footnotesize] at (7.50000000,2.00000000) {$x$};
\node[gfree] at (-1.00000000,6.00000000) {};
\node[gfree] at (4.00000000,0.00000000) {};
\node[gfree] at (7.00000000,2.00000000) {};
\node[gfree] at (2.00000000,3.00000000) {};
\end{tikzpicture}
\end{minipage}\par\smallskip
\begin{tikzpicture}[x=1cm,y=1cm,font=\footnotesize]
\node[ginput] at (0,0) {}; \node[anchor=west] at (.16,0) {candidate};
\node[gcenter] at (2.3,0) {}; \node[anchor=west] at (2.46,0) {cover center};
\node[gkeep] at (5.3,0) {}; \node[anchor=west] at (5.46,0) {representative};
\node[gchosen] at (8.9,0) {}; \node[anchor=west] at (9.06,0) {selected point};
\node[gfree] at (0,-.55) {}; \node[anchor=west] at (.20,-.55) {free center (exchange panel)};
\end{tikzpicture}

\caption[Five of ten sites: cover and exchange]{Cover and exchange for ten sites, $k=5$ and
$\varepsilon=1/2$. (a) The pairs at $A$ and $B$ are magnified in Figure~\ref{fig:epas-grid-run}.
(b) Numerals give the order of the first four farthest-first centers. (c) Blue rings mark the
eight centers of the $2k-2$ cover, with disks of radius $r=\sqrt{13}$. Orange diamonds mark
the optimum $\{p_3,p_4,p_7,p_9,p_{10}\}$. (d) To replace $x=p_3$ (outlined diamond),
exclude the open radius-$r/2$ disks of its four companions. The four free centers
$p_2,p_3,p_6,p_8$ are marked by purple triangles. Their average replacement cost bounds
the energy incident to $x$.}
\label{fig:epas-cover-run}
\end{figure}

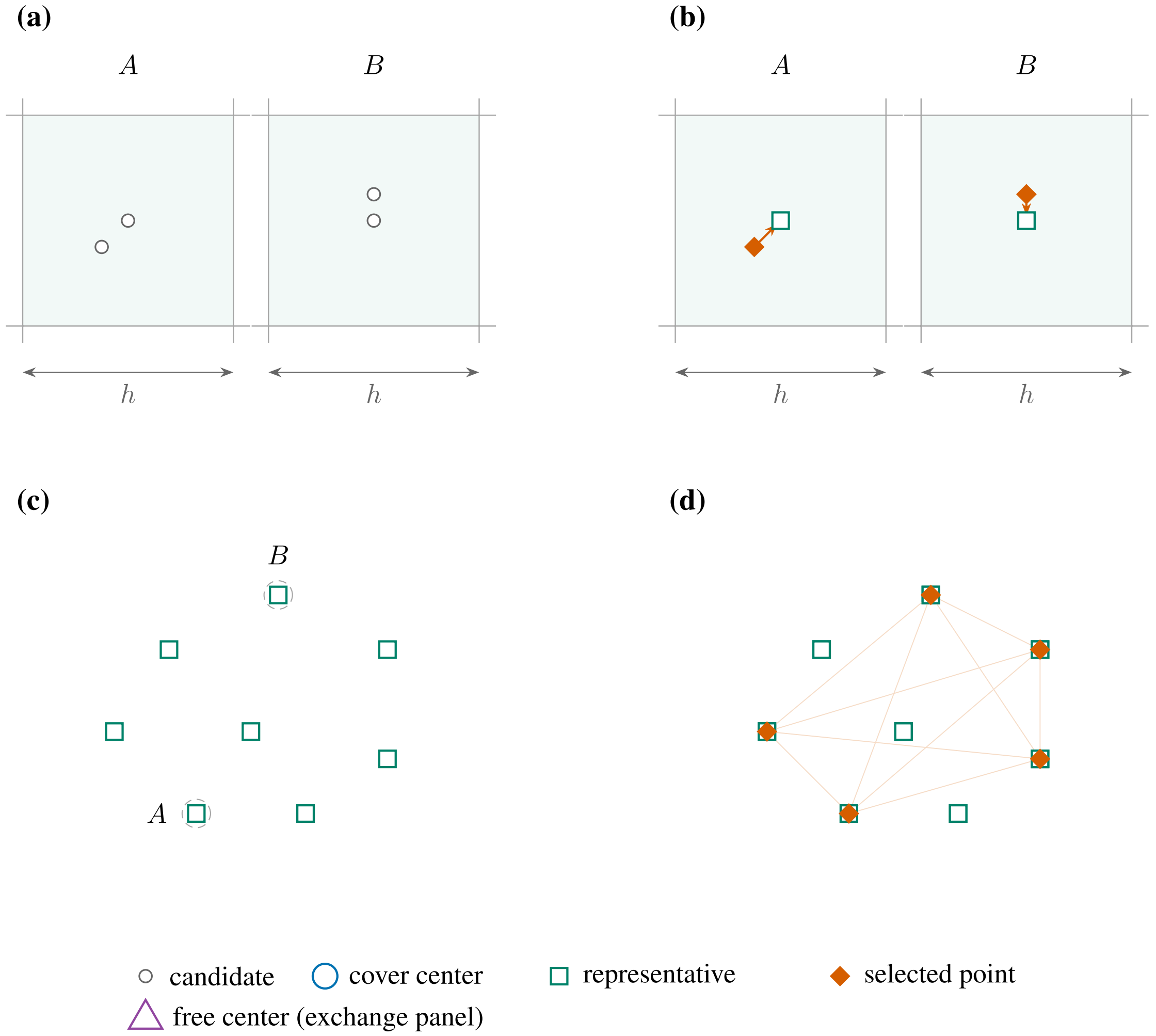
\begin{figure}[!htbp]\centering
\definecolor{epasblue}{RGB}{0,114,178}
\definecolor{epasgreen}{RGB}{0,130,105}
\definecolor{epasorange}{RGB}{213,94,0}
\definecolor{epaspurple}{RGB}{145,70,160}
\tikzset{
ginput/.style={circle,draw=black!60,fill=white,line width=.6pt,inner sep=0pt,minimum size=4.6pt},
gcenter/.style={circle,draw=epasblue,line width=.8pt,inner sep=0pt,minimum size=9pt},
gfree/.style={regular polygon,regular polygon sides=3,draw=epaspurple,line width=.8pt,inner sep=0pt,minimum size=14pt},
gkeep/.style={rectangle,draw=epasgreen,fill=white,line width=.8pt,inner sep=0pt,minimum size=6pt},
gchosen/.style={diamond,draw=epasorange,fill=epasorange,line width=.5pt,inner sep=0pt,minimum size=7pt},
gmove/.style={-{Stealth[length=1.6mm]},epasorange,line width=.8pt},
gcover/.style={draw=epasblue!45,line width=.35pt,fill=epasblue,fill opacity=.035},
gexcluded/.style={draw=epasorange!70,densely dashed,line width=.5pt,fill=epasorange,fill opacity=.07}
}
\begin{minipage}[t]{.49\textwidth}\centering
\begin{tikzpicture}[x=.35cm,y=.35cm]
\path[use as bounding box] (-7,-4.7) rectangle (11,12.3);
\node[anchor=north west,font=\small\bfseries] at (-7,12.3) {(a)};
\begin{scope}[shift={(-2.5,4.0)},x=2.7cm,y=2.7cm]
\fill[epasgreen!5] (-.5,-.5) rectangle (.5,.5);
\draw[black!35,line width=.45pt] (-.5,-.58)--(-.5,.58) (.5,-.58)--(.5,.58) (-.58,-.5)--(.58,-.5) (-.58,.5)--(.58,.5);
\node[font=\footnotesize] at (0,.74) {$A$};
\node[ginput] at (-0.12500000,-0.12500000) {};
\node[ginput] at (0,0) {};
\draw[<->,black!60,>=Stealth,line width=.4pt] (-.5,-.72)--(.5,-.72) node[midway,below,font=\footnotesize] {$h$};
\end{scope}
\begin{scope}[shift={(6.5,4.0)},x=2.7cm,y=2.7cm]
\fill[epasgreen!5] (-.5,-.5) rectangle (.5,.5);
\draw[black!35,line width=.45pt] (-.5,-.58)--(-.5,.58) (.5,-.58)--(.5,.58) (-.58,-.5)--(.58,-.5) (-.58,.5)--(.58,.5);
\node[font=\footnotesize] at (0,.74) {$B$};
\node[ginput] at (0.00000000,0.12500000) {};
\node[ginput] at (0,0) {};
\draw[<->,black!60,>=Stealth,line width=.4pt] (-.5,-.72)--(.5,-.72) node[midway,below,font=\footnotesize] {$h$};
\end{scope}
\end{tikzpicture}
\end{minipage}\hfill\begin{minipage}[t]{.49\textwidth}\centering
\begin{tikzpicture}[x=.35cm,y=.35cm]
\path[use as bounding box] (-7,-4.7) rectangle (11,12.3);
\node[anchor=north west,font=\small\bfseries] at (-7,12.3) {(b)};
\begin{scope}[shift={(-2.5,4.0)},x=2.7cm,y=2.7cm]
\fill[epasgreen!5] (-.5,-.5) rectangle (.5,.5);
\draw[black!35,line width=.45pt] (-.5,-.58)--(-.5,.58) (.5,-.58)--(.5,.58) (-.58,-.5)--(.58,-.5) (-.58,.5)--(.58,.5);
\node[font=\footnotesize] at (0,.74) {$A$};
\draw[gmove,shorten >=1.8pt,shorten <=1.8pt] (-0.12500000,-0.12500000) -- (0,0);
\node[gchosen] at (-0.12500000,-0.12500000) {};
\node[gkeep] at (0,0) {};
\draw[<->,black!60,>=Stealth,line width=.4pt] (-.5,-.72)--(.5,-.72) node[midway,below,font=\footnotesize] {$h$};
\end{scope}
\begin{scope}[shift={(6.5,4.0)},x=2.7cm,y=2.7cm]
\fill[epasgreen!5] (-.5,-.5) rectangle (.5,.5);
\draw[black!35,line width=.45pt] (-.5,-.58)--(-.5,.58) (.5,-.58)--(.5,.58) (-.58,-.5)--(.58,-.5) (-.58,.5)--(.58,.5);
\node[font=\footnotesize] at (0,.74) {$B$};
\draw[gmove,shorten >=1.8pt,shorten <=1.8pt] (0.00000000,0.12500000) -- (0,0);
\node[gchosen] at (0.00000000,0.12500000) {};
\node[gkeep] at (0,0) {};
\draw[<->,black!60,>=Stealth,line width=.4pt] (-.5,-.72)--(.5,-.72) node[midway,below,font=\footnotesize] {$h$};
\end{scope}
\end{tikzpicture}
\end{minipage}\par\medskip\begin{minipage}[t]{.49\textwidth}\centering
\begin{tikzpicture}[x=.35cm,y=.35cm]
\path[use as bounding box] (-7,-4.7) rectangle (11,12.3);
\node[anchor=north west,font=\small\bfseries] at (-7,12.3) {(c)};
\node[ginput] at (0.00000000,0.00000000) {};
\node[ginput] at (4.00000000,0.00000000) {};
\node[ginput] at (7.00000000,2.00000000) {};
\node[ginput] at (7.00000000,6.00000000) {};
\node[ginput] at (3.00000000,8.00000000) {};
\node[ginput] at (-1.00000000,6.00000000) {};
\node[ginput] at (-3.00000000,3.00000000) {};
\node[ginput] at (2.00000000,3.00000000) {};
\node[ginput] at (-0.00195312,-0.00195312) {};
\node[ginput] at (3.00000000,8.00195312) {};
\node[gkeep] at (0.00000000,0.00000000) {};
\node[gkeep] at (4.00000000,0.00000000) {};
\node[gkeep] at (7.00000000,2.00000000) {};
\node[gkeep] at (7.00000000,6.00000000) {};
\node[gkeep] at (3.00000000,8.00000000) {};
\node[gkeep] at (-1.00000000,6.00000000) {};
\node[gkeep] at (-3.00000000,3.00000000) {};
\node[gkeep] at (2.00000000,3.00000000) {};
\draw[black!35,densely dashed] (0,0) circle[radius=.52];
\node[anchor=east,font=\footnotesize] at (-.65,0) {$A$};
\draw[black!35,densely dashed] (3,8) circle[radius=.52];
\node[anchor=south,font=\footnotesize] at (3,8.7) {$B$};
\end{tikzpicture}
\end{minipage}\hfill\begin{minipage}[t]{.49\textwidth}\centering
\begin{tikzpicture}[x=.35cm,y=.35cm]
\path[use as bounding box] (-7,-4.7) rectangle (11,12.3);
\node[anchor=north west,font=\small\bfseries] at (-7,12.3) {(d)};
\node[gkeep] at (0.00000000,0.00000000) {};
\node[gkeep] at (4.00000000,0.00000000) {};
\node[gkeep] at (7.00000000,2.00000000) {};
\node[gkeep] at (7.00000000,6.00000000) {};
\node[gkeep] at (3.00000000,8.00000000) {};
\node[gkeep] at (-1.00000000,6.00000000) {};
\node[gkeep] at (-3.00000000,3.00000000) {};
\node[gkeep] at (2.00000000,3.00000000) {};
\draw[epasorange!22,line width=.35pt] (0.00000000,0.00000000) -- (7.00000000,2.00000000);
\draw[epasorange!22,line width=.35pt] (0.00000000,0.00000000) -- (7.00000000,6.00000000);
\draw[epasorange!22,line width=.35pt] (0.00000000,0.00000000) -- (3.00000000,8.00000000);
\draw[epasorange!22,line width=.35pt] (0.00000000,0.00000000) -- (-3.00000000,3.00000000);
\draw[epasorange!22,line width=.35pt] (7.00000000,2.00000000) -- (7.00000000,6.00000000);
\draw[epasorange!22,line width=.35pt] (7.00000000,2.00000000) -- (3.00000000,8.00000000);
\draw[epasorange!22,line width=.35pt] (7.00000000,2.00000000) -- (-3.00000000,3.00000000);
\draw[epasorange!22,line width=.35pt] (7.00000000,6.00000000) -- (3.00000000,8.00000000);
\draw[epasorange!22,line width=.35pt] (7.00000000,6.00000000) -- (-3.00000000,3.00000000);
\draw[epasorange!22,line width=.35pt] (3.00000000,8.00000000) -- (-3.00000000,3.00000000);
\node[gkeep] at (0.00000000,0.00000000) {};
\node[gkeep] at (4.00000000,0.00000000) {};
\node[gkeep] at (7.00000000,2.00000000) {};
\node[gkeep] at (7.00000000,6.00000000) {};
\node[gkeep] at (3.00000000,8.00000000) {};
\node[gkeep] at (-1.00000000,6.00000000) {};
\node[gkeep] at (-3.00000000,3.00000000) {};
\node[gkeep] at (2.00000000,3.00000000) {};
\node[gchosen] at (0.00000000,0.00000000) {};
\node[gchosen] at (7.00000000,2.00000000) {};
\node[gchosen] at (7.00000000,6.00000000) {};
\node[gchosen] at (3.00000000,8.00000000) {};
\node[gchosen] at (-3.00000000,3.00000000) {};
\end{tikzpicture}
\end{minipage}\par\smallskip
\begin{tikzpicture}[x=1cm,y=1cm,font=\footnotesize]
\node[ginput] at (0,0) {}; \node[anchor=west] at (.16,0) {candidate};
\node[gcenter] at (2.3,0) {}; \node[anchor=west] at (2.46,0) {cover center};
\node[gkeep] at (5.3,0) {}; \node[anchor=west] at (5.46,0) {representative};
\node[gchosen] at (8.9,0) {}; \node[anchor=west] at (9.06,0) {selected point};
\node[gfree] at (0,-.55) {}; \node[anchor=west] at (.20,-.55) {free center (exchange panel)};
\end{tikzpicture}

\caption[Five of ten sites: grid and selection]{Rounding and search on the same ten sites.
(a) The two magnified cells have side $h=1/64$ and belong to the grid with origin
$(-1/128,-1/128)$. Cell $A$ contains $p_1,p_9$ and cell $B$ contains $p_5,p_{10}$.
(b) The first site in each cell is retained (green square). Two sites of the optimum
(orange diamonds) move under rounding, $p_9\mapsto p_1$ and $p_{10}\mapsto p_5$.
The other three stay fixed. (c) The grid retains eight representatives.
(d) Searching their $\binom85=56$ five-site subsets returns $\{p_1,p_3,p_4,p_5,p_7\}$.
Segments show its ten interacting pairs. The certified interval is $[2V/3,V]$.
An independent exhaustive search gives $V/\OPT\approx1.000250627$.}
\label{fig:epas-grid-run}
\end{figure}

Figures~\ref{fig:epas-cover-run}--\ref{fig:epas-grid-run} illustrate $d=s=2$, $n=10$, $k=5$, and
$\varepsilon=1/2$. The $2k$ implementation enumerates directly when $n=2k$; the $2k-2$ variant
below makes both covering and rounding visible even on this small input.

\begin{proposition}[Two fewer covering centers]\label{prop:epas-compact-cover}
For $k\ge2$, the planar construction remains valid with $m=2k-2$ farthest-first centers, $H=H_m$,
and $B=16H/r^2$, using direct enumeration when $n\le m$. It has the same approximation guarantee
and asymptotic bounds as Corollary~\ref{thm:k-epas}.
\end{proposition}
\begin{proof}
For any selected $x$, its $k-1$ companions exclude at most $k-1$ centers, so at least
$m-(k-1)=k-1$ free centers remain. In Lemma~\ref{lem:exchange}, replace $2k$ by $m$: the averaging
factor $(k-1)/|T_x|$ is still at most one. The packing sum is bounded by $16r^{-2}H_m$.
Farthest-first still gives an $r$-separated cover, and the grid, injectivity, and bit arguments
are unchanged. Also $m=O(k)$, $H_m=O(\log(2k))$, and $n\le m$ implies $\binom nk\le4^k$. This
changes the number of covering centers by a constant and preserves the approximation principle.
\end{proof}

\paragraph{Implementation.}
The program \path{planar_riesz_epas.py} uses Python's standard-library \texttt{Fraction}
throughout. Default \texttt{paper} mode follows the $2k$ construction; \texttt{compact} uses
Proposition~\ref{prop:epas-compact-cover}. It maintains nearest squared distances, constructs
dyadic scales by binary lengths, sorts integer cell keys, and tabulates only representative
weights. Input order breaks ties and chooses cell representatives. A translated global grid has
the same proof, with keys $\lfloor(p-o)/h\rfloor$. The solver optimizes over the representatives. A separate exhaustive search provides
the full-input optimum for comparison.

The ten input points, in order, are as follows, with $t=1/512$:
\[
\begin{array}{lllll}
p_1=(0,0),&p_2=(4,0),&p_3=(7,2),&p_4=(7,6),&p_5=(3,8),\\[3pt]
p_6=(-1,6),&p_7=(-3,3),&p_8=(2,3),&p_9=(-t,-t),&p_{10}=(3,8+t),
\end{array}
\]
The two deliberate near pairs require magnification. Starting at $p_1$, the trace gives $\mathcal C=(p_1,p_4,p_6,p_{10},p_2,p_3,p_7,p_8)$, $r^2=13$, $H_8=761/280$, and $B=1522/455$.

The insertion squared distances are
$85,\allowbreak37,\allowbreak5244929/262144,\allowbreak16,\allowbreak13,\allowbreak13,\allowbreak13$.
The final covering radius is actually $\sqrt{2}/512$, but the proof uses the \emph{last insertion
radius} $r=\sqrt{13}$. The dyadic choice $\widehat\delta=1/2$ satisfies
$B\widehat\delta^2\le1<B(2\widehat\delta)^2$, giving $h=1/64$. With origin $o=(-h/2,-h/2)$,
exactly two cells contain pairs; the first-point rule leaves $Q=\{p_1,\ldots,p_8\}$.

\begin{center}\small
\begin{tabular}{@{}lll@{}}\toprule
Stage & Exact result & Role\\\midrule
Cover & $8$ centers, $r^2=13$ & Certifies the rounding scale\\
Grid & $10\to8$ candidates & Merges only the two near pairs\\
Representative search & $\binom85=56$ subsets & Produces the returned solution\\
Independent audit & $\binom{10}5=252$ subsets & Checks the unknown original optimum\\
Certificate & $[2V/3,V]$ & Valid without the exhaustive audit\\\bottomrule
\end{tabular}
\end{center}
The returned set is $S_Q=\{p_1,p_3,p_4,p_5,p_7\}$, with $V=\frac{110421302865713}{413429256648720}\approx0.267086330$. The independent verifier finds the unique original optimum $S^\star=\{p_3,p_4,p_7,p_9,p_{10}\}$,
of energy $0.267019408$ approximately. Its cellwise replacement equals $S_Q$ here, giving the
nonzero-error ratio $1.000250627<1.5$. This equality of the replacement and returned set is not
needed in general. The $2k$ mode instead takes the exact $252$-subset shortcut.

From \path{anc/walkthroughs/}, run the independent audit with
\begin{quote}
\verb|python epas_gallery_run.py|
\end{quote}
This runs both the compact and standard modes, then enumerates the original
ten-site problem independently. It writes \path{epas_gallery_results.json} and checks
78 additional cases covering cardinalities, coordinate signs, scales, translations and
accuracy endpoints, as well as four invalid inputs.

\FloatBarrier

\subsection{A hundred sites spread across the same cover}\label{sec:epas-hundred}
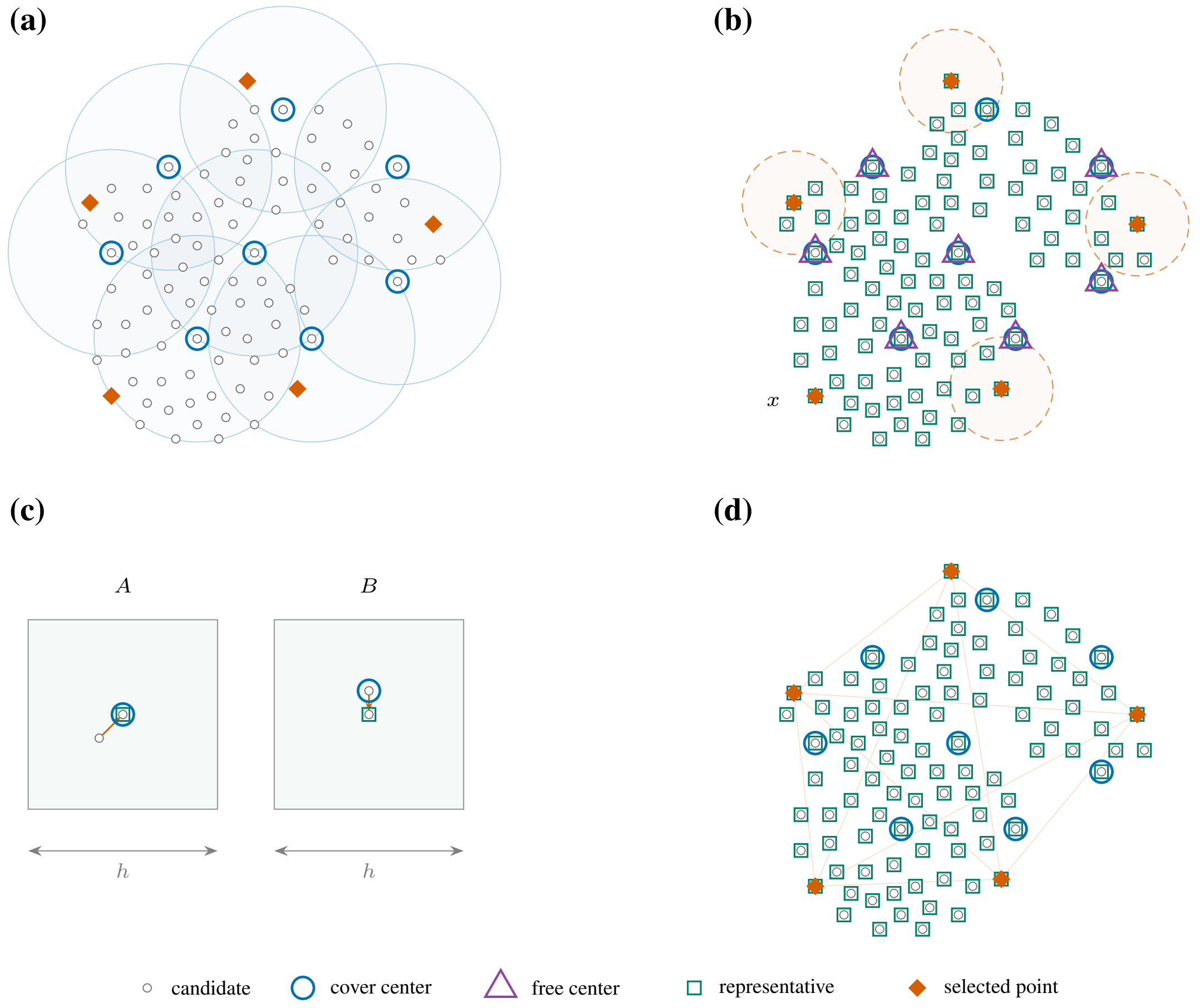
\begin{figure}[!htbp]\centering
\definecolor{epasblue}{RGB}{0,114,178}
\definecolor{epasgreen}{RGB}{0,130,105}
\definecolor{epasorange}{RGB}{213,94,0}
\definecolor{epaspurple}{RGB}{145,70,160}
\tikzset{
spinput/.style={circle,draw=black!55,fill=white,line width=.35pt,inner sep=0pt,minimum size=2.8pt},
spcenter/.style={circle,draw=epasblue,line width=.9pt,inner sep=0pt,minimum size=7.4pt},
spfree/.style={regular polygon,regular polygon sides=3,draw=epaspurple,line width=.8pt,inner sep=0pt,minimum size=12pt},
spkeep/.style={rectangle,draw=epasgreen,line width=.6pt,inner sep=0pt,minimum size=4.4pt},
spchosen/.style={diamond,draw=epasorange,fill=epasorange,line width=.5pt,inner sep=0pt,minimum size=5.6pt},
spcover/.style={draw=epasblue!35,line width=.3pt,fill=epasblue,fill opacity=.018},
spexcluded/.style={draw=epasorange!65,densely dashed,line width=.45pt,fill=epasorange,fill opacity=.035}
}
\begin{minipage}[c]{.49\textwidth}\centering
\begin{tikzpicture}[x=0.34cm,y=0.34cm]
\path[use as bounding box] (-7,-4.4) rectangle (11,12.0);
\node[anchor=north west,font=\small\bfseries] at (-7,12) {(a)};
\draw[spcover] (0.0000000000,0.0000000000) circle[radius=3.6055512755];
\draw[spcover] (7.0000000000,6.0000000000) circle[radius=3.6055512755];
\draw[spcover] (-1.0000000000,6.0000000000) circle[radius=3.6055512755];
\draw[spcover] (3.0000000000,8.0019531250) circle[radius=3.6055512755];
\draw[spcover] (4.0000000000,0.0000000000) circle[radius=3.6055512755];
\draw[spcover] (7.0000000000,2.0000000000) circle[radius=3.6055512755];
\draw[spcover] (-3.0000000000,3.0000000000) circle[radius=3.6055512755];
\draw[spcover] (2.0000000000,3.0000000000) circle[radius=3.6055512755];
\node[spinput] at (0.0000000000,0.0000000000) {};
\node[spinput] at (4.0000000000,0.0000000000) {};
\node[spinput] at (7.0000000000,2.0000000000) {};
\node[spinput] at (7.0000000000,6.0000000000) {};
\node[spinput] at (3.0000000000,8.0000000000) {};
\node[spinput] at (-1.0000000000,6.0000000000) {};
\node[spinput] at (-3.0000000000,3.0000000000) {};
\node[spinput] at (2.0000000000,3.0000000000) {};
\node[spinput] at (-0.0019531250,-0.0019531250) {};
\node[spinput] at (3.0000000000,8.0019531250) {};
\node[spinput] at (2.0000000000,-3.0000000000) {};
\node[spinput] at (-3.0000000000,-2.0000000000) {};
\node[spinput] at (4.0000000000,5.2500000000) {};
\node[spinput] at (-0.7500000000,-3.5000000000) {};
\node[spinput] at (-3.5000000000,0.5000000000) {};
\node[spinput] at (-0.5000000000,2.5000000000) {};
\node[spinput] at (1.5000000000,5.5000000000) {};
\node[spinput] at (4.7500000000,2.7500000000) {};
\node[spinput] at (8.2500000000,4.0000000000) {};
\node[spinput] at (5.2500000000,7.5000000000) {};
\node[spinput] at (-3.0000000000,5.2500000000) {};
\node[spinput] at (2.0000000000,-0.7500000000) {};
\node[spinput] at (6.0000000000,4.2500000000) {};
\node[spinput] at (-1.7500000000,1.0000000000) {};
\node[spinput] at (-1.2500000000,-1.5000000000) {};
\node[spinput] at (0.0000000000,4.2500000000) {};
\node[spinput] at (1.2500000000,7.5000000000) {};
\node[spinput] at (1.5000000000,1.2500000000) {};
\node[spinput] at (0.5000000000,-2.0000000000) {};
\node[spinput] at (3.5000000000,-1.7500000000) {};
\node[spinput] at (3.2500000000,1.7500000000) {};
\node[spinput] at (-1.7500000000,4.0000000000) {};
\node[spinput] at (1.7500000000,9.0000000000) {};
\node[spinput] at (2.7500000000,6.5000000000) {};
\node[spinput] at (5.5000000000,5.7500000000) {};
\node[spinput] at (2.7500000000,4.5000000000) {};
\node[spinput] at (-4.0000000000,4.0000000000) {};
\node[spinput] at (-2.5000000000,-0.5000000000) {};
\node[spinput] at (-2.0000000000,-3.0000000000) {};
\node[spinput] at (0.7500000000,-3.5000000000) {};
\node[spinput] at (2.7500000000,0.5000000000) {};
\node[spinput] at (4.0000000000,7.0000000000) {};
\node[spinput] at (-0.2500000000,1.2500000000) {};
\node[spinput] at (0.2500000000,5.7500000000) {};
\node[spinput] at (0.7500000000,2.7500000000) {};
\node[spinput] at (4.2500000000,4.0000000000) {};
\node[spinput] at (8.5000000000,2.7500000000) {};
\node[spinput] at (-3.0000000000,1.7500000000) {};
\node[spinput] at (-1.7500000000,2.2500000000) {};
\node[spinput] at (-1.2500000000,-0.2500000000) {};
\node[spinput] at (1.2500000000,4.0000000000) {};
\node[spinput] at (6.0000000000,2.7500000000) {};
\node[spinput] at (7.0000000000,3.5000000000) {};
\node[spinput] at (7.2500000000,4.7500000000) {};
\node[spinput] at (-1.7500000000,5.2500000000) {};
\node[spinput] at (0.7500000000,-0.7500000000) {};
\node[spinput] at (5.0000000000,4.7500000000) {};
\node[spinput] at (6.0000000000,6.7500000000) {};
\node[spinput] at (-3.5000000000,-0.7500000000) {};
\node[spinput] at (-2.7500000000,4.2500000000) {};
\node[spinput] at (-1.0000000000,-2.5000000000) {};
\node[spinput] at (-0.7500000000,3.5000000000) {};
\node[spinput] at (-0.7500000000,5.0000000000) {};
\node[spinput] at (-0.2500000000,-1.2500000000) {};
\node[spinput] at (1.0000000000,0.2500000000) {};
\node[spinput] at (1.0000000000,6.5000000000) {};
\node[spinput] at (1.5000000000,-1.7500000000) {};
\node[spinput] at (2.2500000000,2.0000000000) {};
\node[spinput] at (2.5000000000,-2.0000000000) {};
\node[spinput] at (3.0000000000,-0.5000000000) {};
\node[spinput] at (3.0000000000,5.5000000000) {};
\node[spinput] at (4.2500000000,8.0000000000) {};
\node[spinput] at (-2.5000000000,0.5000000000) {};
\node[spinput] at (-2.2500000000,-1.5000000000) {};
\node[spinput] at (-1.0000000000,1.7500000000) {};
\node[spinput] at (0.0000000000,-3.0000000000) {};
\node[spinput] at (0.2500000000,2.0000000000) {};
\node[spinput] at (0.7500000000,4.7500000000) {};
\node[spinput] at (2.0000000000,7.0000000000) {};
\node[spinput] at (2.0000000000,8.0000000000) {};
\node[spinput] at (3.7500000000,1.0000000000) {};
\node[spinput] at (4.5000000000,6.0000000000) {};
\node[spinput] at (5.2500000000,3.5000000000) {};
\node[spinput] at (6.2500000000,5.2500000000) {};
\node[spinput] at (7.5000000000,2.7500000000) {};
\node[spinput] at (-0.7500000000,0.5000000000) {};
\node[spinput] at (-3.7500000000,4.7500000000) {};
\node[spinput] at (-2.2500000000,3.2500000000) {};
\node[spinput] at (-1.7500000000,-2.2500000000) {};
\node[spinput] at (-1.5000000000,3.0000000000) {};
\node[spinput] at (-1.0000000000,4.2500000000) {};
\node[spinput] at (-0.2500000000,-2.2500000000) {};
\node[spinput] at (0.0000000000,3.2500000000) {};
\node[spinput] at (0.5000000000,1.0000000000) {};
\node[spinput] at (1.0000000000,-2.7500000000) {};
\node[spinput] at (1.2500000000,2.0000000000) {};
\node[spinput] at (1.7500000000,0.0000000000) {};
\node[spinput] at (1.7500000000,4.7500000000) {};
\node[spinput] at (1.7500000000,6.2500000000) {};
\node[spinput] at (2.5000000000,1.2500000000) {};
\node[spcenter] at (0.0000000000,0.0000000000) {};
\node[spcenter] at (7.0000000000,6.0000000000) {};
\node[spcenter] at (-1.0000000000,6.0000000000) {};
\node[spcenter] at (3.0000000000,8.0019531250) {};
\node[spcenter] at (4.0000000000,0.0000000000) {};
\node[spcenter] at (7.0000000000,2.0000000000) {};
\node[spcenter] at (-3.0000000000,3.0000000000) {};
\node[spcenter] at (2.0000000000,3.0000000000) {};
\node[spchosen] at (-3.0000000000,-2.0000000000) {};
\node[spchosen] at (8.2500000000,4.0000000000) {};
\node[spchosen] at (3.5000000000,-1.7500000000) {};
\node[spchosen] at (1.7500000000,9.0000000000) {};
\node[spchosen] at (-3.7500000000,4.7500000000) {};
\end{tikzpicture}
\end{minipage}\hfill\begin{minipage}[c]{.49\textwidth}\centering
\begin{tikzpicture}[x=0.34cm,y=0.34cm]
\path[use as bounding box] (-7,-4.4) rectangle (11,12.0);
\node[anchor=north west,font=\small\bfseries] at (-7,12) {(b)};
\draw[spexcluded] (8.2500000000,4.0000000000) circle[radius=1.8027756377];
\draw[spexcluded] (3.5000000000,-1.7500000000) circle[radius=1.8027756377];
\draw[spexcluded] (1.7500000000,9.0000000000) circle[radius=1.8027756377];
\draw[spexcluded] (-3.7500000000,4.7500000000) circle[radius=1.8027756377];
\node[spinput] at (0.0000000000,0.0000000000) {};
\node[spinput] at (4.0000000000,0.0000000000) {};
\node[spinput] at (7.0000000000,2.0000000000) {};
\node[spinput] at (7.0000000000,6.0000000000) {};
\node[spinput] at (3.0000000000,8.0000000000) {};
\node[spinput] at (-1.0000000000,6.0000000000) {};
\node[spinput] at (-3.0000000000,3.0000000000) {};
\node[spinput] at (2.0000000000,3.0000000000) {};
\node[spinput] at (-0.0019531250,-0.0019531250) {};
\node[spinput] at (3.0000000000,8.0019531250) {};
\node[spinput] at (2.0000000000,-3.0000000000) {};
\node[spinput] at (-3.0000000000,-2.0000000000) {};
\node[spinput] at (4.0000000000,5.2500000000) {};
\node[spinput] at (-0.7500000000,-3.5000000000) {};
\node[spinput] at (-3.5000000000,0.5000000000) {};
\node[spinput] at (-0.5000000000,2.5000000000) {};
\node[spinput] at (1.5000000000,5.5000000000) {};
\node[spinput] at (4.7500000000,2.7500000000) {};
\node[spinput] at (8.2500000000,4.0000000000) {};
\node[spinput] at (5.2500000000,7.5000000000) {};
\node[spinput] at (-3.0000000000,5.2500000000) {};
\node[spinput] at (2.0000000000,-0.7500000000) {};
\node[spinput] at (6.0000000000,4.2500000000) {};
\node[spinput] at (-1.7500000000,1.0000000000) {};
\node[spinput] at (-1.2500000000,-1.5000000000) {};
\node[spinput] at (0.0000000000,4.2500000000) {};
\node[spinput] at (1.2500000000,7.5000000000) {};
\node[spinput] at (1.5000000000,1.2500000000) {};
\node[spinput] at (0.5000000000,-2.0000000000) {};
\node[spinput] at (3.5000000000,-1.7500000000) {};
\node[spinput] at (3.2500000000,1.7500000000) {};
\node[spinput] at (-1.7500000000,4.0000000000) {};
\node[spinput] at (1.7500000000,9.0000000000) {};
\node[spinput] at (2.7500000000,6.5000000000) {};
\node[spinput] at (5.5000000000,5.7500000000) {};
\node[spinput] at (2.7500000000,4.5000000000) {};
\node[spinput] at (-4.0000000000,4.0000000000) {};
\node[spinput] at (-2.5000000000,-0.5000000000) {};
\node[spinput] at (-2.0000000000,-3.0000000000) {};
\node[spinput] at (0.7500000000,-3.5000000000) {};
\node[spinput] at (2.7500000000,0.5000000000) {};
\node[spinput] at (4.0000000000,7.0000000000) {};
\node[spinput] at (-0.2500000000,1.2500000000) {};
\node[spinput] at (0.2500000000,5.7500000000) {};
\node[spinput] at (0.7500000000,2.7500000000) {};
\node[spinput] at (4.2500000000,4.0000000000) {};
\node[spinput] at (8.5000000000,2.7500000000) {};
\node[spinput] at (-3.0000000000,1.7500000000) {};
\node[spinput] at (-1.7500000000,2.2500000000) {};
\node[spinput] at (-1.2500000000,-0.2500000000) {};
\node[spinput] at (1.2500000000,4.0000000000) {};
\node[spinput] at (6.0000000000,2.7500000000) {};
\node[spinput] at (7.0000000000,3.5000000000) {};
\node[spinput] at (7.2500000000,4.7500000000) {};
\node[spinput] at (-1.7500000000,5.2500000000) {};
\node[spinput] at (0.7500000000,-0.7500000000) {};
\node[spinput] at (5.0000000000,4.7500000000) {};
\node[spinput] at (6.0000000000,6.7500000000) {};
\node[spinput] at (-3.5000000000,-0.7500000000) {};
\node[spinput] at (-2.7500000000,4.2500000000) {};
\node[spinput] at (-1.0000000000,-2.5000000000) {};
\node[spinput] at (-0.7500000000,3.5000000000) {};
\node[spinput] at (-0.7500000000,5.0000000000) {};
\node[spinput] at (-0.2500000000,-1.2500000000) {};
\node[spinput] at (1.0000000000,0.2500000000) {};
\node[spinput] at (1.0000000000,6.5000000000) {};
\node[spinput] at (1.5000000000,-1.7500000000) {};
\node[spinput] at (2.2500000000,2.0000000000) {};
\node[spinput] at (2.5000000000,-2.0000000000) {};
\node[spinput] at (3.0000000000,-0.5000000000) {};
\node[spinput] at (3.0000000000,5.5000000000) {};
\node[spinput] at (4.2500000000,8.0000000000) {};
\node[spinput] at (-2.5000000000,0.5000000000) {};
\node[spinput] at (-2.2500000000,-1.5000000000) {};
\node[spinput] at (-1.0000000000,1.7500000000) {};
\node[spinput] at (0.0000000000,-3.0000000000) {};
\node[spinput] at (0.2500000000,2.0000000000) {};
\node[spinput] at (0.7500000000,4.7500000000) {};
\node[spinput] at (2.0000000000,7.0000000000) {};
\node[spinput] at (2.0000000000,8.0000000000) {};
\node[spinput] at (3.7500000000,1.0000000000) {};
\node[spinput] at (4.5000000000,6.0000000000) {};
\node[spinput] at (5.2500000000,3.5000000000) {};
\node[spinput] at (6.2500000000,5.2500000000) {};
\node[spinput] at (7.5000000000,2.7500000000) {};
\node[spinput] at (-0.7500000000,0.5000000000) {};
\node[spinput] at (-3.7500000000,4.7500000000) {};
\node[spinput] at (-2.2500000000,3.2500000000) {};
\node[spinput] at (-1.7500000000,-2.2500000000) {};
\node[spinput] at (-1.5000000000,3.0000000000) {};
\node[spinput] at (-1.0000000000,4.2500000000) {};
\node[spinput] at (-0.2500000000,-2.2500000000) {};
\node[spinput] at (0.0000000000,3.2500000000) {};
\node[spinput] at (0.5000000000,1.0000000000) {};
\node[spinput] at (1.0000000000,-2.7500000000) {};
\node[spinput] at (1.2500000000,2.0000000000) {};
\node[spinput] at (1.7500000000,0.0000000000) {};
\node[spinput] at (1.7500000000,4.7500000000) {};
\node[spinput] at (1.7500000000,6.2500000000) {};
\node[spinput] at (2.5000000000,1.2500000000) {};
\node[spkeep] at (0.0000000000,0.0000000000) {};
\node[spkeep] at (4.0000000000,0.0000000000) {};
\node[spkeep] at (7.0000000000,2.0000000000) {};
\node[spkeep] at (7.0000000000,6.0000000000) {};
\node[spkeep] at (3.0000000000,8.0000000000) {};
\node[spkeep] at (-1.0000000000,6.0000000000) {};
\node[spkeep] at (-3.0000000000,3.0000000000) {};
\node[spkeep] at (2.0000000000,3.0000000000) {};
\node[spkeep] at (2.0000000000,-3.0000000000) {};
\node[spkeep] at (-3.0000000000,-2.0000000000) {};
\node[spkeep] at (4.0000000000,5.2500000000) {};
\node[spkeep] at (-0.7500000000,-3.5000000000) {};
\node[spkeep] at (-3.5000000000,0.5000000000) {};
\node[spkeep] at (-0.5000000000,2.5000000000) {};
\node[spkeep] at (1.5000000000,5.5000000000) {};
\node[spkeep] at (4.7500000000,2.7500000000) {};
\node[spkeep] at (8.2500000000,4.0000000000) {};
\node[spkeep] at (5.2500000000,7.5000000000) {};
\node[spkeep] at (-3.0000000000,5.2500000000) {};
\node[spkeep] at (2.0000000000,-0.7500000000) {};
\node[spkeep] at (6.0000000000,4.2500000000) {};
\node[spkeep] at (-1.7500000000,1.0000000000) {};
\node[spkeep] at (-1.2500000000,-1.5000000000) {};
\node[spkeep] at (0.0000000000,4.2500000000) {};
\node[spkeep] at (1.2500000000,7.5000000000) {};
\node[spkeep] at (1.5000000000,1.2500000000) {};
\node[spkeep] at (0.5000000000,-2.0000000000) {};
\node[spkeep] at (3.5000000000,-1.7500000000) {};
\node[spkeep] at (3.2500000000,1.7500000000) {};
\node[spkeep] at (-1.7500000000,4.0000000000) {};
\node[spkeep] at (1.7500000000,9.0000000000) {};
\node[spkeep] at (2.7500000000,6.5000000000) {};
\node[spkeep] at (5.5000000000,5.7500000000) {};
\node[spkeep] at (2.7500000000,4.5000000000) {};
\node[spkeep] at (-4.0000000000,4.0000000000) {};
\node[spkeep] at (-2.5000000000,-0.5000000000) {};
\node[spkeep] at (-2.0000000000,-3.0000000000) {};
\node[spkeep] at (0.7500000000,-3.5000000000) {};
\node[spkeep] at (2.7500000000,0.5000000000) {};
\node[spkeep] at (4.0000000000,7.0000000000) {};
\node[spkeep] at (-0.2500000000,1.2500000000) {};
\node[spkeep] at (0.2500000000,5.7500000000) {};
\node[spkeep] at (0.7500000000,2.7500000000) {};
\node[spkeep] at (4.2500000000,4.0000000000) {};
\node[spkeep] at (8.5000000000,2.7500000000) {};
\node[spkeep] at (-3.0000000000,1.7500000000) {};
\node[spkeep] at (-1.7500000000,2.2500000000) {};
\node[spkeep] at (-1.2500000000,-0.2500000000) {};
\node[spkeep] at (1.2500000000,4.0000000000) {};
\node[spkeep] at (6.0000000000,2.7500000000) {};
\node[spkeep] at (7.0000000000,3.5000000000) {};
\node[spkeep] at (7.2500000000,4.7500000000) {};
\node[spkeep] at (-1.7500000000,5.2500000000) {};
\node[spkeep] at (0.7500000000,-0.7500000000) {};
\node[spkeep] at (5.0000000000,4.7500000000) {};
\node[spkeep] at (6.0000000000,6.7500000000) {};
\node[spkeep] at (-3.5000000000,-0.7500000000) {};
\node[spkeep] at (-2.7500000000,4.2500000000) {};
\node[spkeep] at (-1.0000000000,-2.5000000000) {};
\node[spkeep] at (-0.7500000000,3.5000000000) {};
\node[spkeep] at (-0.7500000000,5.0000000000) {};
\node[spkeep] at (-0.2500000000,-1.2500000000) {};
\node[spkeep] at (1.0000000000,0.2500000000) {};
\node[spkeep] at (1.0000000000,6.5000000000) {};
\node[spkeep] at (1.5000000000,-1.7500000000) {};
\node[spkeep] at (2.2500000000,2.0000000000) {};
\node[spkeep] at (2.5000000000,-2.0000000000) {};
\node[spkeep] at (3.0000000000,-0.5000000000) {};
\node[spkeep] at (3.0000000000,5.5000000000) {};
\node[spkeep] at (4.2500000000,8.0000000000) {};
\node[spkeep] at (-2.5000000000,0.5000000000) {};
\node[spkeep] at (-2.2500000000,-1.5000000000) {};
\node[spkeep] at (-1.0000000000,1.7500000000) {};
\node[spkeep] at (0.0000000000,-3.0000000000) {};
\node[spkeep] at (0.2500000000,2.0000000000) {};
\node[spkeep] at (0.7500000000,4.7500000000) {};
\node[spkeep] at (2.0000000000,7.0000000000) {};
\node[spkeep] at (2.0000000000,8.0000000000) {};
\node[spkeep] at (3.7500000000,1.0000000000) {};
\node[spkeep] at (4.5000000000,6.0000000000) {};
\node[spkeep] at (5.2500000000,3.5000000000) {};
\node[spkeep] at (6.2500000000,5.2500000000) {};
\node[spkeep] at (7.5000000000,2.7500000000) {};
\node[spkeep] at (-0.7500000000,0.5000000000) {};
\node[spkeep] at (-3.7500000000,4.7500000000) {};
\node[spkeep] at (-2.2500000000,3.2500000000) {};
\node[spkeep] at (-1.7500000000,-2.2500000000) {};
\node[spkeep] at (-1.5000000000,3.0000000000) {};
\node[spkeep] at (-1.0000000000,4.2500000000) {};
\node[spkeep] at (-0.2500000000,-2.2500000000) {};
\node[spkeep] at (0.0000000000,3.2500000000) {};
\node[spkeep] at (0.5000000000,1.0000000000) {};
\node[spkeep] at (1.0000000000,-2.7500000000) {};
\node[spkeep] at (1.2500000000,2.0000000000) {};
\node[spkeep] at (1.7500000000,0.0000000000) {};
\node[spkeep] at (1.7500000000,4.7500000000) {};
\node[spkeep] at (1.7500000000,6.2500000000) {};
\node[spkeep] at (2.5000000000,1.2500000000) {};
\node[spcenter] at (0.0000000000,0.0000000000) {};
\node[spcenter] at (7.0000000000,6.0000000000) {};
\node[spcenter] at (-1.0000000000,6.0000000000) {};
\node[spcenter] at (3.0000000000,8.0019531250) {};
\node[spcenter] at (4.0000000000,0.0000000000) {};
\node[spcenter] at (7.0000000000,2.0000000000) {};
\node[spcenter] at (-3.0000000000,3.0000000000) {};
\node[spcenter] at (2.0000000000,3.0000000000) {};
\node[spchosen] at (-3.0000000000,-2.0000000000) {};
\node[spchosen] at (8.2500000000,4.0000000000) {};
\node[spchosen] at (3.5000000000,-1.7500000000) {};
\node[spchosen] at (1.7500000000,9.0000000000) {};
\node[spchosen] at (-3.7500000000,4.7500000000) {};
\node[spfree] at (0.0000000000,0.0000000000) {};
\node[spfree] at (7.0000000000,6.0000000000) {};
\node[spfree] at (-1.0000000000,6.0000000000) {};
\node[spfree] at (4.0000000000,0.0000000000) {};
\node[spfree] at (7.0000000000,2.0000000000) {};
\node[spfree] at (-3.0000000000,3.0000000000) {};
\node[spfree] at (2.0000000000,3.0000000000) {};
\node[anchor=east,font=\scriptsize] at (-3.8000000000,-2.2000000000) {$x$};
\end{tikzpicture}
\end{minipage}\par\medskip\begin{minipage}[c]{.49\textwidth}\centering
\begin{tikzpicture}[x=0.34cm,y=0.34cm]
\path[use as bounding box] (-7,-4.4) rectangle (11,12.0);
\node[anchor=north west,font=\small\bfseries] at (-7,12) {(c)};
\begin{scope}[shift={(-2.6000000000,4.0000000000)},x=2.25cm,y=2.25cm]
\fill[epasgreen!4] (-.5,-.5) rectangle (.5,.5);
\draw[black!40,line width=.4pt] (-.5,-.5) rectangle (.5,.5);
\node[font=\scriptsize] at (0,.68) {$A$};
\draw[-{Stealth[length=1.1mm]},epasorange,line width=.55pt,shorten <=1pt,shorten >=1pt] (-0.1250000000,-0.1250000000) -- (0,0);
\node[spinput] at (0,0) {};
\node[spinput] at (-0.1250000000,-0.1250000000) {};
\node[spkeep] at (0,0) {};
\node[spcenter] at (0,0) {};

\draw[<->,>=Stealth,black!50,line width=.35pt] (-.5,-.72)--(.5,-.72) node[midway,below,font=\scriptsize] {$h$};
\end{scope}
\begin{scope}[shift={(6.0000000000,4.0000000000)},x=2.25cm,y=2.25cm]
\fill[epasgreen!4] (-.5,-.5) rectangle (.5,.5);
\draw[black!40,line width=.4pt] (-.5,-.5) rectangle (.5,.5);
\node[font=\scriptsize] at (0,.68) {$B$};
\draw[-{Stealth[length=1.1mm]},epasorange,line width=.55pt,shorten <=1pt,shorten >=1pt] (0.0000000000,0.1250000000) -- (0,0);
\node[spinput] at (0,0) {};
\node[spinput] at (0.0000000000,0.1250000000) {};
\node[spkeep] at (0,0) {};
\node[spcenter] at (0.0000000000,0.1250000000) {};

\draw[<->,>=Stealth,black!50,line width=.35pt] (-.5,-.72)--(.5,-.72) node[midway,below,font=\scriptsize] {$h$};
\end{scope}
\end{tikzpicture}
\end{minipage}\hfill\begin{minipage}[c]{.49\textwidth}\centering
\begin{tikzpicture}[x=0.34cm,y=0.34cm]
\path[use as bounding box] (-7,-4.4) rectangle (11,12.0);
\node[anchor=north west,font=\small\bfseries] at (-7,12) {(d)};
\draw[epasorange!20,line width=.3pt] (-3.0000000000,-2.0000000000)--(8.2500000000,4.0000000000);
\draw[epasorange!20,line width=.3pt] (-3.0000000000,-2.0000000000)--(3.5000000000,-1.7500000000);
\draw[epasorange!20,line width=.3pt] (-3.0000000000,-2.0000000000)--(1.7500000000,9.0000000000);
\draw[epasorange!20,line width=.3pt] (-3.0000000000,-2.0000000000)--(-3.7500000000,4.7500000000);
\draw[epasorange!20,line width=.3pt] (8.2500000000,4.0000000000)--(3.5000000000,-1.7500000000);
\draw[epasorange!20,line width=.3pt] (8.2500000000,4.0000000000)--(1.7500000000,9.0000000000);
\draw[epasorange!20,line width=.3pt] (8.2500000000,4.0000000000)--(-3.7500000000,4.7500000000);
\draw[epasorange!20,line width=.3pt] (3.5000000000,-1.7500000000)--(1.7500000000,9.0000000000);
\draw[epasorange!20,line width=.3pt] (3.5000000000,-1.7500000000)--(-3.7500000000,4.7500000000);
\draw[epasorange!20,line width=.3pt] (1.7500000000,9.0000000000)--(-3.7500000000,4.7500000000);
\node[spinput] at (0.0000000000,0.0000000000) {};
\node[spinput] at (4.0000000000,0.0000000000) {};
\node[spinput] at (7.0000000000,2.0000000000) {};
\node[spinput] at (7.0000000000,6.0000000000) {};
\node[spinput] at (3.0000000000,8.0000000000) {};
\node[spinput] at (-1.0000000000,6.0000000000) {};
\node[spinput] at (-3.0000000000,3.0000000000) {};
\node[spinput] at (2.0000000000,3.0000000000) {};
\node[spinput] at (-0.0019531250,-0.0019531250) {};
\node[spinput] at (3.0000000000,8.0019531250) {};
\node[spinput] at (2.0000000000,-3.0000000000) {};
\node[spinput] at (-3.0000000000,-2.0000000000) {};
\node[spinput] at (4.0000000000,5.2500000000) {};
\node[spinput] at (-0.7500000000,-3.5000000000) {};
\node[spinput] at (-3.5000000000,0.5000000000) {};
\node[spinput] at (-0.5000000000,2.5000000000) {};
\node[spinput] at (1.5000000000,5.5000000000) {};
\node[spinput] at (4.7500000000,2.7500000000) {};
\node[spinput] at (8.2500000000,4.0000000000) {};
\node[spinput] at (5.2500000000,7.5000000000) {};
\node[spinput] at (-3.0000000000,5.2500000000) {};
\node[spinput] at (2.0000000000,-0.7500000000) {};
\node[spinput] at (6.0000000000,4.2500000000) {};
\node[spinput] at (-1.7500000000,1.0000000000) {};
\node[spinput] at (-1.2500000000,-1.5000000000) {};
\node[spinput] at (0.0000000000,4.2500000000) {};
\node[spinput] at (1.2500000000,7.5000000000) {};
\node[spinput] at (1.5000000000,1.2500000000) {};
\node[spinput] at (0.5000000000,-2.0000000000) {};
\node[spinput] at (3.5000000000,-1.7500000000) {};
\node[spinput] at (3.2500000000,1.7500000000) {};
\node[spinput] at (-1.7500000000,4.0000000000) {};
\node[spinput] at (1.7500000000,9.0000000000) {};
\node[spinput] at (2.7500000000,6.5000000000) {};
\node[spinput] at (5.5000000000,5.7500000000) {};
\node[spinput] at (2.7500000000,4.5000000000) {};
\node[spinput] at (-4.0000000000,4.0000000000) {};
\node[spinput] at (-2.5000000000,-0.5000000000) {};
\node[spinput] at (-2.0000000000,-3.0000000000) {};
\node[spinput] at (0.7500000000,-3.5000000000) {};
\node[spinput] at (2.7500000000,0.5000000000) {};
\node[spinput] at (4.0000000000,7.0000000000) {};
\node[spinput] at (-0.2500000000,1.2500000000) {};
\node[spinput] at (0.2500000000,5.7500000000) {};
\node[spinput] at (0.7500000000,2.7500000000) {};
\node[spinput] at (4.2500000000,4.0000000000) {};
\node[spinput] at (8.5000000000,2.7500000000) {};
\node[spinput] at (-3.0000000000,1.7500000000) {};
\node[spinput] at (-1.7500000000,2.2500000000) {};
\node[spinput] at (-1.2500000000,-0.2500000000) {};
\node[spinput] at (1.2500000000,4.0000000000) {};
\node[spinput] at (6.0000000000,2.7500000000) {};
\node[spinput] at (7.0000000000,3.5000000000) {};
\node[spinput] at (7.2500000000,4.7500000000) {};
\node[spinput] at (-1.7500000000,5.2500000000) {};
\node[spinput] at (0.7500000000,-0.7500000000) {};
\node[spinput] at (5.0000000000,4.7500000000) {};
\node[spinput] at (6.0000000000,6.7500000000) {};
\node[spinput] at (-3.5000000000,-0.7500000000) {};
\node[spinput] at (-2.7500000000,4.2500000000) {};
\node[spinput] at (-1.0000000000,-2.5000000000) {};
\node[spinput] at (-0.7500000000,3.5000000000) {};
\node[spinput] at (-0.7500000000,5.0000000000) {};
\node[spinput] at (-0.2500000000,-1.2500000000) {};
\node[spinput] at (1.0000000000,0.2500000000) {};
\node[spinput] at (1.0000000000,6.5000000000) {};
\node[spinput] at (1.5000000000,-1.7500000000) {};
\node[spinput] at (2.2500000000,2.0000000000) {};
\node[spinput] at (2.5000000000,-2.0000000000) {};
\node[spinput] at (3.0000000000,-0.5000000000) {};
\node[spinput] at (3.0000000000,5.5000000000) {};
\node[spinput] at (4.2500000000,8.0000000000) {};
\node[spinput] at (-2.5000000000,0.5000000000) {};
\node[spinput] at (-2.2500000000,-1.5000000000) {};
\node[spinput] at (-1.0000000000,1.7500000000) {};
\node[spinput] at (0.0000000000,-3.0000000000) {};
\node[spinput] at (0.2500000000,2.0000000000) {};
\node[spinput] at (0.7500000000,4.7500000000) {};
\node[spinput] at (2.0000000000,7.0000000000) {};
\node[spinput] at (2.0000000000,8.0000000000) {};
\node[spinput] at (3.7500000000,1.0000000000) {};
\node[spinput] at (4.5000000000,6.0000000000) {};
\node[spinput] at (5.2500000000,3.5000000000) {};
\node[spinput] at (6.2500000000,5.2500000000) {};
\node[spinput] at (7.5000000000,2.7500000000) {};
\node[spinput] at (-0.7500000000,0.5000000000) {};
\node[spinput] at (-3.7500000000,4.7500000000) {};
\node[spinput] at (-2.2500000000,3.2500000000) {};
\node[spinput] at (-1.7500000000,-2.2500000000) {};
\node[spinput] at (-1.5000000000,3.0000000000) {};
\node[spinput] at (-1.0000000000,4.2500000000) {};
\node[spinput] at (-0.2500000000,-2.2500000000) {};
\node[spinput] at (0.0000000000,3.2500000000) {};
\node[spinput] at (0.5000000000,1.0000000000) {};
\node[spinput] at (1.0000000000,-2.7500000000) {};
\node[spinput] at (1.2500000000,2.0000000000) {};
\node[spinput] at (1.7500000000,0.0000000000) {};
\node[spinput] at (1.7500000000,4.7500000000) {};
\node[spinput] at (1.7500000000,6.2500000000) {};
\node[spinput] at (2.5000000000,1.2500000000) {};
\node[spkeep] at (0.0000000000,0.0000000000) {};
\node[spkeep] at (4.0000000000,0.0000000000) {};
\node[spkeep] at (7.0000000000,2.0000000000) {};
\node[spkeep] at (7.0000000000,6.0000000000) {};
\node[spkeep] at (3.0000000000,8.0000000000) {};
\node[spkeep] at (-1.0000000000,6.0000000000) {};
\node[spkeep] at (-3.0000000000,3.0000000000) {};
\node[spkeep] at (2.0000000000,3.0000000000) {};
\node[spkeep] at (2.0000000000,-3.0000000000) {};
\node[spkeep] at (-3.0000000000,-2.0000000000) {};
\node[spkeep] at (4.0000000000,5.2500000000) {};
\node[spkeep] at (-0.7500000000,-3.5000000000) {};
\node[spkeep] at (-3.5000000000,0.5000000000) {};
\node[spkeep] at (-0.5000000000,2.5000000000) {};
\node[spkeep] at (1.5000000000,5.5000000000) {};
\node[spkeep] at (4.7500000000,2.7500000000) {};
\node[spkeep] at (8.2500000000,4.0000000000) {};
\node[spkeep] at (5.2500000000,7.5000000000) {};
\node[spkeep] at (-3.0000000000,5.2500000000) {};
\node[spkeep] at (2.0000000000,-0.7500000000) {};
\node[spkeep] at (6.0000000000,4.2500000000) {};
\node[spkeep] at (-1.7500000000,1.0000000000) {};
\node[spkeep] at (-1.2500000000,-1.5000000000) {};
\node[spkeep] at (0.0000000000,4.2500000000) {};
\node[spkeep] at (1.2500000000,7.5000000000) {};
\node[spkeep] at (1.5000000000,1.2500000000) {};
\node[spkeep] at (0.5000000000,-2.0000000000) {};
\node[spkeep] at (3.5000000000,-1.7500000000) {};
\node[spkeep] at (3.2500000000,1.7500000000) {};
\node[spkeep] at (-1.7500000000,4.0000000000) {};
\node[spkeep] at (1.7500000000,9.0000000000) {};
\node[spkeep] at (2.7500000000,6.5000000000) {};
\node[spkeep] at (5.5000000000,5.7500000000) {};
\node[spkeep] at (2.7500000000,4.5000000000) {};
\node[spkeep] at (-4.0000000000,4.0000000000) {};
\node[spkeep] at (-2.5000000000,-0.5000000000) {};
\node[spkeep] at (-2.0000000000,-3.0000000000) {};
\node[spkeep] at (0.7500000000,-3.5000000000) {};
\node[spkeep] at (2.7500000000,0.5000000000) {};
\node[spkeep] at (4.0000000000,7.0000000000) {};
\node[spkeep] at (-0.2500000000,1.2500000000) {};
\node[spkeep] at (0.2500000000,5.7500000000) {};
\node[spkeep] at (0.7500000000,2.7500000000) {};
\node[spkeep] at (4.2500000000,4.0000000000) {};
\node[spkeep] at (8.5000000000,2.7500000000) {};
\node[spkeep] at (-3.0000000000,1.7500000000) {};
\node[spkeep] at (-1.7500000000,2.2500000000) {};
\node[spkeep] at (-1.2500000000,-0.2500000000) {};
\node[spkeep] at (1.2500000000,4.0000000000) {};
\node[spkeep] at (6.0000000000,2.7500000000) {};
\node[spkeep] at (7.0000000000,3.5000000000) {};
\node[spkeep] at (7.2500000000,4.7500000000) {};
\node[spkeep] at (-1.7500000000,5.2500000000) {};
\node[spkeep] at (0.7500000000,-0.7500000000) {};
\node[spkeep] at (5.0000000000,4.7500000000) {};
\node[spkeep] at (6.0000000000,6.7500000000) {};
\node[spkeep] at (-3.5000000000,-0.7500000000) {};
\node[spkeep] at (-2.7500000000,4.2500000000) {};
\node[spkeep] at (-1.0000000000,-2.5000000000) {};
\node[spkeep] at (-0.7500000000,3.5000000000) {};
\node[spkeep] at (-0.7500000000,5.0000000000) {};
\node[spkeep] at (-0.2500000000,-1.2500000000) {};
\node[spkeep] at (1.0000000000,0.2500000000) {};
\node[spkeep] at (1.0000000000,6.5000000000) {};
\node[spkeep] at (1.5000000000,-1.7500000000) {};
\node[spkeep] at (2.2500000000,2.0000000000) {};
\node[spkeep] at (2.5000000000,-2.0000000000) {};
\node[spkeep] at (3.0000000000,-0.5000000000) {};
\node[spkeep] at (3.0000000000,5.5000000000) {};
\node[spkeep] at (4.2500000000,8.0000000000) {};
\node[spkeep] at (-2.5000000000,0.5000000000) {};
\node[spkeep] at (-2.2500000000,-1.5000000000) {};
\node[spkeep] at (-1.0000000000,1.7500000000) {};
\node[spkeep] at (0.0000000000,-3.0000000000) {};
\node[spkeep] at (0.2500000000,2.0000000000) {};
\node[spkeep] at (0.7500000000,4.7500000000) {};
\node[spkeep] at (2.0000000000,7.0000000000) {};
\node[spkeep] at (2.0000000000,8.0000000000) {};
\node[spkeep] at (3.7500000000,1.0000000000) {};
\node[spkeep] at (4.5000000000,6.0000000000) {};
\node[spkeep] at (5.2500000000,3.5000000000) {};
\node[spkeep] at (6.2500000000,5.2500000000) {};
\node[spkeep] at (7.5000000000,2.7500000000) {};
\node[spkeep] at (-0.7500000000,0.5000000000) {};
\node[spkeep] at (-3.7500000000,4.7500000000) {};
\node[spkeep] at (-2.2500000000,3.2500000000) {};
\node[spkeep] at (-1.7500000000,-2.2500000000) {};
\node[spkeep] at (-1.5000000000,3.0000000000) {};
\node[spkeep] at (-1.0000000000,4.2500000000) {};
\node[spkeep] at (-0.2500000000,-2.2500000000) {};
\node[spkeep] at (0.0000000000,3.2500000000) {};
\node[spkeep] at (0.5000000000,1.0000000000) {};
\node[spkeep] at (1.0000000000,-2.7500000000) {};
\node[spkeep] at (1.2500000000,2.0000000000) {};
\node[spkeep] at (1.7500000000,0.0000000000) {};
\node[spkeep] at (1.7500000000,4.7500000000) {};
\node[spkeep] at (1.7500000000,6.2500000000) {};
\node[spkeep] at (2.5000000000,1.2500000000) {};
\node[spcenter] at (0.0000000000,0.0000000000) {};
\node[spcenter] at (7.0000000000,6.0000000000) {};
\node[spcenter] at (-1.0000000000,6.0000000000) {};
\node[spcenter] at (3.0000000000,8.0019531250) {};
\node[spcenter] at (4.0000000000,0.0000000000) {};
\node[spcenter] at (7.0000000000,2.0000000000) {};
\node[spcenter] at (-3.0000000000,3.0000000000) {};
\node[spcenter] at (2.0000000000,3.0000000000) {};
\node[spchosen] at (-3.0000000000,-2.0000000000) {};
\node[spchosen] at (8.2500000000,4.0000000000) {};
\node[spchosen] at (3.5000000000,-1.7500000000) {};
\node[spchosen] at (1.7500000000,9.0000000000) {};
\node[spchosen] at (-3.7500000000,4.7500000000) {};
\end{tikzpicture}
\end{minipage}\par\smallskip
\begin{tikzpicture}[font=\scriptsize]
\node[spinput] at (0,0) {};\node[anchor=west] at (.14,0) {candidate};
\node[spcenter] at (1.85,0) {};\node[anchor=west] at (2.03,0) {cover center};
\node[spfree] at (4.2,0) {};\node[anchor=west] at (4.42,0) {free center};
\node[spkeep] at (6.5,0) {};\node[anchor=west] at (6.65,0) {representative};
\node[spchosen] at (9.15,0) {};\node[anchor=west] at (9.32,0) {selected point};
\end{tikzpicture}

\caption[One hundred sites spread across the cover]{One hundred sites with $k=5$, eight
cover centers, $r=\sqrt{13}$ and $h=1/64$. (a) The disks cover all candidates. Orange
diamonds mark the five sites of the independently verified optimum. (b) Green squares mark
the 98 representatives and blue rings the eight centers. For $x=p_{12}$, the four dashed
disks about its companions exclude only $p_{10}$, leaving seven free centers (purple
triangles). (c) The only cells containing two sites are $\{p_1,p_9\}$ and
$\{p_5,p_{10}\}$, shown at a common magnification. All five optimal sites are retained.
(d) The selected sites and their ten pair interactions.}
\label{fig:epas-hundred-supp}
\end{figure}

Keep the first ten candidates and append $90$ points spread across the covered region, preserving
the eight-center traversal, $k=5$, $d=s=2$, and $\varepsilon=1/2$.
Figure~\ref{fig:epas-hundred-supp} plots each candidate at its generated coordinates.

Let $c_1,\ldots,c_8$ be the ordered centers above, and $r_j^2$ their squared insertion distances.
Form the finite pool of points $q=(a/4,b/4)$ with $-28\le a\le44$, $-16\le b\le48$ and $a,b\in\mathbb Z$, excluding existing candidates, and retain only those satisfying $\min_{1\le\ell<j}\|q-c_\ell\|_2^2\le r_j^2$ for every $j=2,\ldots,8$. Exactly $1402$ pool points survive. Append $90$ points, each maximizing its minimum squared
distance to all candidates already present; lexicographic coordinates break ties. This defines the
input. The prefix inequalities prevent added points from outranking prescribed centers; equality
favors the smaller original index. Induction preserves the traversal. The admissible region is
contained in the covering-disk union; both the last insertion radius and final cover radius are
$\sqrt{13}$.

The potential bound $B=1522/455$, dyadic scale $1/2$, grid side $h=1/64$, and origin
$(-1/128,-1/128)$ are unchanged. Recomputing all cell keys yields $96$ singleton cells and the two
cells $\{p_1,p_9\}$ and $\{p_5,p_{10}\}$. Thus $Q=\{p_1,\ldots,p_8,p_{11},\ldots,p_{100}\}$ and $|Q|=98$. The standard uniform grid therefore gives little compression on this spread-out input. A fixed
cover does not imply a fixed occupied-cell count. The comparison is
\begin{center}\small
\begin{tabular}{@{}rrrrr@{}}\toprule
$n$ & $k$ & Original $k$-subsets & $|Q|$ & Representative $k$-subsets\\\midrule
$10$ & $5$ & $252$ & $8$ & $56$\\
$100$ & $5$ & $75\,287\,520$ & $98$ & $67\,910\,864$\\\bottomrule
\end{tabular}
\end{center}
These are compact-EPAS search-space sizes. The larger run tabulates $4753$ representative weights
and uses exact pruning instead of visiting every subset; no measured speedup is asserted.

\paragraph{Selection and certificate.}
Exact optimization on $Q$ returns
\[
 S_Q=\{p_{12},p_{19},p_{30},p_{33},p_{87}\},\qquad
 V=\frac{19091545433244524790715084}{138473515345884275211104385}
   \approx0.137871458.
\]
The EPAS certificate is $[2V/3,V]$. A separate exact search on \emph{all 100 original candidates}
returns the same optimum, certifying $[V,V]$ and error $V/\OPT_{100}-1=0$ for this instance. This
audit does not construct the grid or improve the scheme's worst-case $1.5$ guarantee or asymptotic
search bound.

For $x=p_{12}$, the seven free centers are $T_x=\{p_1,p_2,p_3,p_4,p_6,p_7,p_8\}$; only $p_{10}$ lies within distance strictly less than $r/2$ of one of
$p_{19},p_{30},p_{33},p_{87}$. In contrast to the ten-point example, no selected optimum point is
a cover center. Free centers are determined by these exclusion disks, not by being outside $S_Q$.
The figure plots all centers even when excluded and marks free ones by a second, distinct symbol.

\paragraph{Certified exact pruning.}
The optional \texttt{--search branch\_bound} routine changes only the exact search on the retained
set, leaving the EPAS geometry and default enumerator intact. For rational weights $w_{ij}>0$, set
$K=2^{48}$ and $a_{ij}=\lfloor K w_{ij}\rfloor$. Maintain an exactly evaluated feasible incumbent
of energy $U$. In a node with partial selection $A$, admissible remaining indices $R$, and
$q=k-|A|$ still to select, put $c=\sum_{\{i,j\}\subseteq A}a_{ij}$ and $b_i=\sum_{j\in A}a_{ij}$ for $i\in R$. For $q\ge2$, let $t_i$ sum the $q-1$ smallest values $a_{ij}$ over $j\in R\setminus\{i\}$ and let
$D$ sum the $q$ smallest values $2b_i+t_i$ over $i\in R$. For any completion $T\subseteq R$,
$|T|=q$,
\[
 2K E_2(A\cup T)\ \ge\ 2c+\sum_{i\in T}(2b_i+t_i)
 \ \ge\ 2c+D.
\]
Indeed the attachment terms are charged twice, while each internal completion edge is charged once
at each endpoint; choosing the cheapest available incident edges only lowers these charges.
Consequently $2c+D>2\lceil KU\rceil$ safely discards the node. For $q=1$, use $c+\min_{i\in
R}b_i>\lceil KU\rceil$. Positivity also allows a candidate to be discarded if its partial lower
cost already exceeds $\lceil KU\rceil$. The recursion follows increasing indices, so every
remaining subset is either visited or safely pruned. Complete surviving selections and incumbent
updates use exact rational energy, not rounded weights. Thus termination certifies the exact
optimum; the integer scaling affects pruning strength, not correctness. Greedy completions supply
feasible initial incumbents only.

The representative search visits $29\,539$ partial-tree nodes, and the separate all-candidate
audit visits $32\,039$; these include partial selections and are not full-subset enumeration
counts. Both searches use the same exact routine on different domains. Independent exhaustive
rational minimization validates the pruning implementation on $111$ weighted test cases totaling
$12\,009$ subsets, including tiny weights below $1/K$ and tied optima. The ten-point audit and its
$78$ additional geometric cases also pass.

From \path{anc/walkthroughs/}, run the following three commands. The first
constructs the spread instance, solves it, and audits the full-input optimum;
the second checks the pruning rule against exhaustive search; the third draws the figure.
\begin{quote}
\verb|python epas_100_run.py|\\
\verb|python check_exact_subset_search.py|\\
\verb|python render_epas_spread.py|
\end{quote}
JSON records the full rational input and trace; assertions verify the traversal, geometry, $98$
representatives, and equality of both optima. Maximum replacement squared distance is $1/131072$;
all added points represent themselves.

\FloatBarrier

\subsection{A hundred sites in tight groups}\label{sec:city-details}
The five-of-ten and spread-out hundred-point constructions above determine the coordinates and
their original ordering. Keep the original ten points. From the 90 added spread-out candidates,
admit an anchor $p$ only if all five offsets $p$, $p\pm(1/512,0)$, $p\pm(0,1/512)$ satisfy every
original eight-center traversal inequality strictly. Starting with the original ten points, choose
18 such anchors successively by maximum distance to the already chosen points, breaking ties
lexicographically. Append the five offsets in that fixed order for each anchor. This yields
exactly 100 distinct dyadic candidates; geometrically they represent 26 local groups, with 18
five-site cities, two two-site groups, and six singleton sites. No geographic dataset or demand
model is asserted.

The strict inequalities preserve the first eight farthest-first choices. The full
ten-center traversal next selects $p_{15}$ and $p_{18}$.
The rational implementation gives $h=1/64$ and $M=26$. The representative optimum is
$\{p_5,p_{16},p_{41},p_{51},p_{61}\}$ with energy $V\approx0.159096904279$. An exact search over
all candidates gives $\{p_{10},p_{18},p_{42},p_{55},p_{63}\}$, and $V/\OPT\approx1.000596624154$.
Fixed-cell refinement returns this original optimum. The code records exact fractions for all
energies and checks the free-center set and preservation of the first eight centers. The initial
certificate is $[V/1.5,V]$; after refinement its upper endpoint can be reduced to the refined
energy. Independent exact optimization is an audit, not a step of the EPAS.

From \path{anc/walkthroughs/}, run \texttt{python}\ \path{compressed_hundred_v2.py}. The inputs,
exact energies and audit are recorded in \path{compressed_hundred_v2.json}. Compare the 26
occupied cells here with 98 in the spread-out construction of Appendix~\ref{sec:epas-hundred}.
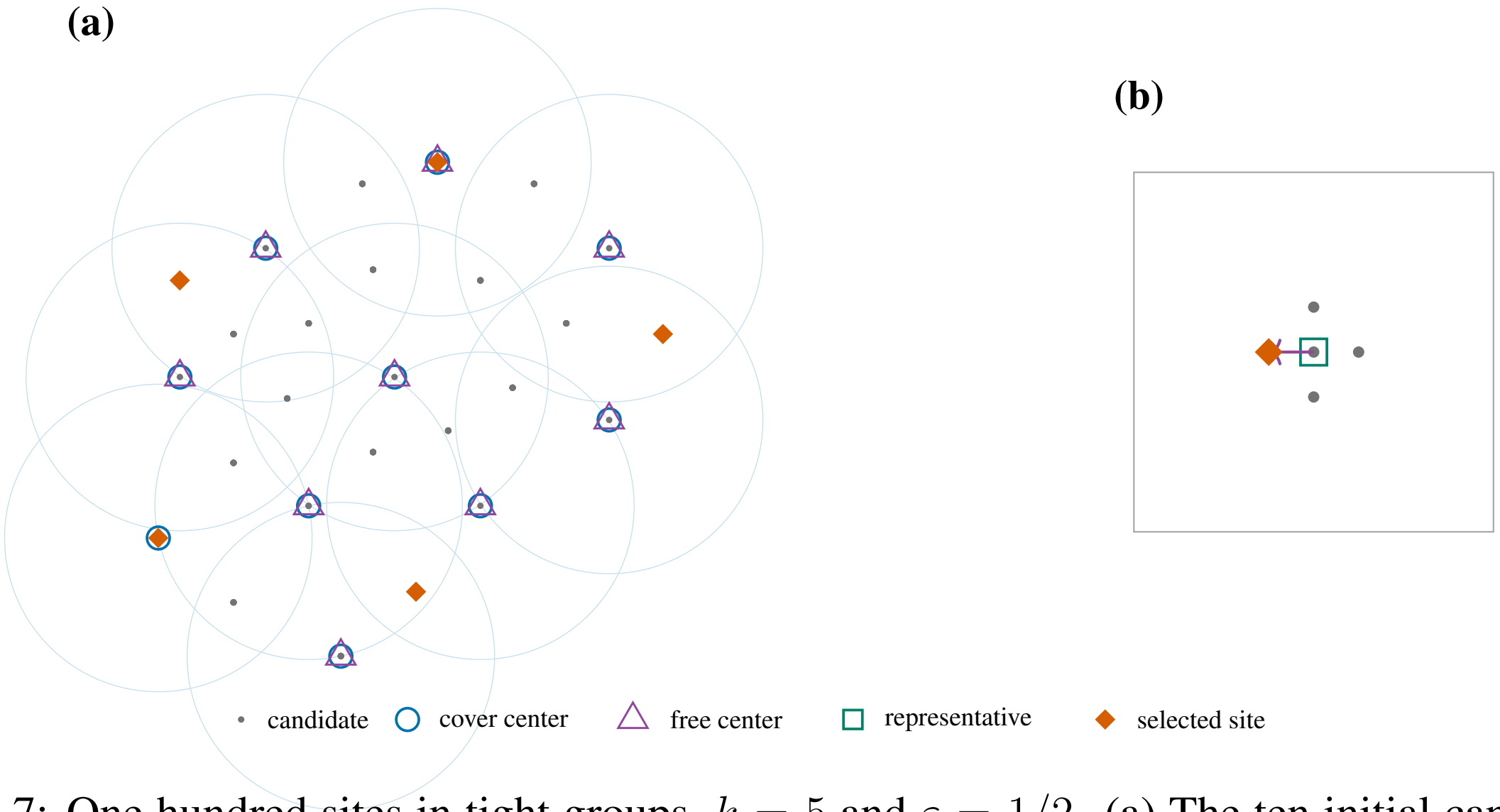
\begin{figure}[!htbp]\centering
\definecolor{cityblue}{RGB}{0,114,178}\definecolor{citygreen}{RGB}{0,130,105}\definecolor{cityorange}{RGB}{213,94,0}\definecolor{citypurple}{RGB}{145,70,160}
\tikzset{citypoint/.style={circle,fill=black!55,inner sep=0pt,minimum size=1.7pt},citycenter/.style={circle,draw=cityblue,line width=.7pt,inner sep=0pt,minimum size=6pt},cityfree/.style={regular polygon,regular polygon sides=3,draw=citypurple,line width=.55pt,inner sep=0pt,minimum size=9pt},citychosen/.style={diamond,draw=cityorange,fill=cityorange,inner sep=0pt,minimum size=5pt},cityrep/.style={rectangle,draw=citygreen,line width=.7pt,inner sep=0pt,minimum size=5pt}}
\begin{minipage}{.61\linewidth}\centering\begin{tikzpicture}[x=.40cm,y=.40cm]
\path[use as bounding box] (-6,-4) rectangle (11,12);
\node[anchor=north west,font=\small\bfseries] at (-6,12) {(a)};
\draw[cityblue!22,line width=.25pt] (0.000000000,0.000000000) circle[radius=3.581365060];
\draw[cityblue!22,line width=.25pt] (7.000000000,6.000000000) circle[radius=3.581365060];
\draw[cityblue!22,line width=.25pt] (-1.000000000,6.000000000) circle[radius=3.581365060];
\draw[cityblue!22,line width=.25pt] (3.000000000,8.001953125) circle[radius=3.581365060];
\draw[cityblue!22,line width=.25pt] (4.000000000,0.000000000) circle[radius=3.581365060];
\draw[cityblue!22,line width=.25pt] (7.000000000,2.000000000) circle[radius=3.581365060];
\draw[cityblue!22,line width=.25pt] (-3.000000000,3.000000000) circle[radius=3.581365060];
\draw[cityblue!22,line width=.25pt] (2.000000000,3.000000000) circle[radius=3.581365060];
\draw[cityblue!22,line width=.25pt] (0.750000000,-3.501953125) circle[radius=3.581365060];
\draw[cityblue!22,line width=.25pt] (-3.501953125,-0.750000000) circle[radius=3.581365060];
\node[citypoint] at (0.000000000,0.000000000) {};
\node[citypoint] at (4.000000000,0.000000000) {};
\node[citypoint] at (7.000000000,2.000000000) {};
\node[citypoint] at (7.000000000,6.000000000) {};
\node[citypoint] at (3.000000000,8.000000000) {};
\node[citypoint] at (-1.000000000,6.000000000) {};
\node[citypoint] at (-3.000000000,3.000000000) {};
\node[citypoint] at (2.000000000,3.000000000) {};
\node[citypoint] at (-0.001953125,-0.001953125) {};
\node[citypoint] at (3.000000000,8.001953125) {};
\node[citypoint] at (0.750000000,-3.500000000) {};
\node[citypoint] at (0.751953125,-3.500000000) {};
\node[citypoint] at (0.748046875,-3.500000000) {};
\node[citypoint] at (0.750000000,-3.498046875) {};
\node[citypoint] at (0.750000000,-3.501953125) {};
\node[citypoint] at (-3.500000000,-0.750000000) {};
\node[citypoint] at (-3.498046875,-0.750000000) {};
\node[citypoint] at (-3.501953125,-0.750000000) {};
\node[citypoint] at (-3.500000000,-0.748046875) {};
\node[citypoint] at (-3.500000000,-0.751953125) {};
\node[citypoint] at (4.000000000,5.250000000) {};
\node[citypoint] at (4.001953125,5.250000000) {};
\node[citypoint] at (3.998046875,5.250000000) {};
\node[citypoint] at (4.000000000,5.251953125) {};
\node[citypoint] at (4.000000000,5.248046875) {};
\node[citypoint] at (-0.500000000,2.500000000) {};
\node[citypoint] at (-0.498046875,2.500000000) {};
\node[citypoint] at (-0.501953125,2.500000000) {};
\node[citypoint] at (-0.500000000,2.501953125) {};
\node[citypoint] at (-0.500000000,2.498046875) {};
\node[citypoint] at (1.500000000,5.500000000) {};
\node[citypoint] at (1.501953125,5.500000000) {};
\node[citypoint] at (1.498046875,5.500000000) {};
\node[citypoint] at (1.500000000,5.501953125) {};
\node[citypoint] at (1.500000000,5.498046875) {};
\node[citypoint] at (4.750000000,2.750000000) {};
\node[citypoint] at (4.751953125,2.750000000) {};
\node[citypoint] at (4.748046875,2.750000000) {};
\node[citypoint] at (4.750000000,2.751953125) {};
\node[citypoint] at (4.750000000,2.748046875) {};
\node[citypoint] at (8.250000000,4.000000000) {};
\node[citypoint] at (8.251953125,4.000000000) {};
\node[citypoint] at (8.248046875,4.000000000) {};
\node[citypoint] at (8.250000000,4.001953125) {};
\node[citypoint] at (8.250000000,3.998046875) {};
\node[citypoint] at (-1.750000000,-2.250000000) {};
\node[citypoint] at (-1.748046875,-2.250000000) {};
\node[citypoint] at (-1.751953125,-2.250000000) {};
\node[citypoint] at (-1.750000000,-2.248046875) {};
\node[citypoint] at (-1.750000000,-2.251953125) {};
\node[citypoint] at (2.500000000,-2.000000000) {};
\node[citypoint] at (2.501953125,-2.000000000) {};
\node[citypoint] at (2.498046875,-2.000000000) {};
\node[citypoint] at (2.500000000,-1.998046875) {};
\node[citypoint] at (2.500000000,-2.001953125) {};
\node[citypoint] at (5.250000000,7.500000000) {};
\node[citypoint] at (5.251953125,7.500000000) {};
\node[citypoint] at (5.248046875,7.500000000) {};
\node[citypoint] at (5.250000000,7.501953125) {};
\node[citypoint] at (5.250000000,7.498046875) {};
\node[citypoint] at (-3.000000000,5.250000000) {};
\node[citypoint] at (-2.998046875,5.250000000) {};
\node[citypoint] at (-3.001953125,5.250000000) {};
\node[citypoint] at (-3.000000000,5.251953125) {};
\node[citypoint] at (-3.000000000,5.248046875) {};
\node[citypoint] at (-1.750000000,1.000000000) {};
\node[citypoint] at (-1.748046875,1.000000000) {};
\node[citypoint] at (-1.751953125,1.000000000) {};
\node[citypoint] at (-1.750000000,1.001953125) {};
\node[citypoint] at (-1.750000000,0.998046875) {};
\node[citypoint] at (6.000000000,4.250000000) {};
\node[citypoint] at (6.001953125,4.250000000) {};
\node[citypoint] at (5.998046875,4.250000000) {};
\node[citypoint] at (6.000000000,4.251953125) {};
\node[citypoint] at (6.000000000,4.248046875) {};
\node[citypoint] at (0.000000000,4.250000000) {};
\node[citypoint] at (0.001953125,4.250000000) {};
\node[citypoint] at (-0.001953125,4.250000000) {};
\node[citypoint] at (0.000000000,4.251953125) {};
\node[citypoint] at (0.000000000,4.248046875) {};
\node[citypoint] at (1.250000000,7.500000000) {};
\node[citypoint] at (1.251953125,7.500000000) {};
\node[citypoint] at (1.248046875,7.500000000) {};
\node[citypoint] at (1.250000000,7.501953125) {};
\node[citypoint] at (1.250000000,7.498046875) {};
\node[citypoint] at (1.500000000,1.250000000) {};
\node[citypoint] at (1.501953125,1.250000000) {};
\node[citypoint] at (1.498046875,1.250000000) {};
\node[citypoint] at (1.500000000,1.251953125) {};
\node[citypoint] at (1.500000000,1.248046875) {};
\node[citypoint] at (3.250000000,1.750000000) {};
\node[citypoint] at (3.251953125,1.750000000) {};
\node[citypoint] at (3.248046875,1.750000000) {};
\node[citypoint] at (3.250000000,1.751953125) {};
\node[citypoint] at (3.250000000,1.748046875) {};
\node[citypoint] at (-1.750000000,4.000000000) {};
\node[citypoint] at (-1.748046875,4.000000000) {};
\node[citypoint] at (-1.751953125,4.000000000) {};
\node[citypoint] at (-1.750000000,4.001953125) {};
\node[citypoint] at (-1.750000000,3.998046875) {};
\node[citycenter] at (0.000000000,0.000000000) {};
\node[citycenter] at (7.000000000,6.000000000) {};
\node[citycenter] at (-1.000000000,6.000000000) {};
\node[citycenter] at (3.000000000,8.001953125) {};
\node[citycenter] at (4.000000000,0.000000000) {};
\node[citycenter] at (7.000000000,2.000000000) {};
\node[citycenter] at (-3.000000000,3.000000000) {};
\node[citycenter] at (2.000000000,3.000000000) {};
\node[citycenter] at (0.750000000,-3.501953125) {};
\node[citycenter] at (-3.501953125,-0.750000000) {};
\node[cityfree] at (0.000000000,0.000000000) {};
\node[cityfree] at (7.000000000,6.000000000) {};
\node[cityfree] at (-1.000000000,6.000000000) {};
\node[cityfree] at (3.000000000,8.001953125) {};
\node[cityfree] at (4.000000000,0.000000000) {};
\node[cityfree] at (7.000000000,2.000000000) {};
\node[cityfree] at (-3.000000000,3.000000000) {};
\node[cityfree] at (2.000000000,3.000000000) {};
\node[cityfree] at (0.750000000,-3.501953125) {};
\node[citychosen] at (3.000000000,8.001953125) {};
\node[citychosen] at (-3.501953125,-0.750000000) {};
\node[citychosen] at (8.251953125,4.000000000) {};
\node[citychosen] at (2.500000000,-2.001953125) {};
\node[citychosen] at (-3.001953125,5.250000000) {};
\end{tikzpicture}\end{minipage}\hfill
\begin{minipage}{.37\linewidth}\centering\begin{tikzpicture}[x=3.35cm,y=3.35cm]
\path[use as bounding box] (-.6,-.7) rectangle (.6,.8);
\node[anchor=north west,font=\small\bfseries] at (-.6,.8) {(b)};
\draw[black!35] (-.5,-.5) rectangle (.5,.5);
\node[citypoint,minimum size=3pt] at (0.0,0.0) {};
\node[citypoint,minimum size=3pt] at (0.125,0.0) {};
\node[citypoint,minimum size=3pt] at (-0.125,0.0) {};
\node[citypoint,minimum size=3pt] at (0.0,0.125) {};
\node[citypoint,minimum size=3pt] at (0.0,-0.125) {};
\node[cityrep,minimum size=7pt] at (0,0) {};
\draw[->,citypurple,line width=.8pt] (0,0) -- (-0.125,0.0);
\node[citychosen,minimum size=7pt] at (-0.125,0.0) {};
\end{tikzpicture}\end{minipage}
\par\smallskip\begin{tikzpicture}[font=\scriptsize]
\node[citypoint] at (0,0) {};\node[anchor=west] at (.1,0) {candidate};
\node[citycenter] at (1.55,0) {};\node[anchor=west] at (1.7,0) {cover center};
\node[cityfree] at (3.65,0) {};\node[anchor=west] at (3.85,0) {free center};
\node[cityrep] at (5.7,0) {};\node[anchor=west] at (5.85,0) {representative};
\node[citychosen] at (8.05,0) {};\node[anchor=west] at (8.2,0) {selected site};
\end{tikzpicture}
\caption{One hundred sites in tight groups, $k=5$ and $\varepsilon=1/2$.
(a) The ten initial candidates and 18 groups of five share a cover with ten farthest-first
centers (blue). Purple triangles mark the nine centers free to replace $p_{10}$ in the
optimum (orange). (b) A magnified cell of side $h=1/64$ contains five candidates.
Refinement moves its representative $p_{16}$ to $p_{18}$. Across the input $M=26$.
The reduced solution is $0.05966\%$ above the full optimum, and the refined solution
attains it. The approximation guarantee is a factor of $1.5$.}
\label{fig:hundred-clusters}
\end{figure}

\section{Reproducing the Swiss settlement experiment}\label{sec:swiss-density}
Choose ten of 5,403 prescribed sites in the Swiss LV95 plane (EPSG:2056).
For integer metre coordinates we sum the 45 unordered inverse squared distances in
$\mathrm{m}^{-2}$. This is a facility \emph{dispersion} example: there are no demand weights,
travel times, or service-area constraints. Figure~\ref{fig:city-v2} shows the hybrid
solution. The source package includes the displaced sites, projected outline, solver and
exact reference results in \path{anc/swiss/}. The optimization can be repeated
offline without downloading any original address data.

\paragraph{Where the candidates come from.}
The clouds follow the built-up areas of the 26 cantonal seats and ten rural settlements.
We used swisstopo's 5 November 2026 building-address directory and the detailed
January 2026 national polygon~\cite{swissaddresses,swissboundary}.
Within each source locality we combined occupied 200~m squares by eight-neighbor
adjacency and kept the component with the most distinct current address positions.
Giving each occupied hectare equal weight, we stratified its centers into 50 groups
by farthest-first initialization and Lloyd iterations, then used the address nearest
each fitted center. The rural pool excluded municipalities already represented and
required 100--2,000 residents, at most 50 residents per square kilometre, and at least
50 occupied hectares in the component of the municipality's most represented postal
locality. We included the eligible settlement nearest the national area centroid, then
chose nine further settlements successively by their distance from existing sites and
settlement anchors. One site in Müstair represents the eastern border region.
Thus a central settlement was deliberately available to the optimizer; selecting it
was optional.

All 1,801 seed positions were displaced by 35--90~m, rounded to whole metres,
and checked against the national boundary and the directory. The released, fixed seed
cloud contains no address identifiers. To increase local density, we generated two
neighbors of each seed in the 36 settlement clouds and a third for two seeds drawn
without replacement, leaving Müstair isolated. Proposals were uniform by area in a
75--250~m annulus around their seed before rounding. We rejected proposals outside
the annulus or country, within 15~m of another accepted site, or within 15~m of any
directory position. All 5,403 resulting coordinates are distinct, inside Switzerland
and at least 15~m from the 3,307,712 directory positions checked. These are
synthetic candidates, not actual building addresses. The generator uses neither
the EPAS grid nor the computed facility locations.

\paragraph{From cover to representatives.}
Set $d=s=2$, $k=10$ and $\varepsilon=1$. Start farthest-first traversal at the
first CSV row, breaking ties by input index. The twentieth insertion has radius
$r=39.558$~km; after that insertion the covering radius is $37.954$~km.
Use the exact values of $B_X=16H_{20}/r^2$ and of the energy $U$ of the first ten
centers, with $r$ expressed in metres. Numerically $B_X=3.67857\times10^{-8}$
and $U=3.42232\times10^{-9}$. The hybrid uses $B_H=\min(B_X,U)$. For each
certificate take the largest dyadic power $z$ with $Bz^2\le1$ and set $h=z/16$.
The grids have the common physical origin $(-128,-128)$~m. For the half-open cells,
the key of $(x,y)$ is $(\lfloor(x+128)/h\rfloor,\lfloor(y+128)/h\rfloor)$;
the first CSV row in each occupied cell is its representative.

\begin{table}[!htbp]\centering\small
\caption{Two certificates on the same 5,403 sites. Gaps are $100(E/\OPT-1)$
relative to the exact full-input optimum. Moves count accepted replacements within
the ten selected cells.}\label{tab:swiss-dense}
\begin{tabular}{@{}lrrrrr@{}}\toprule
Certificate & $h$ (m) & Retained & Reduced gap (\%) & Moves & Refined gap (\%)\\\midrule
Exchange & 256 & 3,365 & 0.10125 & 3 & 0.09449\\
Hybrid & 1,024 & 706 & 0.32553 & 8 & 0\\\bottomrule
\end{tabular}\end{table}

\paragraph{Reading the Hasliberg panel.}
Each of its eight rings is the first site in one occupied hybrid cell; the
150 candidates occupy only these eight cells. The highlighted cell has lower
left corner $(2\,657\,152,1\,178\,496)$~m and contains nine candidates.
The reduced solution uses P1320 at $(2\,657\,976,1\,178\,623)$; refinement moves
to P4949 at $(2\,657\,180,1\,178\,604)$, holding the other nine facilities fixed
for that move. Each sweep visits the reduced solution's cells in index order,
minimizes the exact sum of the nine incident pair costs, and accepts only strict
decreases; the runner stops at a sweep without a change or after ten sweeps.
Eight accepted moves give exactly the ten sites found by the full audit:
P0083 (Appenzell), P1048 (Bellinzona), P1301 (Müstair), P2602 (Schaffhausen),
P3352 (Brusio), P3867 (Basel), P3980 (Genève), P4799 (Sion),
P4949 (Hasliberg Wasserwendi), and P5312 (La Côte-aux-Fées).
Hasliberg is 5.87~km from Switzerland's area centroid. Excluding every candidate
within 30~km of that centroid raises the exact optimum by $1.57265\%$,
so every optimal ten-set contains a central site under this definition.

\paragraph{What is certified.}
Let $V$ be the exact representative optimum and $U_{\rm ref}$ the refined feasible
energy. The hybrid gives $V=2.8215031282\times10^{-9}$ and
$U_{\rm ref}=2.8123482117\times10^{-9}$ to the displayed precision. Before any
full search the EPAS certifies $V/2\le \OPT\le U_{\rm ref}$; the factor two concerns
total energy, not individual distances. A separate exact full search proves
$U_{\rm ref}=\OPT$ here. The hybrid has $\binom{706}{10}\approx7.95\times10^{21}$ reduced subsets. On the representatives, the initial locality bound leaves 69 sets of ten localities for further exact search. The finer exchange grid gives a smaller initial gap, but
its allowed replacements differ and its refinement does not reach the full optimum
(Table~\ref{tab:swiss-dense}). Exact recovery is not a refinement guarantee.

\paragraph{How the independent full-input optimum was computed.}
For each search domain, use a feasible ten-set as an upper bound. Check that the
reciprocal squared diameter of every nonsingleton locality exceeds this bound.
Even the smallest such pair bound on the full input is $1/135429557$~$\mathrm{m}^{-2}$.
Consequently an optimum uses at most one site per locality. During enumeration of
the $\binom{37}{10}=348{,}330{,}136$ possible locality products, bound each edge
from below by the reciprocal of its maximum squared distance, rounded down at scale
$2^{80}$. For a prefix still needing $t$ localities, each possible new locality
must pay its edges to the prefix and at least half the sum of its $t-1$ cheapest
remaining edges when these charges are summed over new localities. The $t$ smallest
such charges bound the completion. Only 91 full-input products survive this stage.

Within a surviving product, round pair weights down at scale $2^{60}$.
For an integer pair matrix $W_{ij}(a,b)$, take each row minimum
$u_i(a)$, then the column minimum of the residual
$u_j(b)=\min_a(W_{ij}(a,b)-u_i(a))$. Since
$u_i(a)+u_j(b)\le W_{ij}(a,b)$, summing the unary terms over pairs and minimizing
separately over each domain gives a lower bound. A candidate's excess above its
domain minimum can also eliminate that candidate. Bounds prune only above the
scaled incumbent's ceiling. Domains that remain are split along their widest
coordinate; surviving singleton products are compared as exact fractions.
Floating-point initial solutions supply feasible bounds only. The code separately
checks a 14-site restriction against exhaustive rational enumeration.

\paragraph{Reproduction.}
From \path{anc/swiss/}, run \texttt{python run.py} with
NumPy, then \texttt{python make\_figure.py}; no network access or source geodata
are needed. The first command verifies the frozen CSV's SHA-256 and computes the
cover, both grids, exact representative optima, accepted moves, full optimum, and
central-exclusion audit. The second generates the inline TikZ from those results
and the projected outline. The result JSON records site indices (zero-based data
rows), exact fractions and search counts. The CSV uses IDs P0001 through P5403.
To regenerate the densification itself, run \texttt{python make\_dense.py} with the supplied
displaced seed cloud and the two official source files specified in the README;
the PCG64 seed is 20261107. The original displacement is frozen to avoid releasing
an address correspondence. Source hashes and validation distances are in
\texttt{generation.json}.

\section{Experimental protocols and results}\label{sec:paired-protocol}
The Swiss instance permits an independent full-input optimum. The synthetic
studies answer a different question: as size, geometry and accuracy change,
when is the reduced problem small enough to solve? We retain all rejected
and timed-out cases in the occupancy statistics; only completed searches
can support claims about refinement gains.

\subsection{Fixed-grid refinement ablation}\label{sec:sweeps}
The exchange-grid study varies input size, group count and width, selected cardinality, accuracy,
and grid translation. It comprises 156 runs: 153 at $\varepsilon=1/2$ and 3 at $\varepsilon=1$. We
attempt exact reduced search only for $M\le300$ and $\binom Mk\le10^{10}$, with a 30-second cap.
Of these runs, 84 completed, 3 timed out and 69 exceeded a size limit. Only completed runs enter
the refinement table.

For each sweep cap we take the corresponding prefix of the same trace, including the time to
confirm a fixed point when the cap permits. Total time is preprocessing plus reduced search plus
refinement; it excludes input generation and separate optimum or heuristic audits. Medians are
taken after forming each run's total. Median preprocessing and reduced-search times were 2.99 and
3.78~ms, respectively. The gain is $100(V-U_t)/V$ for that run, with $V$ its exact reduced
optimum.

\begin{center}\small
\begin{tabular}{rrrrr}\toprule
Sweep cap & Median $E_s$ & Median gain (\%) & Extra (ms) & Total (ms)\\\midrule
0 & 2.913211 & 0.00000 & 0.00 & 31.95\\
1 & 2.908612 & 0.05231 & 2.94 & 40.51\\
2 & 2.908612 & 0.05231 & 5.48 & 50.49\\
3 & 2.908612 & 0.05231 & 5.48 & 50.49\\
5 & 2.908612 & 0.05231 & 5.48 & 50.49\\
10 & 2.908612 & 0.05231 & 5.48 & 50.49\\
\bottomrule\end{tabular}\end{center}
Of the 84 completed runs, 78 first reached their final recorded energy in sweep one and 6 in sweep
two. Nine additional checks use the same generator and seeds $17,29,43$ at $\varepsilon=1/2$. For
$n=20$, $k=4$, uniform inputs kept all 20 sites; eight tight groups kept eight sites for every
seed. Exhaustive search on the six original inputs confirmed that 5 refined solutions attained the
optimum. For $n=1000$, $k=5$, ten tight groups kept ten representatives, with refinement gains of
$0.0238--0.0587\%$. The exact energies and traces are in \path{anc/benchmarks/};
\path{anc/geometry/} contains \path{check_geometry_v3.py} for the polygon traversal, cell
witnesses and twelve metric instances.

\paragraph{Accuracy and cluster geometry.}
The lattice spacing is one; $\rho$ is the cluster half-width in these units. At $n=10^4$, $k=5$,
$c=10$ and $\rho=1/1024$, both $\varepsilon=1/2$ and $1$ retained ten exchange representatives on
every seed: doubling the permissible cell width did not merge any more groups. At $n=10^5$ and
$\varepsilon=1/2$, increasing $\rho$ from $1/4096$ to $1/1024$ and $1/256$ gave $10$, $10$, and
$90$ representatives, respectively. At $\rho=1/4096$ and $c=5$, the $2k$-center traversal must
enter groups more than once; the small final insertion radius left 89{,}215--89{,}315
representatives. At $c=10,25,50$ the same rule retained one site per group on all three seeds.
These exchange counts use origin $(-h/2,-h/2)$ at the exchange width.

On the same five-group inputs, we can see exactly when $M$ reaches $k$. The five unit-lattice
centers each carry a square cloud of half-width $1/4096$. The hybrid gives $h_H=1/128$ on all
three seeds. At origin $(0,0)$ every group straddles two grid lines and occupies four cells.
Translating the origin to $(-1/256,-1/256)$ places every group inside one cell. The $2k$-center
exchange width is $h_X=1/524288$. The translation is an integer multiple of $h_X$ along each axis,
so its occupied-cell count is identical at both origins. Table~\ref{tab:occupancy-budget} gives
the resulting search sizes. Run \texttt{python}\ \path{code/occupancy.py} from
\path{anc/benchmarks/} to reproduce the counts and the unique five-site reduced solutions.

\begin{table}[htbp]\centering\small
\caption{Search size known at the end of preprocessing. The synthetic cases use $k=5$, $\varepsilon=1/2$ and three seeds. The Swiss case uses $k=10$ and $\varepsilon=1$. On the uniform square the half-cell offset and $h=1/256$ give $1{,}025^2$ cells meeting $[0,4)^2$. The last column counts all reduced subsets before pruning.}\label{tab:occupancy-budget}
\begin{tabular}{@{}llrrr@{}}\toprule
Input & Grid & $n$ & $M$ & $\binom Mk$\\\midrule
Five groups & Hybrid, shifted & $10^5$ & $5$ & $1$\\
Five groups & Hybrid, origin zero & $10^5$ & $20$ & $15{,}504$\\
Five groups & Exchange, either origin & $10^5$ & $89{,}166$--$89{,}317$ & $\approx4.7\times10^{22}$\\
Ten groups & Exchange & $10^6$ & $10$ & $252$\\
Uniform square & Exchange & $10^6$ & $644{,}670$--$645{,}259$ & $\approx9.3\times10^{26}$\\
Swiss sites & Hybrid & $5{,}403$ & $706$ & $\approx7.95\times10^{21}$\\\bottomrule
\end{tabular}\end{table}

\subsection{Paired certificate protocol}
The paired study fixes $\varepsilon=1/2$ and uses 114 configurations (99 distinct input arrays).
Each is evaluated on PC, PK, I, X and H. Of 72 configurations per method eligible for subset
search, the common limits are $M\le160$, $\binom Mk\le10^7$, and five seconds. Across the five
methods, 288 searches completed, 72 exceeded a size limit, and none timed out. The other 42
configurations per method (210 runs) measure occupancy only. Neither exclusions nor occupancy
measurements produce an approximate solution.

\paragraph{Generators.}
For $c$ groups, start at $(i\bmod w,\lfloor i/w\rfloor)$, $w=\lceil\sqrt c\rceil$. Irregular
centers add independently sampled uniform offsets in $[-.3,.3]^2$. Group populations are balanced,
except in the unequal family, which uses one point per group plus a multinomial allocation with a
Dirichlet$(1,\ldots,1)$ probability vector. Within each group sample without replacement from a
square lattice of step $4/2^{20}$ and half-width $\rho$. The elongated family stretches offsets by
eight horizontally. Rotated and elongated inputs then rotate by $.37$ radians. Finally round to
denominator $2^{20}$, shuffle, and check distinctness and exact-integer bounds. Uniform controls
sample the square $[0,\sqrt c)^2$ and undergo the same final rounding and validation. This
generator produces synthetic candidate groups, not a model fitted to cities.

The $k=5$, $n=2000$, $c=10$ family varies $\rho\in\{1/4096,1/1024,1/256\}$ across six geometric
types and three seeds. Nine additional origin controls use the same irregular inputs at
$\rho=1/1024$ and physical origins $o=(a h_X,a h_X)$, $a\in\{1/4,1/2,3/4\}$, held fixed across all
methods. They preserve the input hash and all distances. The nine exact cases use $n=24$, $k=4$,
$c=8$, $\rho=1/32$ and irregular, elongated, or uniform input. The 42 occupancy cases use
$n=10^5$, $c=2k$, $\rho=1/256$, $k\in\{8,16,32,64,128,256,512\}$, and lattice or irregular
centers. Different $k$ changes $c$ in that study, so it does not isolate cardinality on an
unchanged input.

\paragraph{Pairing and timing.}
Each method uses the same points and input order, the same traversal prefix, identical
representative selection, common physical origin, and the same rational branch-and-bound routine.
The dyadic scale and hence cell width depend on the certificate. Method execution order rotates
with the seed. Shared traversal and incumbent-evaluation times are logged separately from
method-specific grid and search time; data generation and validation are excluded from those stage
times. Each input-method pair is timed once, so seed ranges are not confidence intervals. Timings
were taken sequentially on an AMD Ryzen 7 5800H, Windows 11, with Python 3.12.2 and NumPy 1.26.4.
At fixed dimension the grid stage uses $O(n)$ storage. The optional exact solver stores pair
weights on the representatives; the separate Swiss audit also materializes full-input squared
distances. These implementations do not operate on a stream.

\paragraph{Reproduction and validation.}
Work in \path{anc/benchmarks/}. The recorded compact data are sufficient to regenerate
Figure~\ref{fig:paired-v2}, Table~\ref{tab:paired-all} and the sweep summary with \verb|python code/report.py|. To rerun the measurements, use \texttt{python}\ \path{code/scale.py},
\texttt{python}\ \path{code/paired.py} and \texttt{python}\ \path{code/validation.py}, then
regenerate the summaries. The raw runs include ordered retained indices; the compact records
retain their hashes and counts, the grid widths, exact energies when available, and timings. The
two search drivers resume by configuration ID. Sixteen small inputs with independently enumerated
optima support the certificate and local-optimum checks in \path{code/checks.py}; together with
input reproducibility checks this gives 157 checks.

\paragraph{Interpretation.}
PC and PK agree for small $k$ because pair count is then the tighter analytic bound. At large $k$
exchange can improve on the evaluated incumbent in both center families, yet the resulting sets
can still be too large for enumeration. Across the 72 cases with a search protocol X retained fewer/equal/more sites
than I in $0/13/59$; among the 42 cases measuring occupancy alone the split was $12/9/21$. H attained the smaller count in every case by grid nesting. The scale
study above and the paired study use different generators and budgets; we report them separately.
\small
\begin{longtable}{@{}llrrrrrrrr@{}}
\caption{All paired configurations at $\varepsilon=1/2$. Each row aggregates seeds $17,29,43$. Entries are median representative counts; Cardinality rows ($n=10^5$) measure occupancy only. $o/h_X$ is the shared physical origin relative to the exchange width.}\label{tab:paired-all}\\
\toprule Family & $\rho$ & $n$ & $k$ & $o/h_X$ & PC & PK & I & X & H\\\midrule\endfirsthead
\toprule Family & $\rho$ & $n$ & $k$ & $o/h_X$ & PC & PK & I & X & H\\\midrule\endhead
lattice & $1/4096$ & 2000 & 5 & $0$ & 40 & 40 & 40 & 40 & 40\\
lattice & $1/1024$ & 2000 & 5 & $0$ & 40 & 40 & 40 & 40 & 40\\
lattice & $1/256$ & 2000 & 5 & $0$ & 40 & 40 & 40 & 43 & 40\\
irregular & $1/4096$ & 2000 & 5 & $0$ & 12 & 12 & 12 & 16 & 12\\
irregular & $1/1024$ & 2000 & 5 & $0$ & 18 & 18 & 16 & 40 & 16\\
irregular & $1/256$ & 2000 & 5 & $0$ & 40 & 40 & 24 & 229 & 24\\
rotated & $1/4096$ & 2000 & 5 & $0$ & 11 & 11 & 11 & 16 & 11\\
rotated & $1/1024$ & 2000 & 5 & $0$ & 15 & 15 & 13 & 43 & 13\\
rotated & $1/256$ & 2000 & 5 & $0$ & 43 & 43 & 25 & 237 & 25\\
unequal & $1/4096$ & 2000 & 5 & $0$ & 12 & 12 & 12 & 16 & 12\\
unequal & $1/1024$ & 2000 & 5 & $0$ & 18 & 18 & 14 & 38 & 14\\
unequal & $1/256$ & 2000 & 5 & $0$ & 38 & 38 & 23 & 208 & 23\\
elongated & $1/4096$ & 2000 & 5 & $0$ & 16 & 16 & 14 & 43 & 14\\
elongated & $1/1024$ & 2000 & 5 & $0$ & 43 & 43 & 26 & 179 & 26\\
elongated & $1/256$ & 2000 & 5 & $0$ & 177 & 177 & 77 & 1094 & 77\\
uniform & $1/4096$ & 2000 & 5 & $0$ & 1948 & 1948 & 1948 & 1998 & 1948\\
uniform & $1/1024$ & 2000 & 5 & $0$ & 1948 & 1948 & 1948 & 1998 & 1948\\
uniform & $1/256$ & 2000 & 5 & $0$ & 1948 & 1948 & 1948 & 1998 & 1948\\
lattice & $1/256$ & 100000 & 8 & $0$ & 115 & 115 & 64 & 115 & 64\\
irregular & $1/256$ & 100000 & 8 & $0$ & 143 & 143 & 64 & 399 & 64\\
lattice & $1/256$ & 100000 & 16 & $0$ & 595 & 595 & 198 & 595 & 198\\
irregular & $1/256$ & 100000 & 16 & $0$ & 793 & 793 & 286 & 793 & 286\\
lattice & $1/256$ & 100000 & 32 & $0$ & 4194 & 4194 & 340 & 1119 & 340\\
irregular & $1/256$ & 100000 & 32 & $0$ & 5077 & 5077 & 572 & 1582 & 572\\
lattice & $1/256$ & 100000 & 64 & $0$ & 31340 & 31340 & 2140 & 2140 & 2140\\
irregular & $1/256$ & 100000 & 64 & $0$ & 32955 & 32955 & 3118 & 9964 & 3118\\
lattice & $1/256$ & 100000 & 128 & $0$ & 83251 & 51398 & 16452 & 4201 & 4201\\
irregular & $1/256$ & 100000 & 128 & $0$ & 83527 & 52989 & 19166 & 19166 & 19166\\
lattice & $1/256$ & 100000 & 256 & $0$ & 97657 & 91133 & 31349 & 8288 & 8288\\
irregular & $1/256$ & 100000 & 256 & $0$ & 97692 & 91294 & 34796 & 34796 & 34796\\
lattice & $1/256$ & 100000 & 512 & $0$ & 99704 & 95533 & 83390 & 16458 & 16458\\
irregular & $1/256$ & 100000 & 512 & $0$ & 99701 & 95511 & 84091 & 54751 & 54751\\
irregular & $1/1024$ & 2000 & 5 & $1/4$ & 19 & 19 & 14 & 38 & 14\\
irregular & $1/1024$ & 2000 & 5 & $1/2$ & 16 & 16 & 14 & 38 & 14\\
irregular & $1/1024$ & 2000 & 5 & $3/4$ & 15 & 15 & 13 & 40 & 13\\
irregular & $1/32$ & 24 & 4 & $0$ & 23 & 23 & 23 & 24 & 23\\
elongated & $1/32$ & 24 & 4 & $0$ & 24 & 24 & 24 & 24 & 24\\
uniform & $1/32$ & 24 & 4 & $0$ & 24 & 24 & 24 & 24 & 24\\
\bottomrule\end{longtable}\normalsize

\subsection{Varying cardinality on fixed inputs}\label{app:fixed-input}
Does a coarser exchange grid at large $k$ translate into a completed solve? We used six fixed
arrays and varied only $k$ within each. The first panel samples 2,048 points in 24 tight
irregular groups (half-width $1/4096$); the second samples 1,024 points around 256 lattice
centers (half-width $1/256$). Each panel uses seeds 17, 29 and 43, accuracy $1/2$, the common
physical origin $(0,0)$ and the same deterministic first-index representative. Coordinates
and their order stay fixed. For each of the 57 array/cardinality pairs we built the incumbent
(I), exchange (X) and hybrid (H) grids, rotating method order. The primary search required
$M\le160$ and $\binom Mk\le10^7$, with a separate five-second wall-clock budget per admitted
solve, including reduced pair-weight construction and excluding worker startup and grid building.

\begin{table}[htbp]\centering\small
\caption{Fixed-input cardinality study at $\varepsilon=1/2$. Each $M$ is the median over three
seeds. The last column in each panel counts completed primary searches out of three for I/X/H.
A dash denotes an untested cardinality. A size exclusion gives an occupancy count but no
reduced optimum. All admitted primary searches completed.}\label{tab:fixed-input}
\begin{tabular}{@{}r rrrc@{\qquad}rrrc@{}}\toprule
&\multicolumn{4}{c}{24 irregular groups ($n=2048$)}&
\multicolumn{4}{c}{256 lattice groups ($n=1024$)}\\
\cmidrule(lr){2-5}\cmidrule(lr){6-9}
$k$&$M_I$&$M_X$&$M_H$&completed&$M_I$&$M_X$&$M_H$&completed\\\midrule
4 & 25 & 31 & 25 & 3/3/3 & 698 & 698 & 698 & 0/0/0\\
5 & 27 & 31 & 27 & 3/3/3 & \multicolumn{4}{c}{---}\\
8 & 27 & 39 & 27 & 3/0/3 & 698 & 698 & 698 & 0/0/0\\
10 & 27 & 39 & 27 & 2/0/2 & \multicolumn{4}{c}{---}\\
11 & 27 & 39 & 27 & 1/0/1 & \multicolumn{4}{c}{---}\\
12 & 27 & 39 & 27 & 1/0/1 & 698 & 698 & 698 & 0/0/0\\
13 & 31 & 2048 & 31 & 0/0/0 & 698 & 698 & 698 & 0/0/0\\
16 & 31 & 2048 & 31 & 0/0/0 & 698 & 698 & 698 & 0/0/0\\
32 & \multicolumn{4}{c}{---} & 698 & 698 & 698 & 0/0/0\\
64 & \multicolumn{4}{c}{---} & 933 & 933 & 933 & 0/0/0\\
96 & \multicolumn{4}{c}{---} & 933 & 933 & 933 & 0/0/0\\
128 & \multicolumn{4}{c}{---} & 1004 & 933 & 933 & 0/0/0\\
192 & \multicolumn{4}{c}{---} & 1004 & 1024 & 1004 & 0/0/0\\
256 & \multicolumn{4}{c}{---} & 1020 & 1024 & 1020 & 0/0/0\\
\bottomrule\end{tabular}
\end{table}

The only strict exchange improvements in this fixed-input study occurred on all three lattice
arrays at $k=128$. In seed order the counts $(M_I,M_X)$ were $(1016,958)$, $(998,933)$ and
$(1004,924)$. To test pruning directly, we launched each I/X/H solve for that cardinality
without the two size gates. All nine reached the five-second limit. This says what the
specified implementation accomplished under that budget; it is not a lower bound on search
time. These experiments do not certify an optimum when search fails.

From \path{anc/benchmarks/}, run \verb|python code/fixed_inputs.py --check| to regenerate
the six arrays, compare them with their frozen integer coordinates, rebuild every grid and
verify exact energies of the returned subsets. Use \verb|--summary| for the recorded table or
\verb|--run| to repeat the timed experiment in a fresh copy. The point and run records are
\path{results/fixed-input-points.json} and \path{results/fixed-input-runs.json}; they include
input and representative hashes, certificate values, grid widths, search sizes and statuses.

\section{Continuous numeral shaped domains: algorithms and certificates}\label{app:numerals}
Figure~\ref{fig:numerals} uses ten frozen rational triangle unions traced from
printed numerals. Each domain admits arbitrary feasible locations, rather than
a supplied finite set of candidate sites. The six includes a completed outline;
its certificate concerns that completed polygon. The ten is disconnected. Holes
are excluded by the triangle lists. We use $s=2$ and $\varepsilon=1$ throughout.
The one-site energy is zero, and its displayed site is a feasible area centroid.
Triangle vertices have rational coordinates. The five and eight use three
working units per original image pixel, so their absolute energies scale by
nine on conversion back to image units; all reported ratios are invariant.

For $k\ge2$, a continuous farthest-first traversal clips triangles by the centres'
Voronoi regions; squared distance attains its maximum at a vertex of each clipped
polygon. If $r$ is the last insertion radius and $U_0$ is the energy of the
first $k$ centres, set $B=\min\{U_0,16H_{2k}/r^2\}$. Choose a dyadic
$\widehat\sigma$ with $\widehat\sigma\le B^{-1/2}<2\widehat\sigma$ and use
$h=\widehat\sigma/16$. For each half-open cell, clip the first intersecting
triangle and retain the mean of the clipped polygon's vertices when it belongs
to that cell. Boundary ownership is checked explicitly. Feasible incumbents may
replace these witnesses. In particular, the two-site diameter endpoints and the
four-site placement from the independent triangle-product calculation are
installed in their cells. The four-site example thus illustrates certification
of a supplied good placement, rather than its independent discovery by the EPAS.

Rational branch and bound solves the full reference set for $k=2,\ldots,8$.
For $9$ and $10$, the reference sets contain 40,021 and 41,475 witnesses, but only
64 coarser-grid witnesses are searched. Those two runs do not claim an exact
reference-grid solve: an independent lower bound $L$ on the continuous optimum
certifies the reduced solution $V\le2L$. Feasible coordinate moves subsequently
decrease energy to $U$. Finite candidate pools contain the current point, and
each accepted move strictly lowers the exact rational energy. Termination of
these pools does not certify local or global optimality over the continuum.
For the coarser searches, refinement first uses the chosen coarse cells and then
the occupied reference cells. Pools include the current point, retained fine-grid
witnesses, clipped-polygon vertices and rational edge samples, as specified by
the code. Ties keep the current point, and sweeps stop when no move is accepted.

The native subset search encloses rational coordinates in integer boxes. Maximum
squared distances between boxes give integer lower bounds on pair costs; a branch
is pruned only when its completion bound exceeds the incumbent threshold. Python
fractions compare the surviving subsets exactly. The search record checks that
excluded and surviving subsets sum to $\binom{M}{k}$. Arithmetic ranges and
128-bit counting limits are checked before the C++17 search. The Python drivers
use a CPU-based C++17/OpenMP search. The workstation is a Lenovo Legion~5
15ACH6H with an AMD Ryzen~7~5800H (eight cores, sixteen logical processors),
16\,GB RAM and Windows~11 Home. Each complete numeral run was observed to
finish within roughly five minutes, often substantially sooner. These were
informal wall-clock observations, not controlled measurements; the archived
records certify the solutions but do not contain per-run timings. We make no
timing comparison between methods.

\begin{table}[htbp]\centering\small
\caption{Numeral results. The certified excess is $100(U/L-1)$, rounded upwards;
refinement gain is $100(V-U)/V$. Counts distinguish the reference grid from the
set actually searched. Exactness for the first three domains has a separate
certificate. No positional-distance guarantee is asserted.}
\label{tab:numerals}
\begin{tabular}{@{}rrrrr@{}}\toprule
$k$ & Reference count & Searched count & Certified excess (\%) & Gain (\%)\\\midrule
1 & --- & --- & exact & 0 \\
2 & 236 & 236 & exact & 0.0000 \\
3 & 1,654 & 1,654 & exact & 1.0778 \\
4 & 5,979 & 5,979 & $\le 0.000101$ & 0.0000 \\
5 & 8,497 & 8,497 & $\le 10.02$ & 0.1531 \\
6 & 10,425 & 10,425 & $\le 5.97$ & 0.3119 \\
7 & 30,664 & 30,664 & $\le 11.78$ & 0.2794 \\
8 & 26,056 & 26,056 & $\le 10.24$ & 0.2706 \\
9 & 40,021 & 64 & $\le 9.31$ & 0.3137 \\
10 & 41,475 & 64 & $\le 11.31$ & 0.2637 \\
\bottomrule\end{tabular}
\end{table}

The two-site certificate is the polygonal diameter. For three sites, an interior
site can move normally away from the line through the other two, increasing both
distances. Thus a minimizer lies on the boundary. All 125,580 triples of boundary
segments (with repetition) are checked. For squared distance $q>0$ and reference
$r>0$, the tangent inequality $1/q\ge2/r-q/r^2$ gives a concave lower bound in
the site coordinates, minimized at segment-product vertices. Integer distance
bounds discard 125,539 triples; exact tangent bounds handle the remaining 41
and match the feasible energy. For four sites the same inequality bounds products
of triangles, with subdivision when necessary, giving $U/L\le10^6/(10^6-1)$.

For the other domains, independent continuous bounds use positive-definite radial
kernels $K(x,y)=(a/(a+\|x-y\|^2))^m$, with $a>0$ and integer $m\ge1$.
Positive definiteness follows by expressing $(a+z)^{-m}$ as a positive Laplace
mixture of Gaussian kernels. For a finite nonnegative reference measure $\mu$,
write $P_\mu(x)=\int K(x,y)\,d\mu(y)$ and
$Q_\mu=\iint K(x,y)\,d\mu(x)\,d\mu(y)$. Applying positive definiteness to
$\sum_{x\in S}\delta_x-\mu$ gives
\[
 \sum_{\{x,y\}\subset S}K(x,y)
 \ge k\min_{x\in D}P_\mu(x)-\tfrac12Q_\mu-\tfrac k2.
\]
A verified minorant $1/z\ge c+\sum_j\alpha_j(a_j/(a_j+z))^{m_j}$,
$\alpha_j\ge0$, therefore bounds energy below. An incumbent restricts relevant
optimum pair distances away from zero. Rational enclosures over squared-distance
bins and triangle subdivisions verify both the minorant and the potential minima.
The saved rational witnesses and checkers supply the numerical certificates;
the fitted weights alone would not constitute a proof.

The self-contained \path{numerals/} folder in \path{riesz-reproducibility-v2.zip} contains the frozen
triangle lists, exact sites, search records, lower-bound witnesses, solver code,
and gallery assets. From that folder, \verb|python verify_numerals.py| checks site
feasibility, exact energies and all reported continuous lower bounds. The README
also gives commands to rerun the discrete searches and regenerate the gallery.
The intervals in Table~\ref{tab:numerals} concern energy on the stored domains;
tracing error relative to a physical printed numeral is outside their scope.

\section{Certified lower bounds from proxies of bounded treewidth}\label{sec:structure-short}
We give the lower-bound construction for rational planar sites and inverse-square energy.

A radius $R>0$ can encode a modeling decision that sites farther apart are sufficiently separated.
Retain their inverse-square interaction only for pairs at distance at most $R$, and let $G_R$ be
the graph of those pairs. Given a tree decomposition of width $t$, storing the selected bag
vertices and the selected count gives an exact proxy optimum in $2^{O(t)}n(k+1)^2$ operations
after polynomial preprocessing. For completeness, use a nice decomposition with an introduce-edge
node for
each retained interaction. A state $(X,j)$ fixes the selected vertices $X$ in
the bag and counts $j$ selected vertices already forgotten. Introducing a vertex
branches on its membership; introducing an edge charges its weight if both
endpoints are in $X$. On forgetting a selected vertex increment $j$. At joins
use matching $X$ and convolve the two forgotten counts: the charged edges
and forgotten vertices of the children are disjoint. At an empty root read
the minimum at $(\varnothing,k)$. For a proxy minimizer $S_R$, its proxy value is a lower bound on
the full optimum and $E_2(S_R)$ an upper bound. Their difference is exactly its omitted tail, at
most $\binom{k}{2}R^{-2}$. This certifies additive error even when the proxy value is zero; a
useful relative guarantee needs a positive lower bound large enough compared with that tail.

To strengthen the lower endpoint in Section~\ref{sec:experiments}, take distance-bin endpoints
$0<R_1<\cdots<R_m=D$ with $D$ at least the input diameter. A pair in bin $(R_{j-1},R_j]$ ($R_0=0$)
receives weight $R_j^{-2}$, at most its true cost. Subtract $D^{-2}$ from each pair weight and
form a graph of the strictly positive residuals. The last bin creates no edges. Adding
$\binom{k}{2}D^{-2}$ to the exact minimum residual cost gives a lower bound $B$ on the original
optimum. The same treewidth dynamic program applies with the supplied decomposition; sharper bins
may increase its width. No such proxy was needed for the independently verified Swiss optimum.

\section{Reproducing the computations}\label{app:reproduce}
The \href{https://github.com/emmerichmtm/PCGA-Riesz/releases/tag/arxiv-v2}{versioned archive} \texttt{riesz-reproducibility-v2.zip} contains the implementation, Swiss site coordinates, exact solution records, benchmark generators and recorded runs. Its self-contained \path{numerals/} folder supplies the continuous-domain study in Appendix~\ref{app:numerals}. Ordered benchmark arrays are regenerated from the bundled seeds and generators; no secondary archive is required for replay. Synthetic coordinates are integer numerators divided by $2^{20}$; the Swiss CSV gives integer metres in LV95 (EPSG:2056). The archive also includes the manuscript source under \path{paper/} and SHA-256 checksums.

With Python 3.12 and NumPy 1.26.4, run \verb|python anc/reproduce.py --swiss|
from the extracted archive root.
The offline driver uses a temporary copy. It rebuilds benchmark summaries, checks
all 171 fixed-input grids and returned energies, and repeats the small examples.
The Swiss audit recomputes both reduced optima, the full-input optimum and eight
hybrid refinement moves. Allow about 1~GB of free memory; omit \texttt{--swiss}
for the shorter replay. For the numeral certificates, run
\verb|python verify_numerals.py| from \path{numerals/}.

Appendix paths refer to the archive root. Study READMEs give seeds, budgets and
timing commands; run benchmarks sequentially. Exact scores and cell membership
are deterministic, but timings and timeouts depend on the machine.

In \path{paper/}, run pdf\LaTeX{}, Bib\TeX{}, then pdf\LaTeX{} twice on the
job \texttt{main\_arxiv\_v2}. Keep all supplied files together. The PRIMEarxiv and
\texttt{plainurl} styles are unmodified; the bibliography is in
\path{references.bib}. Apart from the numeral gallery, figures are inline.
The root README gives exact build commands.

\end{document}